\documentclass[twoside]{articlek}

\usepackage[normal]{caption}

\renewcommand{\refname}{REFERENCES}

\def\BibTeX{{\rm B\kern-.05em{\sc i\kern-.025em b}\kern-.08em
		T\kern-.1667em\lower.7ex\hbox{E}\kern-.125emX}}

\usepackage{graphicx}

\renewcommand{\refname}{REFERENCES}

\def\BibTeX{{\rm B\kern-.05em{\sc i\kern-.025em b}\kern-.08em
		T\kern-.1667em\lower.7ex\hbox{E}\kern-.125emX}}

\usepackage{graphicx}

\begin{document}
%
%\title{}
% subtitle is optionnal
%
%%%\subtitle{Do you have a subtitle?\\ If so, write it here}

\begin{flushleft}
	{\large\bf Aspects on Identification, Decay Properties and Nuclear Structure of the Heaviest Nuclei
		}
	\vspace*{25pt}

	{\bf Fritz Peter He\ss berger$^{1,2,}\footnote{E-mail: \texttt{f.p.hessberger@gsi.de}}$} \\
	\vspace{5pt}
	{$^1$GSI - Helmholtzzentrum f\"ur Schwerionenforschung GmbH, Planckstra\ss e 1, 64291 Darmstadt, Germany\\
		$^2$Helmholtz Institut Mainz, Staudingerweg 18, 55128 Mainz, Germany}\\

\end{flushleft}
          
Version: April, 27, 2022\\

\begin{abstract}\noindent
Synthesis of new elements at the upper border of the charts of nuclei and investigation of their decay properties and nuclear structure
has been one of the main research topics in low energy nuclear physics since more than five decades. Main items are the quest for the
heaviest nuclei that can exist and the verification of the theoretical predicted spherical proton and neutron shells at Z\,=\,114, 120 or 126
and N\,=\,172 or 184.\\
The scope of the present paper is to illustrate some technical and physical aspects in investigation of the heaviest nuclei 
('superheavy nuclei') and to critical
discuss some selected results, which from a strict scientific point of view are not completely clear so far, making partly also suggestions
for alternative interpretations.\\
A complete review of the whole field of superheavy element research, however, is out of the scope of this paper.
\end{abstract}
% \maketitle
%
\vspace{10 mm}
\section{1. Introduction}
\label{intro}
First extensions of the nuclear shell model \cite{Goep48,Haxel49} into regions far beyond the heaviest known doubly magic nucleus,
$^{208}$Pb ({\it{Z}} = 82, {\it{N}} = 126), performed more than fifty years ago lead to the prediction of spherical proton and neutron
shells at {\it{Z}} = 114 and {\it{N}} = 184 \cite{Sobi66,Meld67}. Nuclei in the vicinity of the crossing of both shells were 
expected to be extremely stabilized against spontaneous fission by fission barriers up to about 10 MeV. 
Particulary for the doubly magic nucleus a fission barrier of 9.6 MeV and hence a partial fission half-life of 10$^{16}$ years
\cite{Nils69}, in a preceding study even 2$\times$10$^{19}$ years \cite{Nils68}, were expected. In an allegorical picture these
nuclei were regarded to form an island of stability, separated from the peninsula of known nuclei (the heaviest, safely identified element at that time was lawrencium
(Z\,=\,103)) 
by a sea of instability and soon were denoted as 'superheavy' (see {\it{e.g}} \cite{Fric70}).
The theoretical predictions initiated tremendous efforts from experimental side to produce these superheavy nuclei and to investigate their decay properties as well as 
their nuclear and atomic structure and their chemical properties. The major and so far only successful method to synthesize transactinide elements ({\it{Z}} $>$ 103) were complete fusion reactions.\\
These efforts were accompanied by pioneering technical developments a) of accelerators and ion sources to deliver stable heavy ion beams of high intensity, b) of targets
being able to stand the high beam intensities for long irradiation times ($>$ several weeks), c) for fast and efficient separation of products from complete fusion reactions
from the primary beam and products from nuclear reactions others than complete fusion, d) of detector systems to measure the different decay modes ($\alpha$ - decay, EC - decay, spontaneous 
fission and acompanying $\gamma$ radiation and conversion electrons), e) of data analysis techniques, and f) for modelling measured particle spectra by advanced simulations, {\it{e.g.}} GEANT4 \cite{Agost03}.
Despite all efforts it took more than thirty years until the first serious results on the production of elements {\it{Z}} $\ge$ 114
(flerovium) were reported \cite{Ogan99,Ogan99a}. However, these first results could not be reproduced independently and are still ambiguous \cite{Hess13}.\\
Nevertheless, during the past twenty years synthesis of elements {\it{Z}} = 113 to {\it{Z}} = 118 has been reported and their discovery was approved 
by the International Union for Pure and Applied Chemistry (IUPAC) \cite{BarK11,KarB16,KarB16a}. The decay data reported for the isotopes of 
elements {\it{Z}} $\ge$ 113 that have been claimed to be identified indicate the existence of a region of shell stabilized nuclei towards {\it{N}} = 184, 
but the center has not been reached so far. Data on the strength of the possible shells is still scarce.\\
Tremendous efforts have also been undertaken from the theoretical side to make predictions on stability ('shell effects'), fission barriers, Q$_{\alpha}$ - values, decay modes, 
halflives, spin and parity of the ground-state as well as of low lying excited states, {\it{etc.}}. For about thirty years the calculations were performed using 
macroscopic-microscopic approaches based on the nuclear drop model \cite{Weiz35} and the Strutinsky shell correction method \cite{Struti67}. Although predicted
shell correction energies ('shell effects') 
disagreed considerably the models agreed in {\it{Z}}\,=\,114 and {\it{N}}\,=\,184 as proton and neutron shell closures (see {\it{e.g.}} \cite{Smolan95,Moller95}). 
The situation changed by the end of the 1990ties when for the first time results 
using self-consistent models like Skyrme-Hartree-Fock-Bogoliubov (SHFB) calculations or relativistic mean-field models (RMF) 
were published \cite{Rutz97,Ben03}.
Most of the calculations predict {\it{Z}}\,=\,120 as proton shell closure, while others predict {\it{Z}}\,=\,114 (SkI4) or {\it{Z}}\,=\,126 (SkP, SkM*). Skyrme force based calculations agree in {\it{N}}\,=\,184 as neutron shell closure, while the RMF calculations favour {\it{N}}\,=\,172. As a common feature all these parametrizations and also the macroscopic - microscopic calculations result in a wide area of high shell effects.
That behavior is different to that at known shell closures, {\it{e.g.}} {\it{Z}}\,=\,50, {\it{N}}\,=\,50,\,82, where the region of high shell effects (or high 2p\,- and 2n\,- separation energies) is strongly localized. It is thus not evident if the concept of nuclear shells as known from the lighter nuclei is still reasonable in the region of superheavy nuclei. It might be wiser to speak of regions of high shell stabilization instead.
On the other hand, it has been already  discussed extensively by Bender et al.\cite{Bend99}
that the proton number {\it{Z}} and the neutron number {\it{N}}, where the shell closure occurs strongly depend on details in the description of the underlying forces, specifically on the values for the effective masses {\it{m$^{*}$}} and the strength of the spin - orbit interaction. It also 
has been emphasized in \cite{Bend99} that the energy gap between the spin - orbit partners
2f$_{5/2}$ and 2f$_{7/2}$ determines whether the proton shell occurs at Z\,=\,114 or Z\,=\,120.
Under these circumstances predictions of shell closures at different proton ({\it{Z}}) and/or neutron ({\it{N}}) numbers by different models may be regarded rather as a feature of 'fine tuning' of the models than as a principle disagreement. 
Having this in mind superheavy elements represent an ideal laboratory for investigation of the nuclear ('strong') force.
More detailed knowledge of properties and structure of superheavy heavy nuclei is thus undoubtedly decisive for 
deeper understanding of basic interactions. Therefore investigations of decay properties and structure of superheavy nuclei will become in future even more important than synthesis of new elements. \\  
One has, however, to keep in mind, that the theoretically predicted high density of nuclear levels in a narrow energy interval above the ground-state may lead to complex $\alpha$-decay patterns, while on the other hand often only little numbers of decay events are observed. Therefore it is tempting to take the average of the measured decay data, which finally results to assign the
measured data as due to one transition in case of the $\alpha$-decay energies, or due to originating from one nuclear level in the case of life-times. Thus fine structure in the $\alpha$ decay or the existence of isomeric levels might be overseen. Rather, a critical analysis and assessment of the measured data is required.\\
As already indicated above the expression 'superheavy nuclei' or 'superheavy elements' had been originally suggested for the nuclei in the vicinity of the crossing of the spherical proton and neutron shells at Z\,=\,114 and N\,=\,184. The establishment of deformed proton and neutron shells at Z\,=\,108 and N\,=\,162 \cite{Cwiok83,Moller86,Patyk91,Patyk91a} resulted in the existence of a ridge between the 'peninsula' of known nuclei and the 'island of stability'. Thus it became common to denote all purely shell stabilized nuclei as 'superheavy', 
{\it{i.e.}} nuclei with liquid drop fission barriers lower than the zero - point motion energy ($<$0.5 MeV). \\
The region of superheavy nuclei that shall be treated in this review is shown in fig. 1.

\begin{figure*}
%\begin{figure*}
\resizebox{1.0\textwidth}{!}{%
  \includegraphics{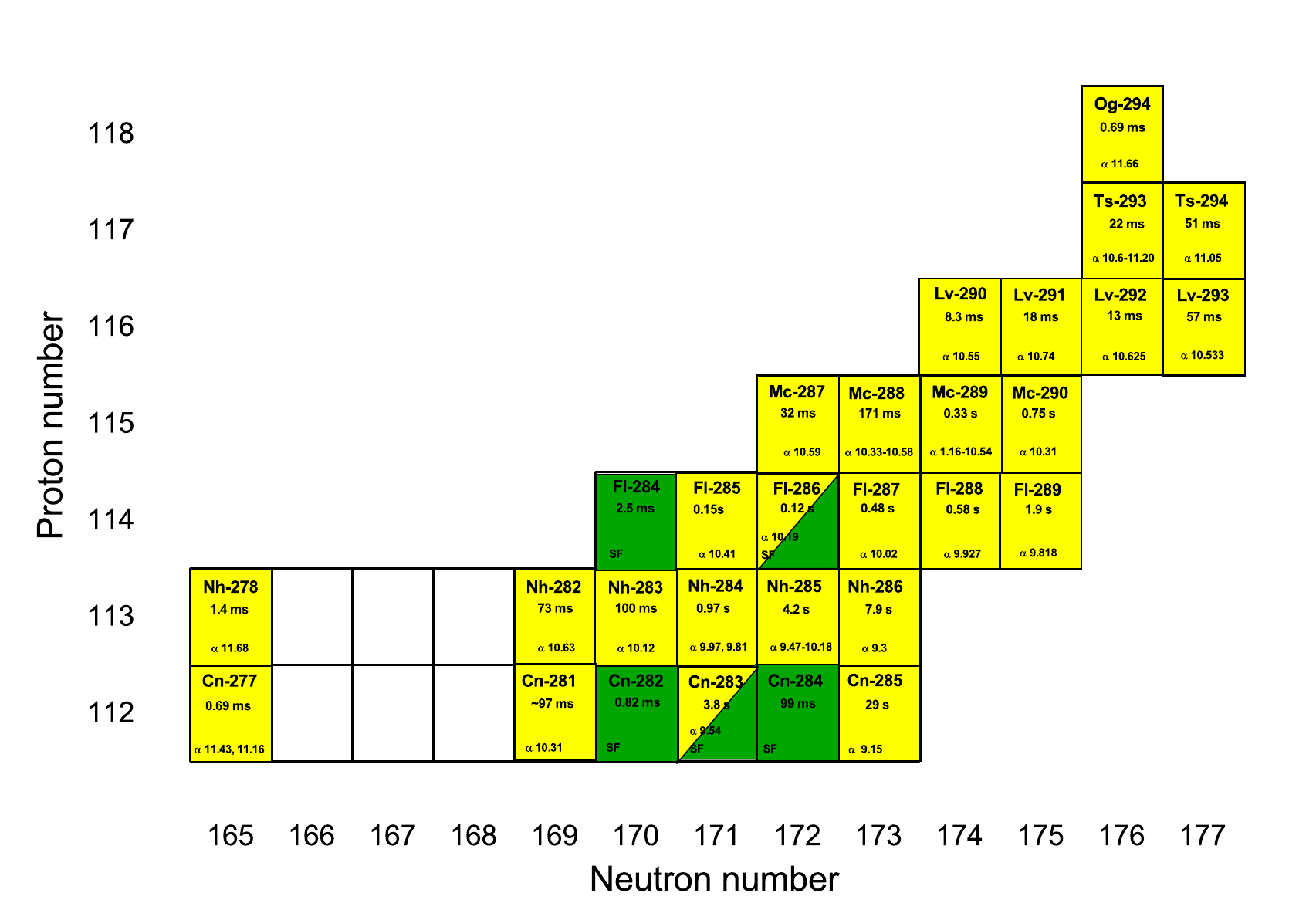}
}
% If not, use
%\vspace{5cm}       % Give the correct figure height in cm
\caption{Excerpt from the charts of nuclei for the region {\it{Z}}\,$\ge$\,112.}
\label{fig:1}       % Give a unique label
\end{figure*}

\section{2. Experimental Approach}

Complete fusion reactions of suited projectile and target nuclei has been so far the only successful method to produce nuclei with atomic numbers Z $>$ 103 (see e.g. \cite{Mun16,Oga16}). 
Under these considerations a separation method has been developped to take into account the specific features of this type of nuclear reactions.
Due to momentum conservation the velocity of the fusion product, in the following denoted as 'compund nucleus' (CN)\footnote{it is common to denote the primary fusion products which represent in mass and atomic number the sum of projectile and target as 'compound nucleus' (CN) which is highly excited.The final product after deexcitation by prompt emission of nucleons and/or $\alpha$ particles is denoted as 'evaporation residue' (ER).}    
can be written as\\
\\
                    v$_{CN}$ = (m$_{p}$ / (m$_{p}$ + m$_{t}$)) x v$_{p}$ \\
\\
where {\it{m$_{p}$}}, {\it{m$_{t}$}} denote the masses of the projectile and target nucleus, and {\it{v$_{p}$}} the velocity of the projectile. This simply means that a) fusion products are
emitted in beam direction (with an angular distribution around zero degree determined by particle emission from the highly excited CN and by scattering in the target foil)
and b) that CN are slower than the projectiles. It seemed therefore straightforward to use the velocity difference for separation of the fusion products from the projectiles 
and products from nuclear reactions others than complete fusion. Such a method has the further advantage of being fast, as the separation is performed in-flight 
without necessity to stop the products. So separation time is determined by the flight-time through the separation device. In the region of transactinide nuclei this
separation technique has been applied for the first time at the velocity filter SHIP at GSI, Darmstadt (Germany) \cite{MuF79} for investigation of evaporation residue production in the 
reactions $^{50}$Ti + $^{207,208}$Pb, $^{209}$Bi \cite{Hess85,Hess85a} and for the identification of element Z = 107 (bohrium) 
in the reaction $^{54}$Cr + $^{209}$Bi \cite{Muenz81}. Separation times in these cases were in the order of 2 $\mu$s.\\
As an alternative separation technique gas-filled separators have been developped using the different magnetic rigidities B$\rho$ of fusion products and projectiles 
as the basis for separation. Early devices used in investigation of tranfermium nuclei were SASSY \cite{GhY88} and SASSY II \cite{Ghi88} at LNBL Berkeley, USA and
HECK \cite{NiA95} at GSI.\\
Due to their simpler conception and more compact construction, which allows for separation times below 1 $\mu$s gas-filled separators meanwhile have 
become a wide spread tool for investigation of heaviest nuclei and are operated in many laboratories, e.g. RITU (University of Jyv\"askyl\"a, Finland) \cite{LeA95},
BGS (LNBL Berkeley, USA) \cite{NiG98}, DGFRS (JINR, Dubna, Russia) \cite{LaL93}, GARIS (RIKEN, Wako, Japan) \cite{MoY92}, SHANS (IMP, Lanzhou, China) \cite{Zhang13}, 
TASCA (GSI, Darmstadt, Germany) \cite{SeB08}.\\
Separation, however, is only one side of the medal. The fusion products have also to be identified safely. Having in mind that the essential decay modes of superheavy
nuclei are $\alpha$ decay and spontaneous fission (SF) detection methods suited for these items have been developped. After it was shown that suppression of the
projectile beam by SHIP was high enough to use silicon detectors \cite{Schmidt78} an advanced detection set-up for investigation of heaviest nuclei was built \cite{Hofm84}.
It consisted of an array of seven position-sensitive silicon detectors ('stop detector'), suited for $\alpha$ spectroscopy and registration of heavy particles (fission products, 
evaporation residues (ER), scattered projectiles etc.). To obtain a discrimination between particles passing the velocity filter and being stopped in the detector 
(ER, projectiles, products from few nucleon transfer) and radioactive decays ($\alpha$ decay, SF) and to obtain a separation of ER and scattered projectiles or 
transfer products a time-of-flight detector was placed in front of the stop detector \cite{Hess82}. Also the possibility to measure $\gamma$ rays emitted
in coincidence with $\alpha$ particles was considered by placing a Ge- detector behind the stop detector.
This kind of detection system has been improved in the course of the years at SHIP \cite{Hof95, Ack18} and was also adopted in modified versions and improved by other 
research groups in other laboratories; examples are GREAT \cite{Page03}, GABRIELA \cite{Haus06} and TASISpec \cite{Anders10}.\\
The improvements essentially comprise the following items:\\
a) the detector set-ups were upgraded by adding a box-shaped Si - dector arrangement, placed upstream and facing the 'stop detector',
allowing with high efficiency registration of $\alpha$ particles and fission 
products escaping the 'stop detector'. This was required as the ranges of $\alpha$ particles and fission fragments in silicon are larger than the range of ER; so about
half of the $\alpha$ particles and 30\,-\,50 $\%$ of the fission fragments will leave the 'stop detector' releasing only part of their kinetic energy in it.\\
b) the 'old' Si detectors where positions were determined by charge division were replaced by pixelized detectors allowing for a higher position resolution and thus 
longer correlation times.\\
c) effort was made to reduce the detector noise to have access to low energy particles (E\,$<$\,500 keV) like conversion electrons (CE).\\
d) digital electronics was introduced to enable dead time free data aquisition and to have access to short-lived acitivities with halflives 
lower than some microseconds. \\
e) detecor geometry and mechanical components were optimized to use several Ge detectors and to minimize scattering and absorption of $\gamma$ rays 
in the detector frames to increase the efficiency for $\gamma$ ray detection.\\
As it is not the scope of this work to present experimental techniques and set-ups in detail we refer for this item
to a recent review paper \cite{AckT17}.\\
Another technical aspect concerns the targets. As production cross-sections for heaviest elements are small, highest available beam currents
have to been used. Consequently a technology was desired to avoid destruction of the targets, which led to the development of rotating
target wheels \cite{Marx79}; performance of the wheels and target quality were continuously improved. \\
As an alternative method to produce superheavy nuclei (SHN) recently the idea of using multinucleon transfer reactions was
resumed, see {\it{e.g.}} \cite{Zag11,Zag13}. Indeed, intensive studies of those reactions with respect to SHN production, {\it{e.g.}}
$^{238}$U + $^{238}$U or $^{238}$U + $^{248}$Cm, had been performed at the UNILAC at GSI already  about forty years ago. A summary 
of these efforts is given in \cite{Kratz13}. Heaviest nuclides that could be identified in these experiments were isotopes of mendelevium ({\it{Z}}\,=\,101). A drawback of these studies, however, was the use of radiochemical methods which restricted 
isotope identification to those with 'long' halflives, T$_{1/2}$ $>$$>$ 1 min, giving no excess to short-lived nuclei in the
region Z\,$\le$102. To proceed in studying those reactions new types of separators are required, taking into account not
only short halflives down to the $\mu$s - range, but also a broad angular distribution of the reaction products. 
A more detailed discussion of this feature, however, is beyond the scope of this review.\\

\begin{figure}
	%\begin{figure*}
	\resizebox{0.9\textwidth}{!}{%
		\includegraphics{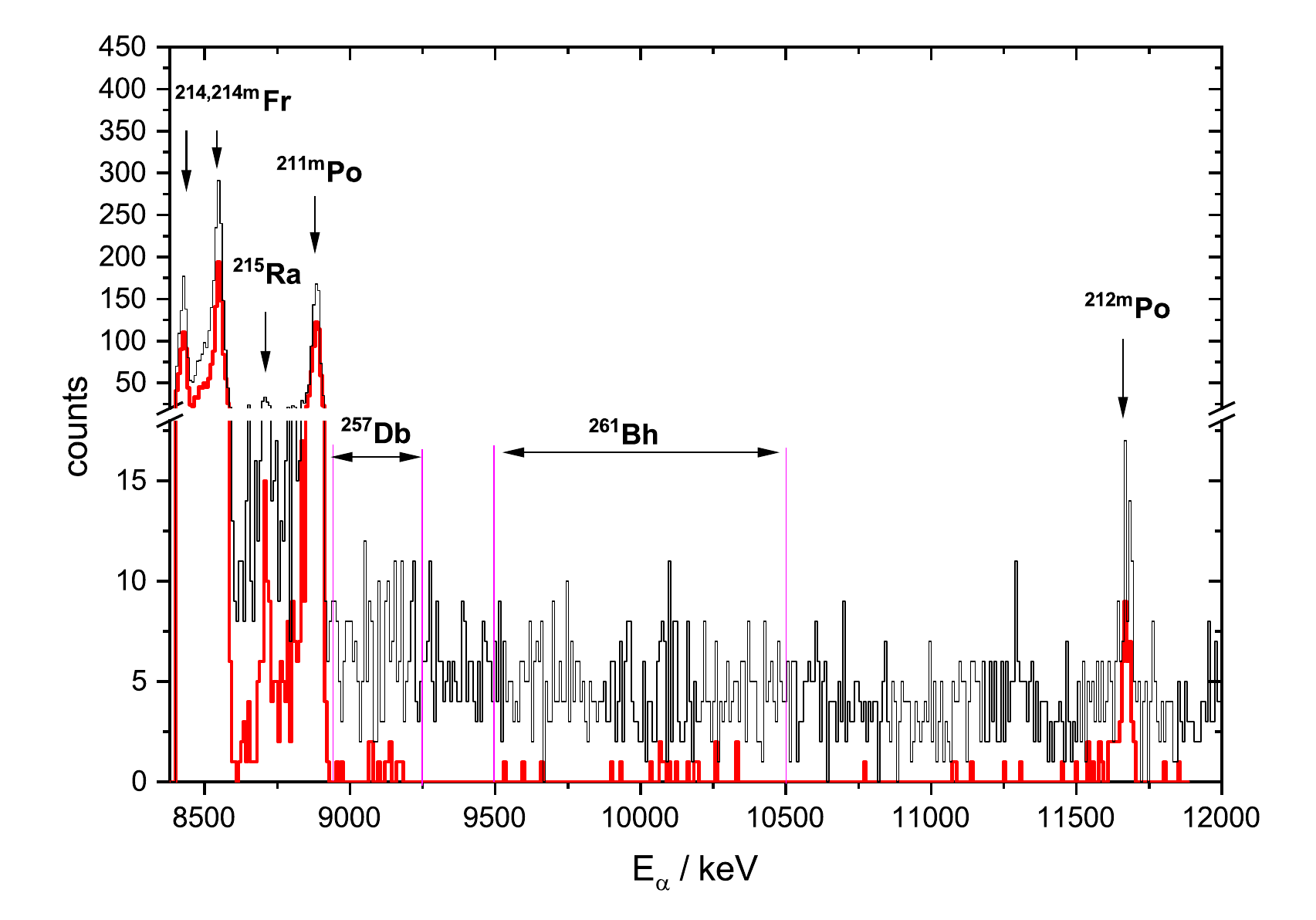}
	}
	% If not, use
	%\vspace{5cm}       % Give the correct figure height in cm
	\caption{Spectra of particles registered in an irradiation of $^{209}$Bi with $^{54}$Cr at SHIP \cite{HesA10a}; black line:
		particles registered in anticoincidence to the time-of-flight detectors, red line: particles registered during the beam-off period.}
	\label{fig:2}       % Give a unique label
\end{figure}
% ..
\begin{figure}
	\vspace{-0.8cm}
	%\begin{figure*}
	\resizebox{0.9\textwidth}{!}{%
		\includegraphics{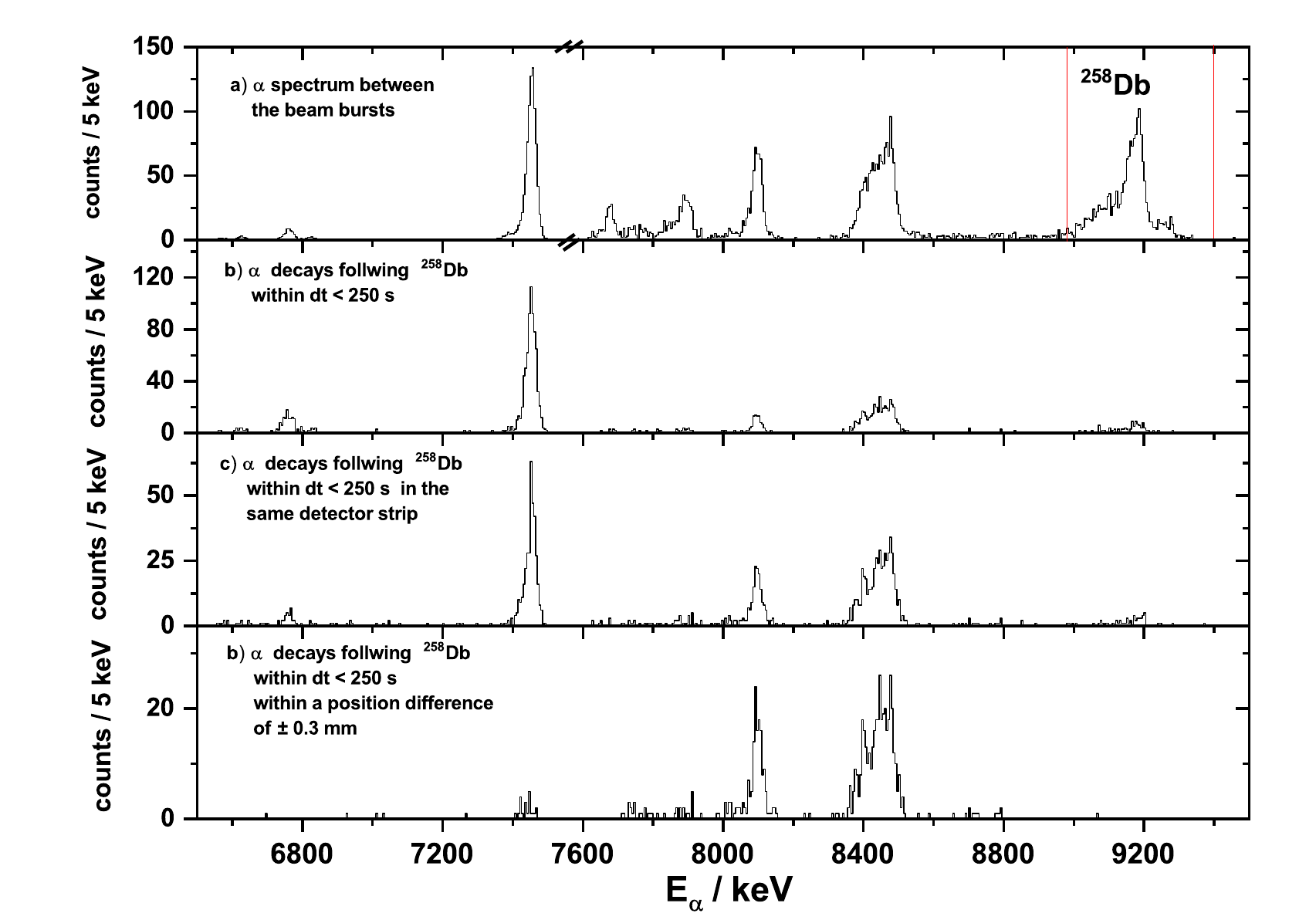}
	}
	% If not, use
       % Give the correct figure height in cm
	\caption{$\alpha$ - $\alpha$ correlation analysis of the radioactive decay chain starting from $^{258}$Db. a) Spectrum of
		$\alpha$ events observed in an irradiation of $^{209}$Bi with $^{50}$Ti at SHIP \cite{Vost15} between the beam bursts; the region below 7.6 MeV has been downscaled by a factor of five for better presentation; b) spectrum of the first $\alpha$ decays following the $\alpha$ decay of 
		$^{258}$Db within 250 s; c) same as b), but requiring that both $\alpha$ decays occur in the same detector strip; d) spectrum of
		correlated (daughter) $\alpha$ events, requiring a maximum position difference of $\pm$0.3 mm and a time difference $\Delta$t $<$ 250 s.}
	\label{fig:3}       % Give a unique label
\end{figure}

\vspace{10 mm}

\section{3. Data Selection}
Within the commonly used techniques particles passing the in-flight separator are implanted into a silicon-detector set-up.
As separation of the ER from 'unwanted' particles (scattered projectiles, scattered target nuclei, products from few nucleon
tansfer etc.) is not clean, there will be always a cocktail of particles registered, forming a background covering the energy
range of the $\alpha$ decays of the nuclei to be investigated, and usually also the energy range of the spontaneous fission
products. In cases of small production cross sections, typically $<$1$\mu$b, $\alpha$ decays of the ER are usually not visible in the particle spectra. Further cleaning procedures are required. An often applied procedure is the use of transmission detctors in front 
of the 'stop - detector' and requiring an anticoincidence between events registered in the stop detector and the transmission detector.
In practise the efficiency of the latter is never excatly 100 $\%$, therefore there will still be a residual background in the spectra.
In cases of a pulsed beam, one can restrict to the time intervals between the pulses. \\
An example is shown in fig. 2. where the $\alpha$ spectrum taken in an irradiation of $^{209}$Bi with $^{54}$Cr (271 MeV) at SHIP is 
presented \cite{HesA10a}.
The black line represents the particle spectrum taken in anti-coincidence with the time-of-flight (TOF) detectors; clearly products 
($^{211m,212m}$Po, $^{214,214m}$Fr, $^{215}$Ra) stemming from few nucleon transfer reactions are visible, the ER, $^{261}$Bh and its 
daughter product $^{257,257m}$Db, however, are buried under the background. In cases of pulsed beams a further purification is achieved by requiring a 'beam - off' condition. Thus the $\alpha$ decays of $^{261}$Bh, $^{257,257m}$Db become visible (red line in 
fig. 2). Such a restriction, however, is not desirable in many cases as it restricts identification to nuclei having liftimes in the order of the pulse-lengths or longer.\\
A possible way out is the use of genetic correlations between registered events; these may be correlations of the type ER - $\alpha$,
$\alpha$ - $\alpha$, ER - SF, $\alpha$ - SF etc.

\vspace{15 mm}
\section{4. Data Treatment}
\subsection{\bf{4.1 Genetic Correlations}}
To establish genetic relationships between mother and daughter $\alpha$ decays is presently a standard method to identify 
unknown isotopes or to assign individual decay energies to a certain nucleus.\\
Originally it was developped at applying the He - jet technique for stopping the reaction products and transport them to the detection system. 
As the reaction products were deposited on the surface of the detector, depending on the direction of emission 
of the $\alpha$ particle the latter could be either registered in the detector, but the residual nucleus was kicked off the detector 
by the recoil of the emitted $\alpha$ particle or the residual nucleus was shallowly implanted into the detector, while the $\alpha$ particle 
was emitted in opposite direction and did not hit the detector. To establish correlations sophisticated detector arrangements were required (see {\it{i.e.}} \cite{Ghiorso82}).
The technique of stopping the reaction products in silicon surface barrier detectors after in-flight separation from the projectile beam simplified the procedures considerably \cite{Schmidt78,Schmidt79}. 
Due to implantation into the detector by $\approx$(5\,-\,10) $\mu$m the residual nucleus was not kicked out of the detector by the recoil of the emitted $\alpha$ particle 
and therefore decays of the implanted nucleus and all daughter products occured in the same detector; 
so it was sufficient to establish chronological relationship between $\alpha$ events measured within the same detector \cite{Schmidt79}. 
The applicability of this method was limited by the decay rate in the detector, as the time sequence of decays became accidential if the search time for correlations exceeded the 
average time distance between two decays. The application of this technique was improved by using position sensitive silicon detectors \cite{Hofm84,Hofm79}. 
These detectors deliver the position of implantation as an additional parameter. The position resolution is typically around 300 $\mu$m (FWHM), 
while the range of $\alpha$ particles of (5\,-\,10) MeV is (40\,-\,80) $\mu$m \cite{North70} and the dislocation of the residual nucleus due to the $\alpha$ recoil is $<$1 $\mu$m. 
Thus all subsequent decays of a nucleus will occur at the same position (within the detector resolution). The probability to observe random correlations is reduced significantly 
by this procedure. \\
In these set-ups position signals were produced by charge division between an upper (top) and a lower (bottom) detector electrode (see {\it{e.g.}} \cite{Hofm12} for an advanced version of such a detector set-up).
In modern set-ups (see {\it{e.g.}} \cite{Anders10}) these position sensitve detectors have been replaced by pixeled detectors having vertical strips 
(a typical width is 1 mm) on one side and horizontal strips of the same width on the other side. The position is then given by the coordinates 
representing the numbers of the horizontal and vertical strips. One advantage of the pixeled detectors is a somewhat better position resolution;
taking strip widths of each 1 mm, one obtains a pixel size of 1 mm$^{2}$; for the SHIP - type detector \cite{Hofm12} (5 mm wide strips, position resolution 0.3 mm (FWHM)) 
taking in the analysis three times the FWHM one obtains an effective pixel size of
4.5 mm$^{2}$ (3 x 0.3 x 5 mm$^{2}$). More stringent, however, is the fact that the position resolution for a pixeled detector is given solely by 
the strip numbers and is thus independent of the energy deposit of the particle and of the range of the particle
(as long as it does not exceed the strip width). \\
In position sensitive detectors low energy particles ($\alpha$ particles escaping the detector, 
conversion electrons) deliver small signals often influenced by the detector noise and nonlinearities of the used amplifiers and ADCs, 
which significantly lowers the position resolution. In many cases signals are missing at all, as they are lower than the detection threshold. Another drawback is that at electron energies of around 300 keV the range in silicon becomes $\approx$300 $\mu$m and thus reaches the detector resolution, which then requires to enhance the position window for correlation search.\\
But also one drawback of the pixeled detectors should be at least mentioned: 
due to the small widths of the strips (typically 1 mm) already for a notable fraction of the implanted particles the energy
signal is split between two strips making sophisticated data analysis algorithms necessary to reconstruct the energy of the particles.
Also, the energy split between two strips also introduces some ambiguities in the determination of the position.\\

An illustrative example for the benefit of including the position into the correlation search is given in fig. 3. Here the $\alpha$ spectrum obtained in an irradiation of $^{209}$Bi with $^{50}$Ti at SHIP \cite{Vost15} (using the same set-up as in \cite{Hofm12})  between the beam bursts is shown in fig. 3a. Besides $^{258}$Db (9.0\,--\,9.4 MeV), produced in the reaction $^{209}$Bi($^{50}$Ti,n)$^{258}$Db, and its decay products
$^{254}$Lr (8.35\,--\,8.50 MeV) and $^{254}$No (8.1 MeV, EC decay daughter of $^{254}$Lr) also $\alpha$ lines from $^{212g,212m}$At (7.68,7.90 MeV and $^{211}$Po (7.45 MeV) are present; these activities were produced by few nucleon transfer reactions; in addition also the $\alpha$ line of $^{215}$Po (6.78 MeV) is visible, stemming from $\alpha$ decay of $^{223}$Ra (T$_{1/2}$ = 11.43 d), produced in a preceeding experiment. 
In fig. 3b the spectrum of the first $\alpha$ particles following an $\alpha$ decay of $^{258}$Db (energy region is marked by the red lines in fig. 3a is shown. Besides the daughter products $^{254}$Lr and $^{254}$No, strong random correlations with $^{211}$Po, $^{215}$Po, $^{258}$Db, $^{212g,212m}$At are observed; the random correlation can be significantly suppressed if in addition the occurence of both $\alpha$ decays in the same detector strip is required, as seen in fig. 3c; the result of the position correlation analysis finally is shown in fig. 3d. Here, in addition the occurence of both $\alpha$ events within a position difference of $\pm$0.3 mm is required. The background of $\alpha$ decays is completely gone, and also details 
in the energy distribution of the $\alpha$ events are visible; the $\alpha$ events at (7.7\,-\,7.9) MeV stem from decay of $^{250}$Md ($\alpha$ - decay daughter of $^{254}$Lr) and those at 7.45 MeV are here from $^{250}$Fm ($\alpha$-decay daughter of $^{254}$No, EC - decay daughter of $^{250}$Md). The events at (8.7\,-\,8.8) MeV are from decay of $^{253g,253m}$Lr, the $\alpha$ decay daughters of $^{257g,257m}$Db, which was produced to a small amount in the reaction $^{209}$Bi($^{50}$Ti,2n)$^{257}$Db.

\begin{figure}
%\begin{figure*}
\resizebox{0.9\textwidth}{!}{%
  \includegraphics{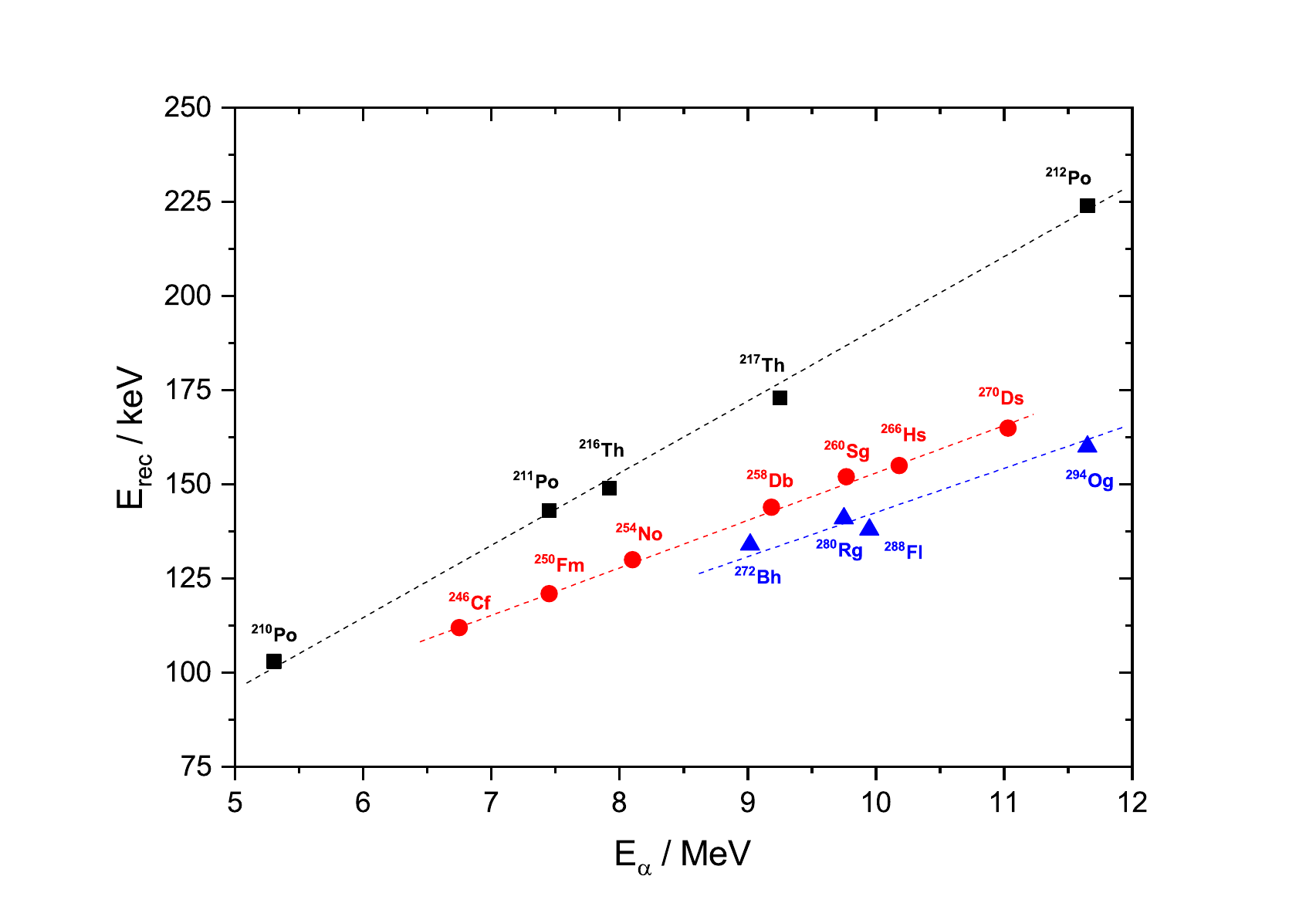}
}
% If not, use
%\vspace{5cm}       % Give the correct figure height in cm
\caption{Recoil energies transferred to the residual nuclei by $\alpha$ - decay of heavy nuclei. The lines are to guide the eye.}
\label{fig:4}       % Give a unique label
\end{figure}

\begin{figure}
	%\begin{figure*}
	\resizebox{0.9\textwidth}{!}{%
		\includegraphics{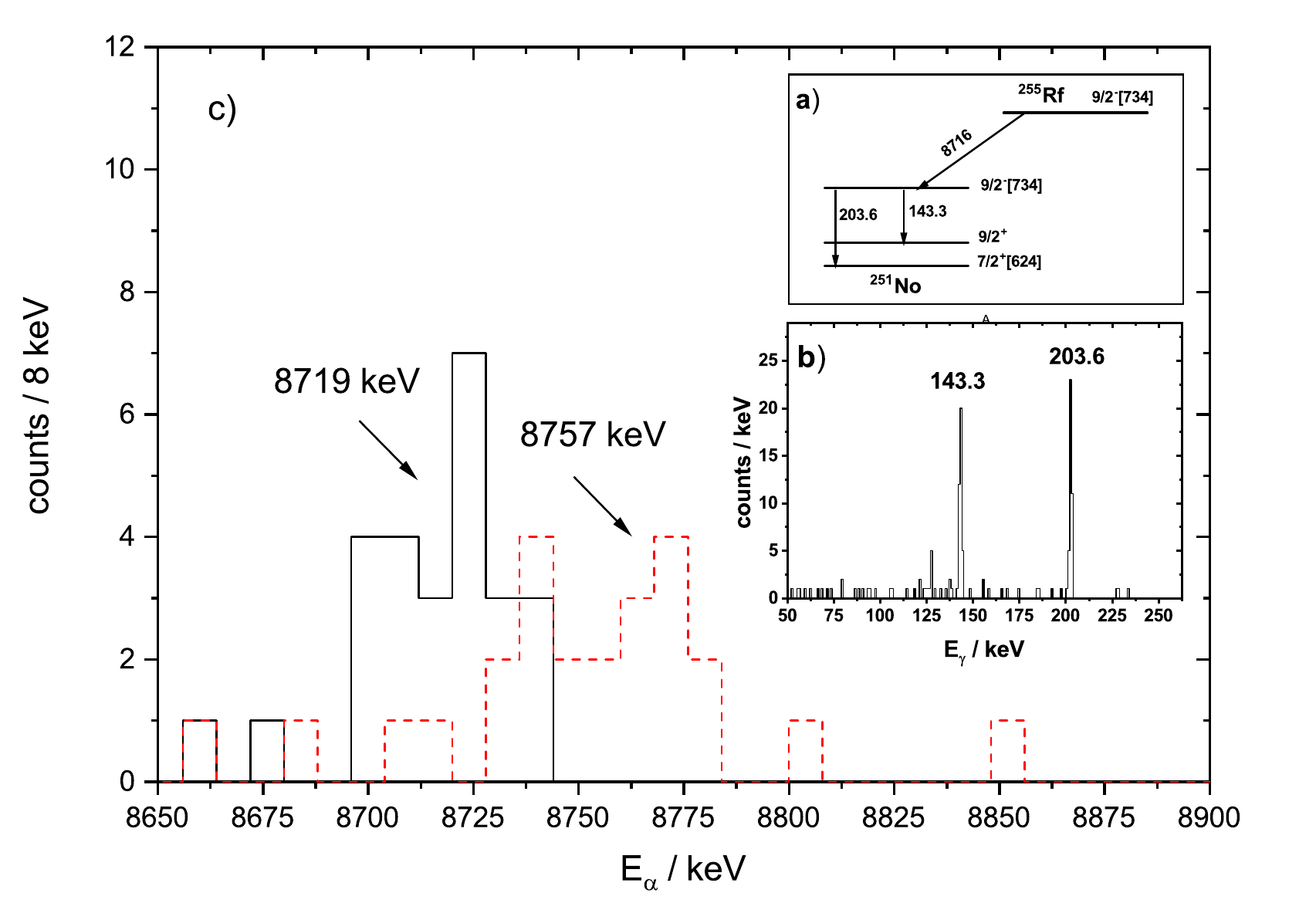}
	}
	% If not, use
	%\vspace{5cm}       % Give the correct figure height in cm
	\caption{Energy summing of $\alpha$ particles and conversion electrons (CE); a) decay scheme of $^{255}$Rf  \cite{Hessb06};
		 b) spectrum of $\gamma$ rays emitted in coincidence 
		with $\alpha$ decays of $^{255}$Rf; c) energy distribution of $\alpha$ particles in coincidence with the E\,=\, 203.6 keV (full line) or E\,=\,143.3 keV $\gamma$-line (dashed line).}
	\label{fig:5}       % Give a unique label
\end{figure}

\vspace{10 mm}
\subsection{\bf{4.2 Summing of $\alpha$-particle and Recoil - energies}}

Implantation of ER into a silicon detector has consquences for measuring the energies of $\alpha$ particles.
One item concerns summing of the $\alpha$ particle energy and the energy transferred by the $\alpha$ particle to the  
residual nucleus, which will be in the following denoted as recoil energy {\it{E$_{rec}$}}. \\
The total decay energy {\it{Q}} (for a ground-state to ground-state transition) is given by \\

Q = (m$_{mother}$ - m$_{daughter}$ - m$_{\alpha}$) $\times$ c$^{2}$ \\

This energy splits in two components \\

Q = E$_{\alpha}$ + E$_{rec}$  =  (1 + m$_{\alpha}$/m$_{daughter}$) $\times$ E$_{\alpha}$ \\

Here {\it{m$_{mother}$}}, {\it{m$_{daughter}$}} denote the masses of the mother and daughter nucleus\footnote{strictly spoken, the atomic mass, not the mass of a bare
	nucleus}, {\it{E$_{\alpha}$}} the kinetic energy of the 
the $\alpha$ particle.\\

Evidently the recoil energy {\it{E$_{rec}$ = (m$_{\alpha}$/m$_{daughter}$) $\times$ E$_{\alpha}$}} is stronly dependent on the mass
of the daughter nucleus and the kinetic energy of the $\alpha$ - particle.\\
This behavior is shown in fig. 4, where for some isotopes {\it{E$_{rec}$}} is plotted versus {\it{E$_{\alpha}$}}. The black squares represent 
the results for $^{210,211,212}$Po and $^{216,217}$Th, which are often produced in reactions using lead or bismuth targets by nucleon transfer or in 
so called 'calibration reactions' (reactions used to check the performance of the experimental set-up), the red dots are results for 'neutron deficient' isotopes in the range {\it{Z}}\,=\,(98-110), the blue triangles, finally, results for neutron rich SHN produced so far in irradiations of actinide targets with $^{48}$Ca. Evidently the recoil energies for the polonium and thorium isotopes
are by 15-30 keV higher than for the  {\it{Z}}\,=\,(98-110) - isotopes, while the differences between the latter and the
'neutron rich SHN' are typically in the order of 10 keV; specifically striking is the difference of {\it{$\Delta$E$_{rec}$}}\,=\,65 keV between $^{212}$Po and $^{294}$Og, both having nearly the same $\alpha$ - decay energy.\\
In practise, however, the differences are less severe:
the measured energy of the $\alpha$ particle is not simply the sum of both contributions as due to the high ionisation density
of the heavy recoil nucleus part of the created charge carriers will recombine and thus only a fraction of them will contribute to the
hight of the detector signal, hence\\

E$_{\alpha}$(measured) = E$_{\alpha}$ + a$\times$E$_{rec}$ \\

with a\,$<$\,1, giving the fraction of the contribution of the recoil energy, which can be considered to be in the
order of a\,$\approx$\,0.3 \cite{Eyal82}. One should, however, keep in mind, that this analysis was performed 
for nuclei around {\it{A}}\,=\,150. As ionization density increases for heavier nuclei (larger {\it{Z}}) the recombination 
might be larger for SHN, thus a\,$<$\,0.3. Nevertheless different recoil contributions should be considered when
calibrations are performed.\\
Further discussion of this item is found in \cite{Hofm12}.

\subsection{\bf{4.3 Summing of $\alpha$ particle and conversion electron (CE) energies}}

One more problem is connected with energy summing of $\alpha$ particles and conversion electrons (CE) in cases
where excited levels are populated decaying towards the ground state by internal conversion, leading to a shift of
the measured $\alpha$ energies towards higher values \cite{HessH87}. \\
An illustrative example is shown in fig. 5 where the decay of $^{255}$Rf is presented.
The decay scheme is shown in fig. 5a; $\alpha$ decay populates the 9/2$^{-}$[734] - level in $^{251}$No, which 
then decays by $\gamma$ emission either into the 7/2$^{+}$[624] ground-state (E$_{\gamma}$ = 203.6 keV ) or
into the 9/2$^{+}$ state (E$_{\gamma}$ = 143.3 keV (fig. 5b) \cite{Hessb06}. 
The M1 - transition 9/2$^{+}$ $\rightarrow$ 7/2$^{+}$ is highly converted. In fig 5c we present the energy distributions
of $\alpha$ particles either in coincidence with the E$_{\gamma}$ = 203.6 keV (black line) or the
E$_{\gamma}$ = 143.3 keV (red line). We observe a shift in the $\alpha$ energies by $\Delta$E\,=\,38 keV, which is even 
larger than the CE energy (31 keV)\cite{KibB07}, indicating that not only the CE contribute to the energy shift but also the 
energy released during deexcitation of the atomic shell (e.g. Auger electrons).

\subsection{\bf{4.4 $\alpha$  particles escaping the detector}}
As the implantation depth into the detector is typically $\leq$10 $\mu$m and thus considerably 
smaller than the range of $\alpha$ - particles in silicon ($>$50 $\mu$m at E\,$>$\,8 MeV) only part of them
(50 \,-\,60 $\%$) will be registered with full enery.\\
So if one observes on the basis of a small number of events, besides a 'bulk' at a mean energy {\it{E}} also $\alpha$ particles  with energies
of {\it{E\,-\,$\Delta$E}} will be registered. So it is a priori not possible to state if these events represent decays into higher lying daughter
levels or if they are just $\alpha$ particles of energy {\it{E}} escaping the detector with an energy {\it{E\,-\,$\Delta$E}}.
However, some arguments can be given on the basis of the probability to observe the latter events.
As an illustrative example  
the $\alpha$ spectrum of $^{253}$No \cite{Hess12} is given in fig. 6.\\

\begin{figure}
	%\begin{figure*}
	\vspace{-1cm}
	\resizebox{1.0\textwidth}{!}{%
		\includegraphics{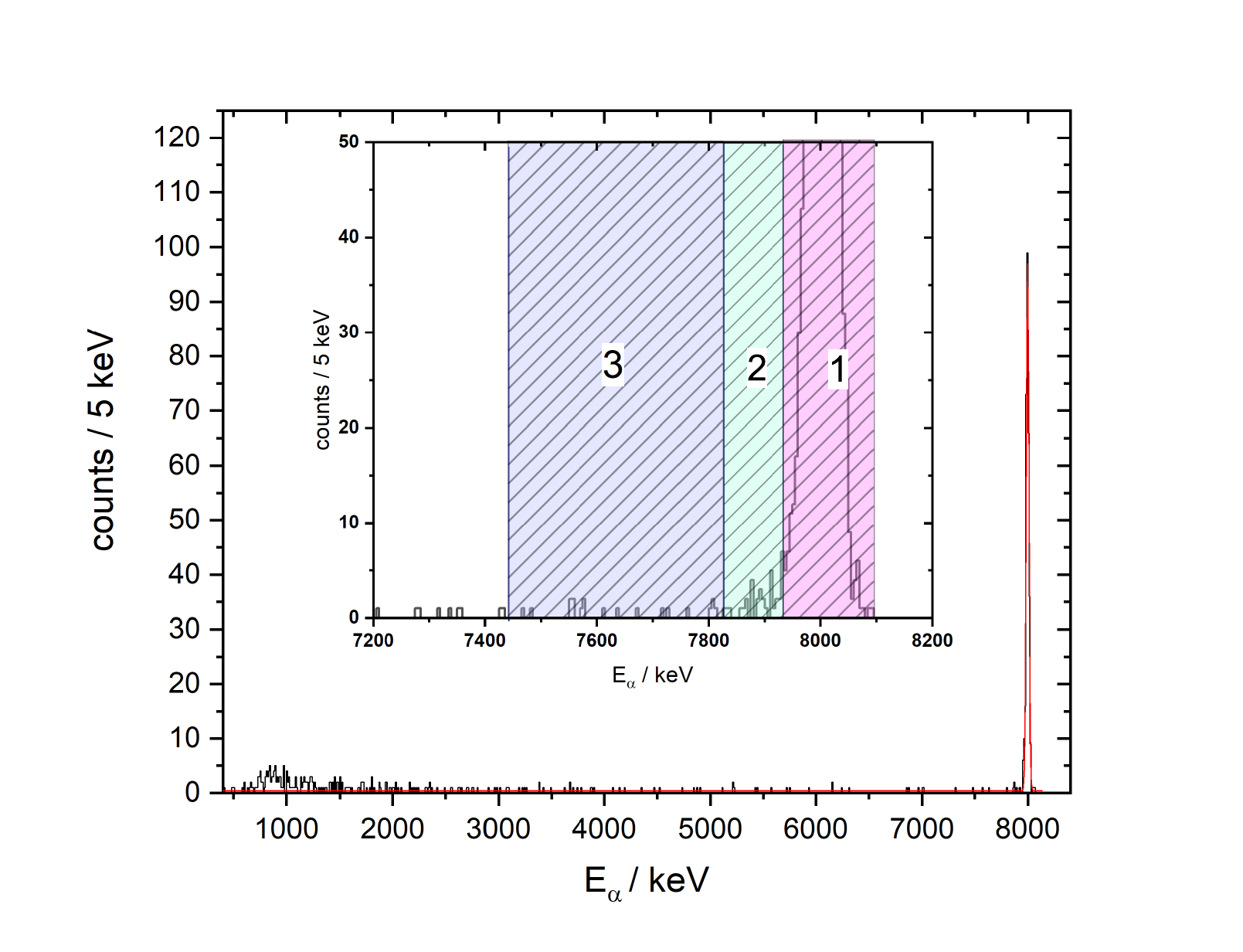}
	}
	% If not, use
	%\vspace{5cm}       % Give the correct figure height in cm
	\caption{alpha - decay spectrum of $^{253}$No (in coincidence with the 279.5 keV $\gamma$ - transition).
		The insert shows the region (7200\,-\,8200) keV in expanded scale.}
	\label{fig:6}       % Give a unique label
\end{figure}

Here the $\alpha$ decays in coincidence with the 279.5 keV $\gamma$ line are shown, which represents the transition of the
9/2$^{-}$[734] level in $^{249}$Fm populated by the $\alpha$ decay, into 7/2$^{+}$[624] ground-state. In that case one obtains a clean
$\alpha$ spectrum of a single transition not disturbed by energy summing with CE.\\
Besides the 'peak' at E$_{\alpha}$\,=\,8005 keV a bulk of events at {\it{E}}\,$<$\,2 MeV is shown. About
55 $\%$ of the $\alpha$ particles are registered in the peak, about 32$\%$ are found at {\it{E}}\,$<$\,2 MeV; the rest 
(13$\%$) is distributed in the energy range in between. In the insert the energy range between {\it{E}}\,=(7.2-8.2) MeV 
is expanded. It is clearly seen that the number of $\alpha$ particles in the range, taken here somewhat arbitrarily
as {\it{E$_{mean}$}}\,-\,570 keV is small. The 'peak' is here defined as the energy region {\it{E}}\,$>$\,7935 keV, as at this energy 
the number of events (per 5 keV) has been dropped to 5$\%$ of the number in the peak maximum (region 1).
The ratio of events in the energy interval (7835,7935) keV (region 2) is about 1.2\,$\%$ of that in the 'peak' (region 1),
while the ratio of events in the energy interval (7435,7835) keV (region 3) is about 0.8 $\%$.
These small numbers indicate that at low number of observed total decays, it is quite
unlikely that
events with energies some hundred keV lower than the 'bulk' energy represent $\alpha$ particles from the 'bulk' leaving the detector with nearly full energy loss.
They rather stem from decays into excited daughter levels (but possibly influenced by energy summing with CE)\footnote{We briefly want to point
to an detector effect that may pretend lower energies. In cases where the detector has already suffered from radiation damages the charge
collection may be incomplete and so the signal might be lower than that for a 'full energy event' even if the $\alpha$ particles was
completely stopped in the detector.}.

\subsection{\bf{4.5 Compatibility of $\alpha$ energy measurements in the region of SHN}}
As the numbers of decays observed in specific experiments is usually quite small it seems of high 
interest to merge data of different experiments to enhance statistics to possibly extract details
on the decay properties, {\it{e.g.}} fine structure in the $\alpha$ decay. One drawback concerning
this item is possible energy summing between $\alpha$ particles and CE as discussed above; another problem
is the compatibility of the decay energies measured in the different experiments and thus a consequence of
calibration. This is not necessecarily a trivial problem as shown in fig. 7, where the $\alpha$ energies obtained 
for $^{272}$Bh in three experiments are shown. $^{272}$Bh was produced in irradiations of $^{243}$Am with $^{48}$Ca 
within the decay chain of $^{288}$Mc, via $^{243}$Am($^{48}$Ca,3n)$^{288}$Mc $^{\alpha}_{\rightarrow}$
$^{284}$Nh $^{\alpha}_{\rightarrow}$ $^{280}$Rg $^{\alpha}_{\rightarrow}$ $^{276}$Mt $^{\alpha}_{\rightarrow}$ $^{272}$Bh $^{\alpha}_{\rightarrow}$ $^{268}$Db.
This decay chain has been investigated so far at three different separators, DGFRS at FLNR, Dubna, Russia \cite{OgA13},
TASCA at GSI, Darmstadt, Germany \cite{RuF13}, and BGS at LNBL, Berkeley, USA \cite{GaG15}.
The energy distributions of the odd-odd nuclei occuring in the decay chain of $^{288}$Mc are in general quite
broad indicating decays into different daughter levels accompanied by energy summing of $\alpha$ particles and CE.
Solely for $^{272}$Bh a 'quite narrow' line is observed. The results of the different experiments are compared in
fig. 7. To avoid ambiguities due to worse energy resolution of 'stop + box' events we restricted to events with full energy release 
in the 'stop' detector. 
Evidently there are large discrepancies in the $\alpha$ energies: the DGFRS experiment \cite{OgA13} delivers a mean value 
{\it{E$_{\alpha}$(DGFRS)}}\,=\,9.022\,$\pm$\,0.012 MeV, the TASCA experiment \cite{RuF13} {\it{E$_{\alpha}$(TASCA)}}\,=\,9.063\,$\pm$\,0.014 MeV, and
the BGS experiment {\it{E$_{\alpha}$(BGS)}}\,=\,9.098\,$\pm$\,0.022 MeV, hence differences
{\it{E$_{\alpha}$(TASCA)}} - {\it{E$_{\alpha}$(DGFRS)}}\,=\,41 keV,
{\it{E$_{\alpha}$(BGS)}} - {\it{E$_{\alpha}$(TASC)}}\,=\,35 keV, 
{\it{E$_{\alpha}$(BGS)}} - {\it{E$_{\alpha}$(DGFRS)}}\,=\,76 keV, which are by far larger than  calibration uncertainties in the range of 
10-20 keV, which might be expected usually. That is a very unsatisfying situation. 

\begin{figure}
%\begin{figure*}
\vspace{-1cm}
\resizebox{0.9\textwidth}{!}{%
  \includegraphics{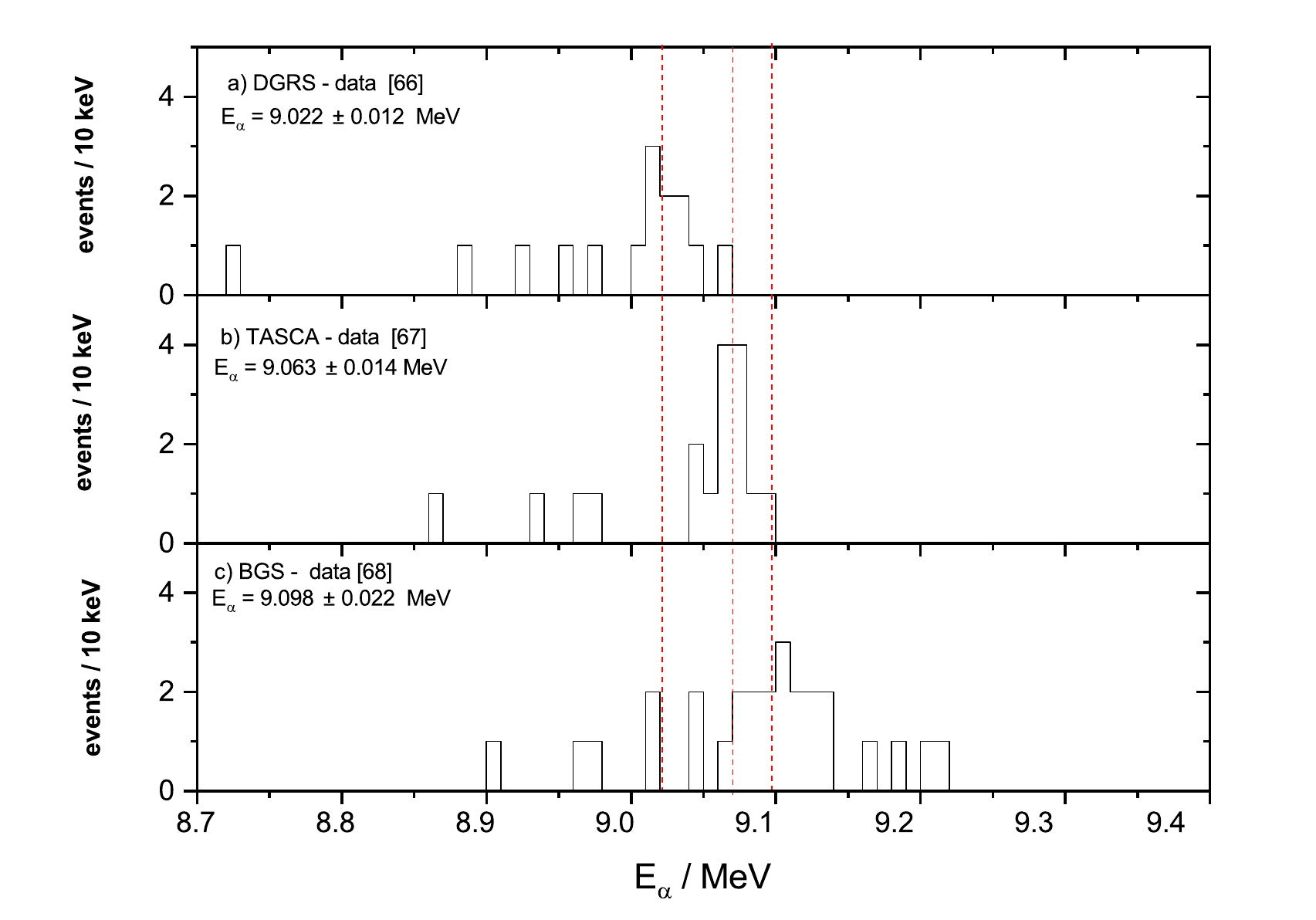}
}
% If not, use
%\vspace{5cm}       % Give the correct figure height in cm
\caption{Energy distributions of $^{272}$Bh; a) DGFRS experiment \cite{OgA13}; b) TASCA experiment \cite{RuF13}; c) BGS experiment \cite{GaG15}}
\label{fig:7}       % Give a unique label
\end{figure}

\newpage
\section{5. Discovery of elements Z\,=\,107 (bohrium) to Z\,=\,112 (copernicium) and their approvement by the IUPAC}
The elements {\it{Z}}\,=\,107 to {\it{Z}}\,=\,112 where first synthesized at the velocity filter SHIP, GSI, in the period 1981 - 1996. 
The corresponding isotopes where identified
after implantation into arrangements of silicon detectors by registering their $\alpha$ decay chains. Identification was based on
decay properties ($\alpha$ energies, halflives) of at least one member of the decay chain, that had been either known
from literature or had been synthesized and investigated at SHIP in preceeding experiments.
The latter is the main difference to elements {\it{Z}}\,$\ge$\,114, where the decay chains are not connected to the region of known
and safely identified isotopes. Nevertheless, the elements {\it{Z}}\,=\,107 - {\it{Z}}\,=\,112 depict in some cases already the difficulties to unambiguously 
identify an isotope on the basis of only a few observed decays and also the problems evaluaters in charge to approve discovery of
a new element are faced with.\\
In order not to overtop the banks in the following only the reports of the Tranfermium Working Group of the IUPAC and IUPAP (TWG)
(for elements bohrium to meitnerium) or the IUPAC/IUPAP Joint Working Party (JWP) (for elements darmstadtium to copernicium) 
concerning the GSI new element claims are considered. Other claims on discovery of one ore more of these elements are not discussed.

\subsection{\bf{5.1 Element 107 (Bohrium)}}
The first isotope of element 107, $^{262}$Bh, was synthesized in 1981 in the reaction $^{209}$Bi($^{54}$Cr,n)$^{262}$Bh
\cite{Muenz81}. Altogether six decay chains where observed at that time. Prior to approval of the discovery by the IUPAC two more 
experiments were performed. The complete results are reported in \cite{Muenz89}: two states of $^{262}$Bh decaying by $\alpha$ emission
$^{262g}$Bh ({\it{T$_{1/2}$}}\,=\,102\,$\pm$\,26 ms (15 decays)) and $^{262m}$Bh ({\it{T$_{1/2}$}}\,=\,8.0\,$\pm$\,2.1 ms (14 decays)) as well as the 
neighbouring isotope $^{261}$Bh (10 decays) were observed. Thus approval of the discovery of element 107 was based on a 
'safe ground', and it was stated by the TWG \cite{Wilk93}: 'This work ({\it\cite{Muenz81}}) is considered sufficiently convincing and was
confirmed in 1989 {\it\cite{Muenz89}}.'

\subsection{\bf{5.2 Element 108 (Hassium)}}
Compared to bohrium the data for hassium on which the discovery was approved was scarce. In the first experiment performed in 1984
\cite{Muenz84a} three decay chains of $^{265}$Hs were observed in an irradiation of $^{208}$Pb with $^{58}$Fe. 
In two cases a full energy event of $^{265}$Hs was followed by an escape event of $^{261}$Sg, while in one case an escape event
of $^{265}$Hs was followed by a full energy event of $^{261}$Sg. The $\alpha$ particle from the granddaughter $^{257}$Rf was measured
in all three cases with full energy.\\
In a follow-up experiment only one decay chain of the neighbouring isotope $^{264}$Hs was observed 
in an irradiation of $^{207}$Pb by $^{58}$Fe \cite{Muenz86}. The chain consisted of two escape events followed by an SF, which was 
attributed to $^{256}$Rf on the basis of the decay time.
Nevertheless discovery of element 108 was approved on the basis of these data and it was stated by the TWG \cite{Wilk93}: 'The Darmstadt work 
in itself is sufficiently convincing to be accepted as a discovery.'

\subsection{\bf{5.3 Element 109 (Meitnerium)}}
Discovery of element 109 was connected to more severe problems. In the first experiment at SHIP, performed in summer 1982, only one decay chain 
shown in fig. 8. was observed \cite{Muenz82} in an irradiation of $^{209}$Bi with $^{58}$Fe.

\begin{figure}
	%\begin{figure*}
	\resizebox{0.8\textwidth}{!}{%
		\includegraphics{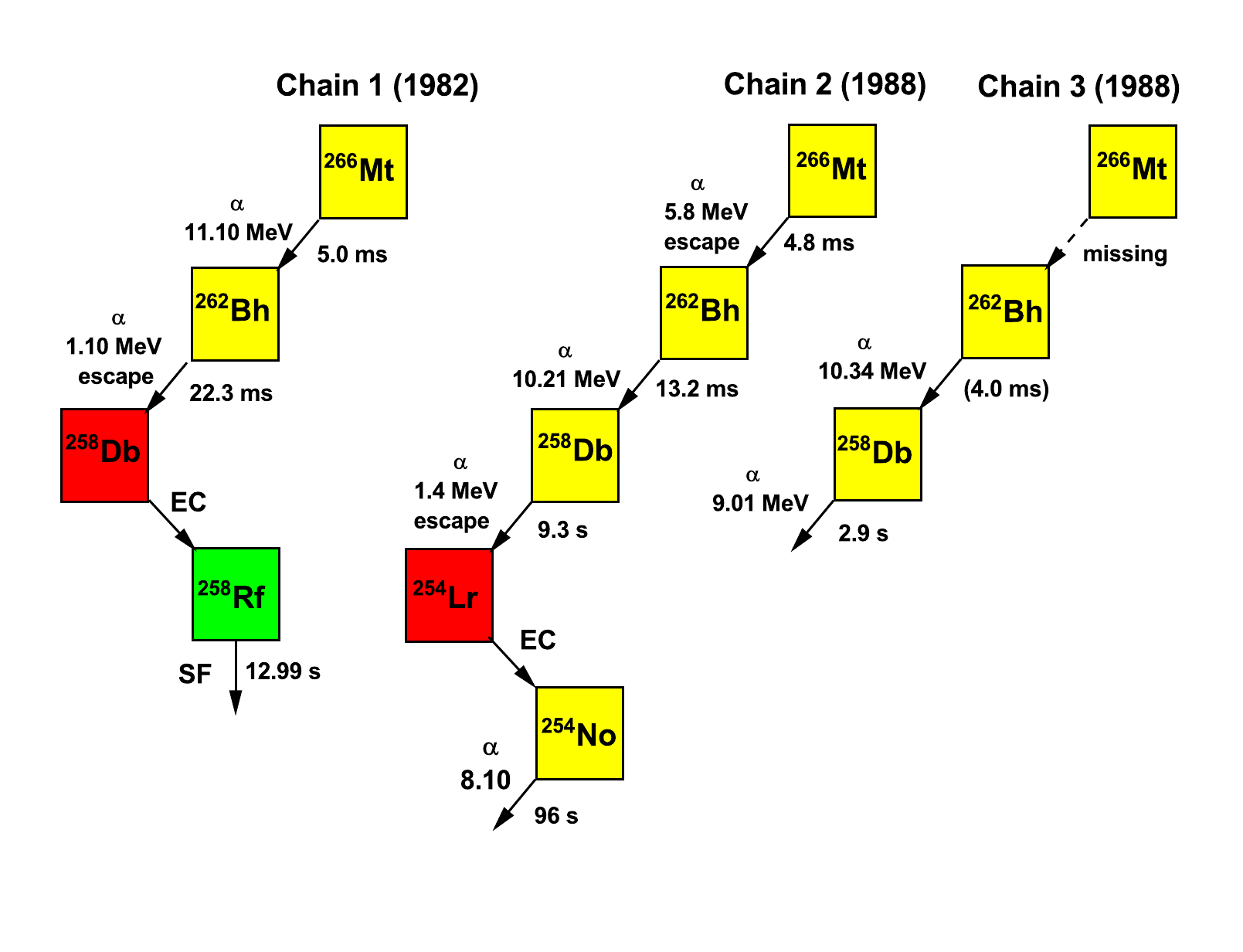}
	}
	% If not, use
	%\vspace{5cm}       % Give the correct figure height in cm
	\caption{alpha - decay chains observed in SHIP experiments in 1982 and 1988 and attributed to the decay of $^{266}$Mt
	\cite{Muenz82,Muenz88}.}
	\label{fig:8}       % Give a unique label
\end{figure}

It started with an  alpha - event with full energy, followed by an escape event and was terminated by an SF event. The latter was attributed to
$^{258}$Rf produced by EC decay of $^{258}$Db. A thorough investigation of the data showed that the probability for the event sequence to be random
was $<$10$^{-18}$ \cite{Muenz84}. Among all possible 'starting points' (energetically possible evaporation residues) $^{266}$Mt was the most likely
one \cite{Muenz84}. In a second experiment performed early in 1988 (january 31st to february 13th) two more decay chains, also shown in fig. 8
where observed \cite{Muenz88}; chain number 2 consisted of four $\alpha$ events, two with full energy, and two escape events, attributed to $^{266}$Mt (first chain member)
and $^{258}$Db (third chain member); the two full energy events were attributed to $^{262}$Bh (second chain member) and $^{254}$No (forth chain member),
which was interpreted to be formed by EC decay of $^{254}$Lr. The third chain consisted of two $\alpha$ decays, which were assigned to $^{262}$Bh and 
$^{258}$Db on the basis of the measured energies, while the $\alpha$ particle from the decay of $^{266}$Mt was not observed. The non-registration of $^{266}$Mt could have different reasons:\\
a) $^{262}$Bh was produced directly via the reaction $^{209}$Bi($^{58}$Fe,$\alpha$n)$^{262}$Bh. This possibility was excluded as this reaction channel 
was assumed to be considerably smaller than the 1n - deexcitation channel. And indeed in a later experiment, performed after the approval of element 109
by the IUPAC, twelve more decay chains of $^{266}$Mt were observed, but no signature for an $\alpha$n - deexcitation channel was found \cite{HofH97}. \\
b) $^{266}$Mt has a short-lived isomer decaying in-flight during separation. This interpretation seemed unlikely as in case of $\alpha$ emission
the recoil of the $\alpha$ particle would have kicked the residual nucleus out of its trajectory, so it would not have reached the detector placed in the focal 
plane of SHIP; similary in case of decay by internal transitions one could expect that emission of Auger electrons following internal conversion would have
changed the charge state of the atom, so it would been also kicked out of its trajectory.\\
c) a short-lived isomer may decay within 20 $\mu$s after implantation of the ER, i.e. during the dead time of the data acquisition system, and thus not be
recorded. Also for this interpretation no arguments were found in the later experiment \cite{HofH97}.\\
d) the $\alpha$ particle from the decay of $^{266}$Mt escaped with an energy loss $<$670 keV, which was the lower detection limit in this experiment. 
This was seen as the 
most reasonable case.\\
To summarize: the three chains presented strong evidence for having produced an isotope of element 109, but it still may be discussed if the presented
data really showed an unambiguous proof. However, the TWG did not share those concerns, leading to the assessment 'The result is convincing even though 
originally only one event was observed' and came to the conclusion 'The Darmstadt work \cite{Muenz82} gives confidence that element 109 has been observed'
\cite{Wilk93}'.

\subsection{\bf{5.4 Element 110 (Darmstadtium)}}
After a couple of years of technical development, including installation of a new low energy RFQ - IH acceleration structure coupled to an ECR ion source 
of the UNILAC
and construction of a new detector set-up experiments on synthesis of new elements were resumed in 1994. 
The focal plane detector was surrounded by a 'box' formed of six silicon detectors allowing to measure with a probability of
about 80$\%$ the full energy of $\alpha$ - particles escaping the 'stop' detector as the sum 
{\it{E\,=\,$\Delta$E(stop)\,+\,E$_{residual}$(box)}}.
The first isotope of element 110, $^{269}$Ds, was synthesized 1994 in the reaction $^{208}$Pb($^{62}$Ni,n)$^{269}$Ds \cite{Hof95}; four decay chains were  observed.
In three of the four chains $\alpha$ decays with full energy were observed down to $^{257}$Rf or $^{253}$No, respectivly.
For the forth decay chain of $^{269}$Ds only an energy loss signal was measured, while $\alpha$ decays of $^{265}$Hs and $^{261}$Sg were
registered with full energy. Further members of the decay chain ($^{257}$Rf, $^{253}$No, etc.) were not recorded. In a later
re-analysis of the data this chain could not be reproduced any more, similary to the case of decay chains
reported from irradiations of $^{208}$Pb with $^{86}$Kr  at the BGS, Berkeley, and interpreted to start from an isotope of 
element 118 \cite{Ninov99,Ninov02}.
This deficiency, however, did not concern the discovery of element 110 and it was stated by the JWP of the IUPAC/IUPAP 'Element 110 has been discovered by 
this collaboration' \cite{Karol01}.\\

\subsection{\bf{5.5 Element 111 (Roentgenium)}}
The first experiment on synthesis of element 111 was performed in continuation of the element 110 discovery experiment. 
In an irradiation of $^{209}$Bi with $^{64}$Ni three $\alpha$ decay chains were observed. They were assigned to $^{272}$Rg, produced in the reaction
$^{209}$Bi($^{64}$Ni,n)$^{272}$Rg \cite{Hofm95a}. Two of these chains ended with $\alpha$ decay of $^{260}$Db; for the first member, $^{272}$Rg, 
only a {\it{$\Delta$E}} signal was registered. In the third chain $\alpha$ decay was observed down to $^{256}$Lr and all $\alpha$ particles from the decay 
chain members ($^{272}$Rg, $^{268}$Mt, $^{264}$Bh, $^{260}$Db, $^{256}$Lr) were registered with
'full' energy. It should be noted that also $^{268}$Mt and $^{264}$Bh had not been observed before.
The JWP members, however, were quite cautious in that case \cite{Karol01}. It was remarked that the $\alpha$ energy of $^{264}$Bh in chain 1 was quite different to
the values of chain 2 and 3 and that the $\alpha$ energy {\it{E}}\,=\,9.146 MeV of $^{260}$Db in chain 2 was in fact in-line with the literature value (given {\it{e.g.}} in
\cite{Fire96}), but was quite different to the value {\it{E}}\,=\,9.200 MeV in chain 3. Further it was noted, that the time difference $\Delta$t($^{262}$Db\,-\,$^{256}$Lr)\,=\,66.3 s, was considerably longer than the known half-life of $^{256}$Lr ({\it{T$_{1/2}$}}\,=\,28$\pm$3 s \cite{Fire96}). So it was stated (JWP assessment): 'The results of this study are definitely of high quality but there is insufficient internal redundancy to warrant certitude at this stage. Confirmation by further results is needed to assign priority of discovery to this collaboration' \cite{Karol01}. In a further experiment at SHIP three more decay chains were observed which confirmed the previous
results \cite{Hofm02}, leading to the JWP statement:'Priority of discovery of element 111 by Hofmann et al. collaboration in \cite{Hofm95a} has been confirmed
owing to the additional convincing observations in \cite{Hofm02}' \cite{Karol03}.\\
For completeness it should be noted that the SHIP results were later confirmed in experiments performed at the GARIS separator at RIKEN, Wako (Japan), where
the same reaction was used \cite{MoM04a}, and at the BGS separator at LBNL Berkeley (USA), where, however, a different reaction, $^{208}$Pb($^{65}$Cu,n)$^{272}$Rg 
was applied \cite{FoG04}. 

\vspace{15mm}
\subsection{\bf{5.6 Element 112 (Copernicium)}}
Concerning discovery of element 112 the situation was even more complicated. In a first irradiation of $^{208}$Pb with $^{70}$Zn performed at SHIP early in 
1996 two decay chains interpreted to start from $^{277}$Cn were reported \cite{Hofm96}. In chain 1 $\alpha$ decays down to $^{261}$Rf, in chain 2 alpha decays down 
to $^{257}$No were observed. Both chains showed severe differences in chain members $\alpha$(1) and  $\alpha$(2). (In the following $\alpha$(n) in chains 1 and 2 will be 
denoted as $\alpha$(n1) and $\alpha$(n2), respectively.)\\
The $\alpha$ energies for $^{277}$Cn differed by 0.22 MeV, while the 'lifetimes' {\it{$\tau$}} (time differences between ER implantation and $\alpha$ decay were comparable)
with {\it{E$_{\alpha 11}$}}\,=\, 11.65 MeV, {\it{$\tau_{\alpha 11}$}}\,=\,400 $\mu$s and  {\it{E$_{\alpha 12}$}}\,=\, 11.45 MeV, 
{\it{$\tau_{\alpha 12}$}}\,=\,280 $\mu$s.  
For {\it{$\alpha$(2)}} ($^{273}$Ds) the discrepancies were more severe: {\it{E$_{\alpha 21}$}}\,=\,9.73 MeV, {\it{$\tau_{\alpha 21}$}}\,=\,170 ms and  {\it{E$_{\alpha 22}$}}\,=\, 11.08 MeV, {\it{$\tau_{\alpha 12}$}}\,=\,110 $\mu$s. It seemed thus likely that the $\alpha$ decays of $^{273}$Ds (and thus also of $^{277}$Cn) occurred from different levels.
This was commented in the JWP report \cite{Karol01} as 'Redundancy is arguably and unfortunately confounded by the effects of isomerism. The two observed alphas
from $^{277}$112 involve different states and lead to yet two other very different decay branches in $^{273}$110. (...) The first two alpha in the chains show
no redundancy.' It was further remarked that the energy of $^{261}$Rf in chain 2 ({\it{E}}\,=\,8.52 MeV) differed by 0.24 MeV from the literature value \cite{Fire96}.
Indeed it was later shown by other research groups \cite{DvB08,HaK11,HaK12} that two longlived states decaying by $\alpha$ emission exist in $^{261}$Rf with one
state having a decay energy and a halflife of {\it{E$_{\alpha}$}}\,=\,8.51$\pm$0.06 MeV and {\it{T$_{1/2}$}}\,=\,2.6$^{+0.7}_{-0.5}$s \cite{HaK12} in-line with the data from 
chain 2 ({\it{E$_{\alpha 52}$}}\,=\,8.52 MeV, {\it{$\tau_{\alpha 52}$}}\,=\,4.7 s). But this feature was not known when the TWG report \cite{Karol01} was written.
Consequently it was stated 'The results of this study are of characteristically high quality, but there is insufficient internal redundancy to warrant conviction
at this state. Confirmation by further experiments is needed to assign priority of discovery to this collaboration.'\\
One further experiment was performed at SHIP in spring 2000, where one more decay chain was observed, which resembled chain 2, but was terminated by a fission 
event \cite{Hofm02}.The latter was remarkable, as the fission branch of $^{261}$Rf was estimated at that time as {\it{b$_{sf}$}}\,$<$\,0.1. But also here later
experiments \cite{DvB08,HaK11,HaK12} established a high fission branch for the 2.6 s - activity with the most recent value 
{\it{b$_{sf}$}}\,=\,0.82$\pm$0.09 \cite{HaK12}.\\
Then, during preparing the manuscript \cite{Hofm02} a 'desaster' happed: in a re-analysis of the data from 1996 chain 1 could not be
reproduced, similary to the case of one chain in the element 110 synthesis experiment (see above) and of decay chains
reported from irradiations of $^{208}$Pb with $^{86}$Kr  at the BGS, Berkeley, and interpreted to start from an isotope of 
element 118 \cite{Ninov99,Ninov02}. It was shown that this chain had been created spuriously \cite{Hofm02}. At least this finding could
explain the inconsistencies concerning the data for $^{277}$Cn and $^{273}$Ds in chains 1 and 2.
On this basis the JWP concluded \cite{Karol03}: 'In summary, though there are only two chains, and neither is completely characterized on its own merit.
Supportive, independent results on intermediates remain less then completely compelling at that stage.'\\
In the following years two more experiments at SHIP using the reaction $^{70}$Zn + $^{208}$Pb were performed without observing one more chain 
\cite{Hess20}, however, decay studies of $^{269}$Hs and $^{265}$Sg confirmed specifically the data for $^{261}$Rf \cite{DvB08}, while the decay chains
of $^{277}$Cn were reproduced in an irradiation of $^{208}$Pb with $^{70}$Zn at the GARIS separator at RIKEN, Wako (Japan) \cite{Morita05,Morita07}.\\
On this basis the JWP concluded in their report from 2009 \cite{Barber09}: 'The 1996 collaboration of Hofmann et al. \cite{Hofm96} combined with the
2002 collaboration of Hofmann et al. \cite{Hofm02} are accepted as the first evidence for synthesis of element with atomic number 112 being
supported by subsequent measurements of Morita \cite{Morita05,Morita07} and by assignment of decay properties of likely hassium imtermediates 
\cite{DvB08,Duell02,Turl03} in the decay chain of $^{277}$112'.\\

\section{6. Some Critical assessments of decay chains starting from elements Z$\ge$112 and discussion of decay data of the chain members}

The experiments on synthesis of the new elements with {\it{Z}}\,=\,113 to {\it{Z}}\,=\,118 reflect the extreme difficulties connected
with identification of new elements on the basis of observing their decay when only very few nuclei are produced and decay chains end
in a region where no isotopes had been identified so far or their decay properties are only known scarcely.
Nevertheless discovery of elements {\it{Z}}\,=\,113 to {\it{Z}}\,=\,118 has been approved by IUPAC and discovery priority 
was settled \cite{BarK11,KarB16,KarB16a}, and also names have been proposed and accepted so far:\\
 {\it{Z}}\,=\,113: Nihonium (Nh)\\
 {\it{Z}}\,=\,114: Flerovium (Fl)\\
 {\it{Z}}\,=\,115: Moscovium (Mc)\\
 {\it{Z}}\,=\,116: Livermorium (Lv)\\
 {\it{Z}}\,=\,117: Tennessine (Ts)\\
 {\it{Z}}\,=\,118: Oganesson (Og)\\
Still there remain a couple of open questions and ambiguities concerning decay properties of several isotopes, which may have feedback 
to their final assignment. In the following we will discuss some selected cases and point to open problems that need to be clarified
in further experiments.\\ 
For illustrating the following discussion an excerpt of the charts of nuclei covering
the region {\it{Z}}\,$\ge$\,112 and {\it{N}}\,=\,(165\,-\,178)
is shown in fig. 1.\\

\subsection{\bf{6.1 Ambiguities in the assignment of decay chains - case of $^{293}$Lv - $^{291}$Lv} }
As already briefly mentioned in sect. 3, the continuous implantation of nuclei, the overlap of low energy particles
passing the separator with the $\alpha$ - decay energies of the expected particles and efficiencies lower than 100 $\%$ of detectors used
for anti-coincidence to discriminate between 'implantation of nuclei' and 'decays in the detector' introduces a problem of background.
It might be severe, if only very few decay chains are observed, since at a larger number of events single chains containing a member that does not fit to the rest of the data can be easily removed.\\

 \begin{figure}
  %\begin{figure*}
  \vspace{-1cm}
  \resizebox{0.99\textwidth}{!}{%
    \includegraphics{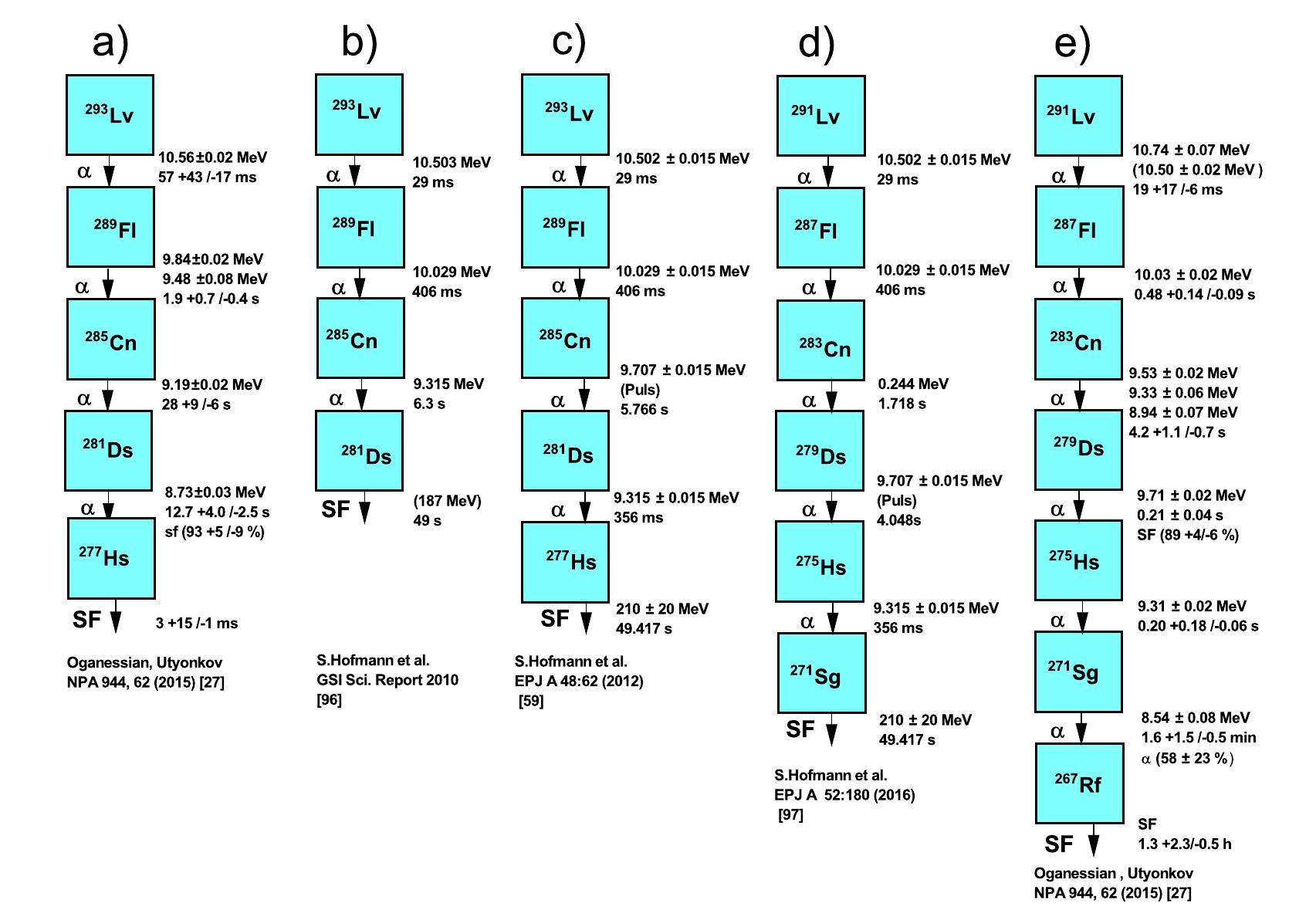}
  }
  % If not, use
  %\vspace{5cm}       % Give the correct figure height in cm
  \caption{ Assignments and reassignments of a decay chain observed in an irradiation of 
  $^{248}$Cm with $^{48}$Ca; a) decay data reported for $^{293}$Lv and its daughter products \cite{Oga16}; b) decay chain as assigned in \cite{HofH10}; c)decay chain as assigned in \cite{Hofm12}; d) decay chain as assigned in \cite{HofH16};
  e) decay data reported for $^{291}$Lv and its daughter products \cite{Oga16}.}
  \label{fig:9}       % Give a unique label
  \end{figure}
 
 An example to illustrate those related difficulties is given in fig. 9. 
 The decay chain was observed at SHIP, GSI, in an irradiation of $^{248}$Cm with $^{48}$Ca at
 a bombarding energy E$_{lab}$ = 265.4 MeV \cite{Hofm12}. A first analysis of the data resulted in an implantation of an ER,
 followed by three $\alpha$ decays. The chain was terminated by a spontaneous fission event \cite{HofH10} as shown in fig. 9b.
 It was tentatively assigned to the decay of $^{293}$Lv.
 After some further analysis, one more $\alpha$ decay (an event that occured during the beam-on period) placed at the position of $^{285}$Cn 
 was included into the chain, but still assigned tentatively to the decay of $^{293}$Lv \cite{Hofm12} as shown in fig. 9c.
 However, except for $^{293}$Lv the agreement of the decay properties ($\alpha$ energies, lifetimes) of the chain members with
 literature data \cite{Oga16}, shown in fig. 9a, was rather bad. 
 Therefore in a more recent publication \cite{HofH16} the assigment was revised, including a low energy signal of 0.244 MeV
 registered during the beam-off period, but without position signal, into the chain at the place of $^{283}$Cn. The chain is now
 assigned to the decay of $^{291}$Lv (fig. 9d). A comparison with the literature data \cite{Oga16}, presented in fig. 9e, shows a good agreement 
 in $\alpha$ decay energies for $^{287}$Fl, $^{279}$Ds, $^{275}$Hs and in 'lifetimes' (i.e. time differences between consecutive decay events)
 for $^{291}$Lv, $^{287}$Fl, $^{283}$Ds, $^{275}$Hs, $^{271}$Sg. Stringent differences, however, are obtained for the $\alpha$ energy of
 $^{291}$Lv and the lifetime of $^{279}$Ds (the event observed in the beam-on period). 
 The differences in the $\alpha$ decay energies of 240 keV principally can be explained by decay in different daughter levels. As in 
 \cite{Oga16} only three decays are reported, it might be that the decay of the lower energy simply was not observed in the experiments 
 from which the data in \cite{Oga16} were obtained. Such an explanation is principally reasonable.
 For {\it{E$_{\alpha}$}}\,=\,10.74 MeV one obtains a theoretical $\alpha$ decay halflife of {\it{T$_{\alpha}$}}\,=\,32 ms using the formula suggested by
 Poenaru \cite{PoI80} using the parameter modification suggested in \cite{Rur83} which has been proven to reproduce $\alpha$ decay halflives in the region of heaviest nuclei very well \cite{Hess16a}.
 The value is indeed in good agreement with the reported half-life of {\it{T$_{\alpha}$}}\,=\,19$^{+17}_{-6}$ ms \cite{Oga16}. 
 For {\it{E$_{\alpha}$}}\,=\,10.50 MeV one obtains {\it{T$_{\alpha}$}}\,=\,139 ms. This means, that one expects some 20$\%$ intensity for an $\alpha$ transition 
 with an energy lower by about 250 keV, provided that $\alpha$ decay hindrance factors are comparable for both transitions.
 More severe, however, seems the lifetime of $^{279}$Ds, which is a factor of twenty longer than the reported half-life of 
 {\it{T$_{\alpha}$}}\,=\,0.21$\pm$0.04 s. The probability to observe an event after twenty halflives is only $\approx$10$^{-6}$.
 To conclude: it is certainly alluring to assign this chain  to $^{291}$Lv, the assignment, however, is not unambiguous.
 As long as it is not confirmed by further data, it should be taken with caution.\\
   
\subsection{\bf{6.2 Ambiguities in the assignment of decay chains - case of $^{289,288}$Fl}}
The observation of a decay chain, registered in an irradiation of $^{244}$Pu with $^{48}$Ca at {\it{E$_{lab}$}}\,=\,236 MeV 
at the Dubna Gasfilled Separator (DGFRS), assigned to start
from $^{289}$Fl was reported by Oganessian et al. \cite{Ogan99a}. The data presented in \cite{Ogan99a} are shown in fig. 10a.
In a follow-up experiment two more decay chains with different decay chracteristics were observed and attributed to the
neighbouring isotope $^{288}$Fl \cite{Ogan00a}, while in an irradiation of $^{248}$Cm with $^{48}$Ca at {\it{E$_{lab}$}}\,=\,240 MeV
one decay chain, shown in fig. 10c was registered \cite{Ogan00b}. Decay properties of members 2, 3 and 4 of the latter chain were 
consistent with those for $^{288}$Fl, $^{284}$Ds, and $^{280}$Hs. Consequently the chain (fig. 10c) was interpreted to start from
$^{292}$Lv.\\ 

\begin{figure}
 %\begin{figure*}
 \vspace{-1cm}
 \resizebox{0.99\textwidth}{!}{%
   \includegraphics{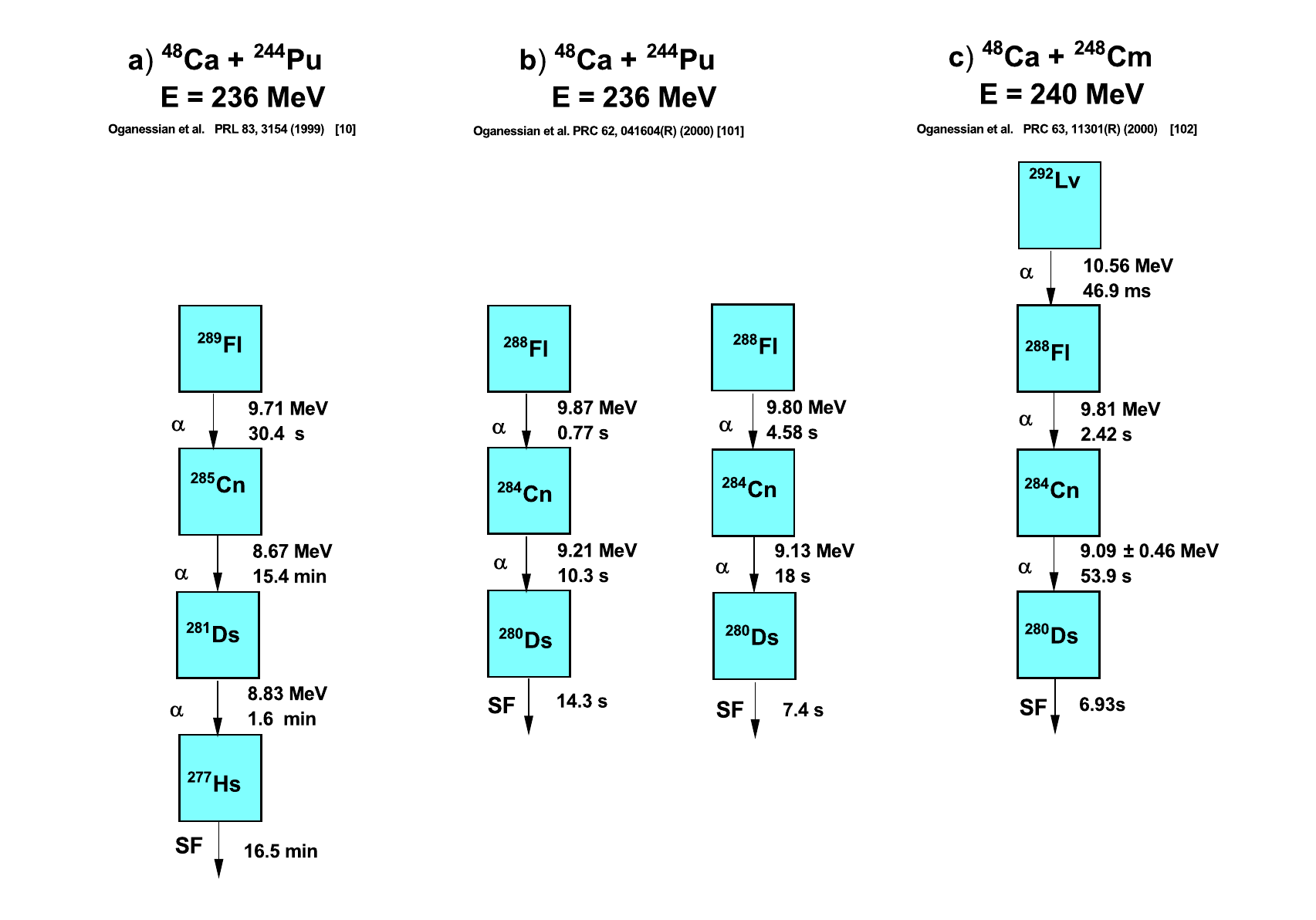}
 }
 % If not, use
 %\vspace{5cm}       % Give the correct figure height in cm
 \caption{ Decay chains observed at the DGFRS in irradiations of $^{244}$Pu or $^{248}$Cm with $^{48}$Ca and assigned to decays
 starting from $^{289}$Fl ((a) \cite{Ogan99a}), $^{288}$Fl ((b) \cite{Ogan00a}), and $^{292}$Lv ((c) \cite{Ogan00b}).}
 \label{fig:12}       % Give a unique label
 \end{figure}
 
 However, results from later irradiations of $^{244}$Pu with $^{48}$Ca were interpreted in a different way \cite{Ogan04}.
 The activity previously attributed to $^{288}$Fl was now assigned to $^{289}$Fl, while no further events having the characteristics of the chain
 originally attributed to $^{289}$Fl \cite{Ogan99a} were observed. It was now considered as a possible candidate for
 $^{290}$Fl \cite{OganUt04}. But this chain was not mentioned later as a decay chain stemming from a
 flerovium isotope \cite{Ogan07}. However, a new activity, consisting of an $\alpha$ decay of {\it{E$_{\alpha}$}}\,=\,9.95$\pm$0.08 MeV and 
 {\it{T$_{1/2}$}}\,=\,0.63$^{+0.27}_{-0.14}$s followed by a fission activity of
 {\it{T$_{1/2}$}}\,=\,98$^{+41}_{-23}$ms was observed. It was assigned to the decay sequence 
 $^{288}$Fl $^{\alpha}_{\rightarrow}$  $^{284}$Ds $^{SF}_{\rightarrow}$. These 'new' results were consistent with those obtained in later 
 irradiations of $^{244}$Pu with $^{48}$Ca \cite{Gates11} and  $^{248}$Cm with $^{48}$Ca \cite{Hofm12,Kaji17} in other labs.\\
 As a summary the 'old' and new results for $^{288,289}$Fl are compared in fig. 11. The 'new' results are taken from the recent review \cite{Oga16}, the 'old' results for $^{288}$Fl are the mean values from the three decays reported in \cite{Ogan00a,Ogan00b} as evaluated by the author.\\
 It should be noticed for completeness, that Kaji et al. \cite{Kaji17} observed also a chain consisting of three $\alpha$ particles
 terminated by a fission event. The chain was not regarded as unambiguous and so $\alpha_{3}$ and the SF event were only tentatively
 assigned to $^{284}$Cn ($\alpha_{3}$) and $^{280}$Ds (SF). In a more recent decay study of $^{288}$Fl using the production reaction
 $^{244}$Pu($^{48}$Ca,4n)$^{288}$Fl a small $\alpha$ - decay branch (b$_{\alpha}$$\approx$0.02) and spontaneous fission of $^{280}$Ds were
 confirmed \cite{Sarm21}. 
 
 \begin{figure}
  %\begin{figure*}
  \resizebox{0.99\textwidth}{!}{%
    \includegraphics{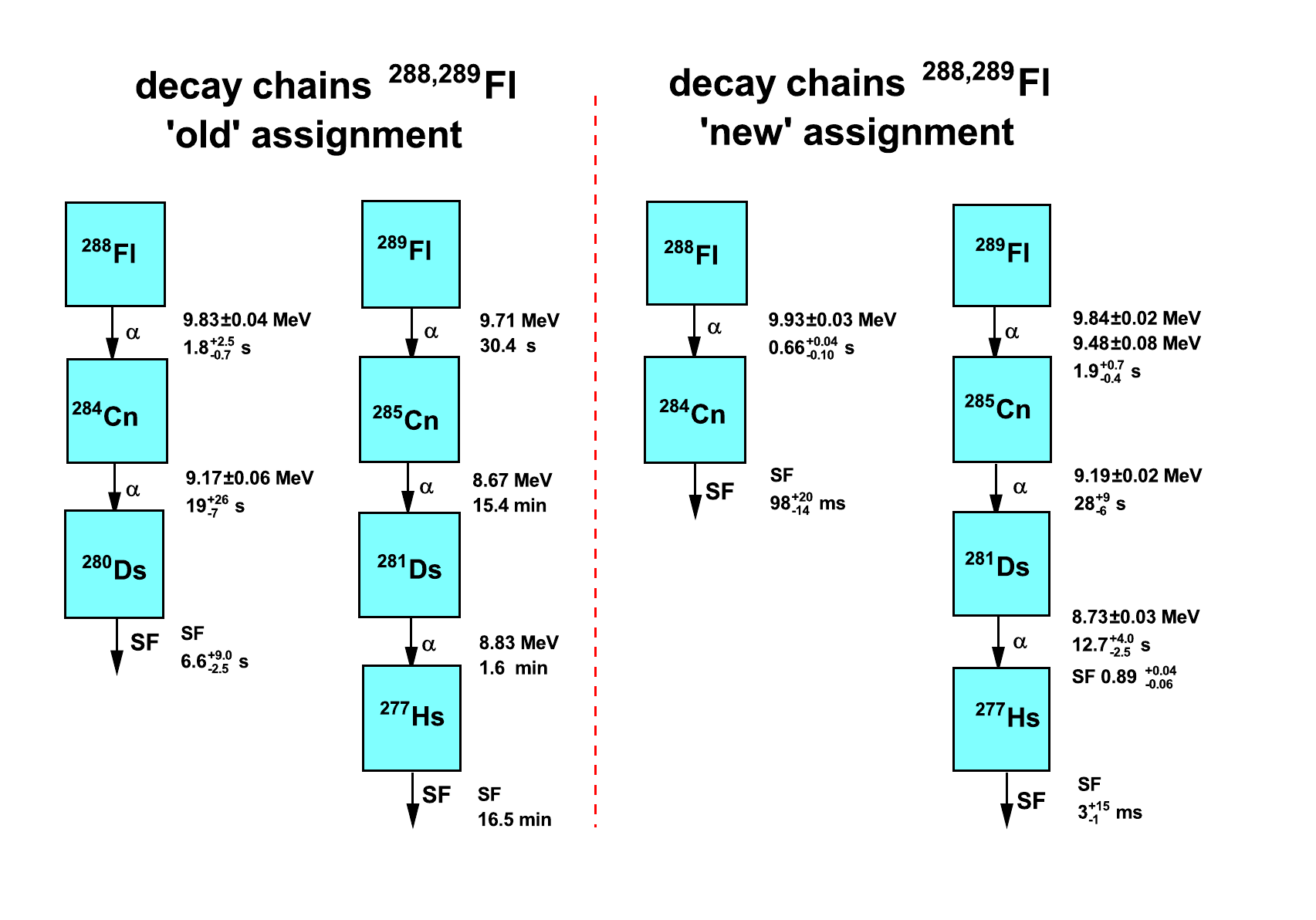}
  }
  % If not, use
  %\vspace{5cm}       % Give the correct figure height in cm
  \caption{ 'Old' and 'new' decay chains for $^{289}$Fl and $^{288}$Fl.}
  \label{fig:13}       % Give a unique label
  \end{figure}

 \begin{figure}
  %\begin{figure*
  \vspace{-1cm}
  \resizebox{0.99\textwidth}{!}{%
    \includegraphics{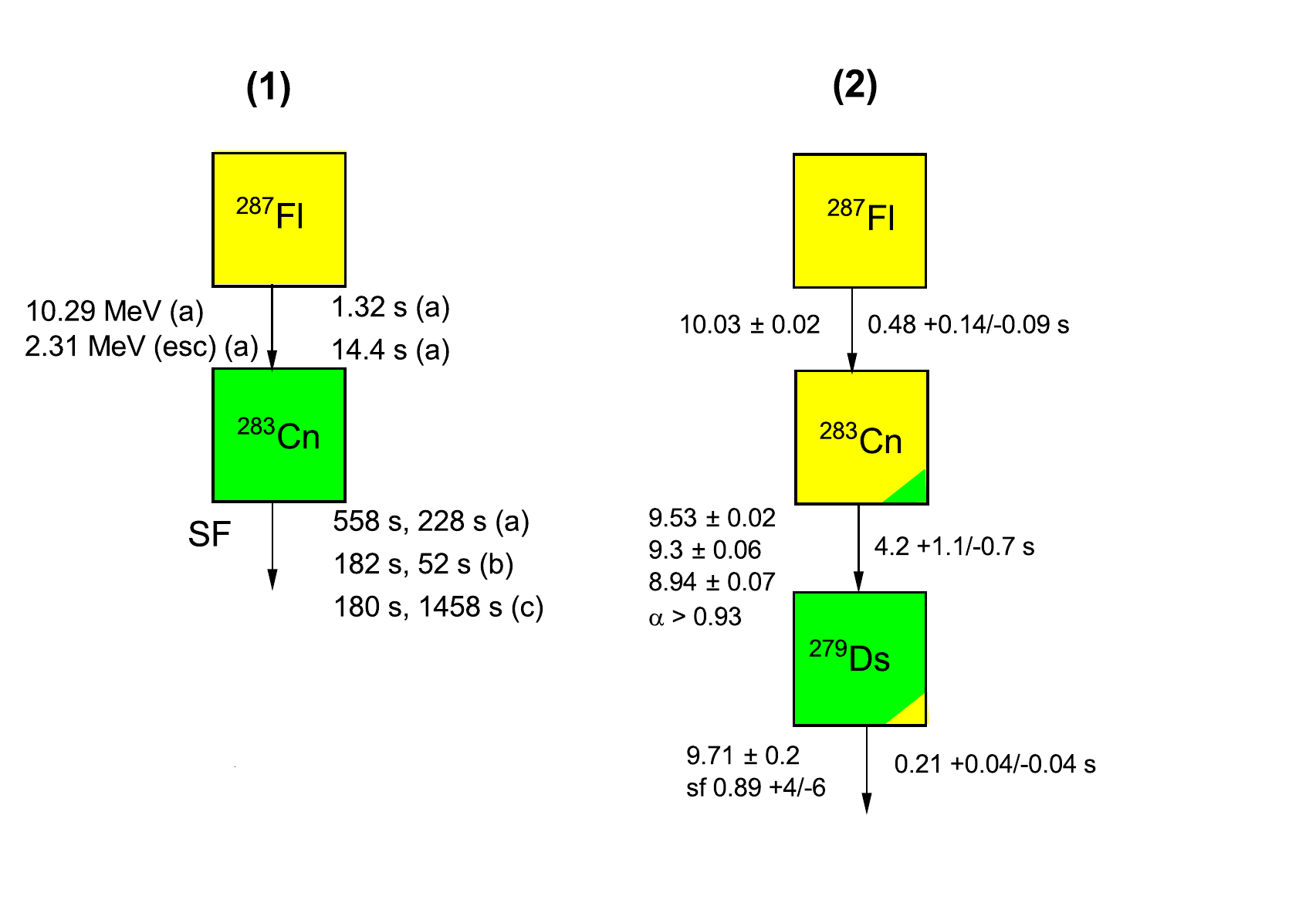}
  }
  % If not, use
  %\vspace{5cm}       % Give the correct figure height in cm
  \caption{ 'Old' and 'new' decay chains for $^{287}$Fl and $^{283}$Cn; fig. 12.1: results observed at VASSILISSA: (a) data from
  $^{48}$Ca + $^{242}$Pu \cite{Ogan99}; (b) \cite{OgaY99}, (c) \cite{OgaY04} from $^{48}$Ca + $^{238}$U irradiation; 
  fig. 12.2: results observed at DGFRS \cite{Ogan07}, SHIP \cite{HofA07}, GARIS II \cite{Kajia17}, BGS \cite{StavG09} in 
  irradiations of $^{238}$U or $^{242}$Pu with $^{48}$Ca. See text for details.}
  \label{fig:14}       % Give a unique label
  \end{figure}

\subsection{\bf{6.3 Ambiguities in the assignment of decay chains - case of $^{287}$Fl - $^{283}$Cn }}
A couple of weeks after submission of  \cite{Ogan99a} (recieved march 9,1999) another paper was submitted by Yu.Ts. Oganessian et al.
reporting on synthesis of a flerovium isotope with mass number {\it{A}}\,=\,287 \cite{Ogan99} (received april 19, 1999). The experiment had been performed
at the energy filter VASSILISSA at FLNR-JINR Dubna, and the decay chains (shown in fig. 12a) were observed in bombardments of $^{242}$Pu with
$^{48}$Ca at E$_{lab}$\,=\,230\,-\,235 MeV. Two chains consisting of an $\alpha$ - decay (in one case only an 'escape' $\alpha$ - particle was registered) followed by 
spontaneous fission were observed. Although lifetimes of the SF events were longer than those of  two SF events correlated to ER observed in a preceding irradiation of $^{238}$U with 
$^{48}$Ca at VASSILISSA \cite{OgaY99} (fig. 12b) they were attributed to the same isotope, $^{283}$Cn and the $\alpha$ decays were attributed to $^{287}$Fl. In a later 
irradiation of $^{238}$U with $^{48}$Ca at the same set-up two more SF events attributed to $^{283}$Cn were observed \cite{OgaY04} (fig. 12c). The production cross section was
$\sigma$\,=\,3.0$^{+4.0}_{-2.0}$ pb in fair agreement with the value $\sigma$\,=\,5.0$^{+6.3}_{-3.2}$pb obtained in the first experiment \cite{OgaY99}.\\
Both acitivities, however, could not be reproduced in irradiations of $^{238}$U, $^{242}$Pu with $^{48}$Ca performed at the Dubna Gas-filled Separator (DGFRS) 
\cite{Ogan04b} (see fig. 12). $^{287}$Fl was here interpreted as an $\alpha$ emitter of  {\it{E$_{\alpha}$}}\,=\,10.02$\pm$0.06 MeV, {\it{T$_{1/2}$}}\,=\,0.51$^{+0.18}_{-0.10}$ s, 
$^{283}$Cn as an $\alpha$ emitter of {\it{E$_{\alpha}$}}\,=\,9.162$\pm$0.06 MeV, {\it{T$_{1/2}$}}\,=\,4.0$^{+1.3}_{-0.70}$ s. 
Most of the decay chains were terminated by SF 
of $^{279}$Ds, except in two cases: in one decay chain, observed in the irradiation of $^{242}$Pu also $\alpha$ decay of $^{279}$Ds and $^{275}$Hs was observed and 
the chain was terminated by SF of $^{271}$Sg; in one chain observed in the irradiation of $^{238}$U also $\alpha$ decay of $^{271}$Sg was observed and the chain was terminated by SF $^{267}$Rf. 
The previously observed chains at VASSILISSA were suspected to represent a less probable decay mode \cite{OganUt04}, but not listed any more in later
publications (see e.g. \cite{Ogan07}).
The 'DGFR - results' were in-line with data for $^{283}$Cn and $^{287}$Fl data later obtained in the reactions $^{238}$U($^{48}$Ca,3n)$^{283}$Cn investigated at SHIP, GSI Darmstadt \cite{HofA07} and at GARIS II, RIKEN, Wako \cite{Kajia17}, as well as in the reaction  $^{242}$Pu($^{48}$Ca,3n)$^{287}$Fl investigated at BGS, LNBL Berkeley  \cite{StavG09}, while the 
'VASSILISSA - events' were not observed. \\
It should be noted, however, that due to a more sensitive detector system used in \cite{HofA07} than that used in \cite{Ogan04b} 
in cases where the $\alpha$ decay of $^{283}$Cn was denoted as 'missing' in \cite{Ogan04b},
since fission was directly following $\alpha$ decay of $^{287}$Fl, the $\alpha$ decay of $^{283}$Cn probably was not missing, 
but fission occured from $^{283}$Cn \cite{HofA07}.\\
The discrepancy between the 'DGFRS results' and the 'VASSILiSSA results' could not be clarified so far, but it should be noted that the latter ones were not
considered any more in later reviews of SHE synthesis experiments at FLNR - JINR Dubna \cite{Oga16,Ogan07}.\\
However, the 'VASSILISSA results' again were discussed in context with a series of events, registered in an irradiation of $^{248}$Cm with $^{54}$Cr at SHIP, which
were regarded as a signature for a decay chain starting from an isotope of element 120 \cite{HofH16}. 
It should be noted that a critical re-inspection of this sequence of events showed that it does not fulfil the physics criteria for a 'real' decay chain and the probability to be a real chain
is p $<$$<$ 0.01 \cite{HesA17}. Nevertheless the discussion in \cite{HofH16} can be regarded as  an illustrative example of trying to match doubtful data
from different experiments. Therefore it will be treated her in some more detail.\\

\begin{table}
	\caption{Comparison of the 'event sequence' from the $^{54}$Cr + $^{248}$Cm irradiation at SHIP \cite{HofH16} and the VASSILISSA results
		for $^{287}$Fl and $^{283}$Cn. Data taken from \cite{HofH16}.}
	\label{tab:1}       % Give a unique label
	% For LaTeX tables use
	\begin{tabular}{lllll}
		\hline\noalign{\smallskip}
		\noalign{\smallskip}\hline\noalign{\smallskip}
		Isotope & E$_{\alpha}$(SHIP) / MeV & $\Delta$t & E$_{\alpha}$(VASSILISSA) / MeV  & T$_{1/2}$ / s \\ 
		\hline\noalign{\smallskip}     
		($^{299}$120)  & 13.14$\pm$0.030 &  5.4 s* &   &  \\ 
		($^{295}$Og)  & 11.814$\pm$0.040 &  261.069 ms &  &  \\
		($^{291}$Lv)  & 10.698$\pm$0.030 &   18.378 ms &  &  \\ 
		($^{287}$Fl)  & 0.353 (10.14$^{+0.09}_{-0.27}$)** & 20.1 s & 10.29$\pm$0.02 & 5.5$^{+9.9}_{-2.1}$ s   \\ 
		($^{283}$Cn  & SF & 701 s  & SF & 308$^{+212}_{-89}$ s \\ 
	\end{tabular}

     * time difference to the closed possible evaporation residue\\
     ** energy calculated from $\Delta$t = 20.1 s for a hindrance factor HF\,=\,104 \cite{HofH16}.
	% Or use
	\vspace*{0.cm}  % with the correct table height
\end{table}  

\begin{figure}
	%\begin{figure*}
	\vspace{-2cm}
	\resizebox{0.95\textwidth}{!}{%
		\includegraphics{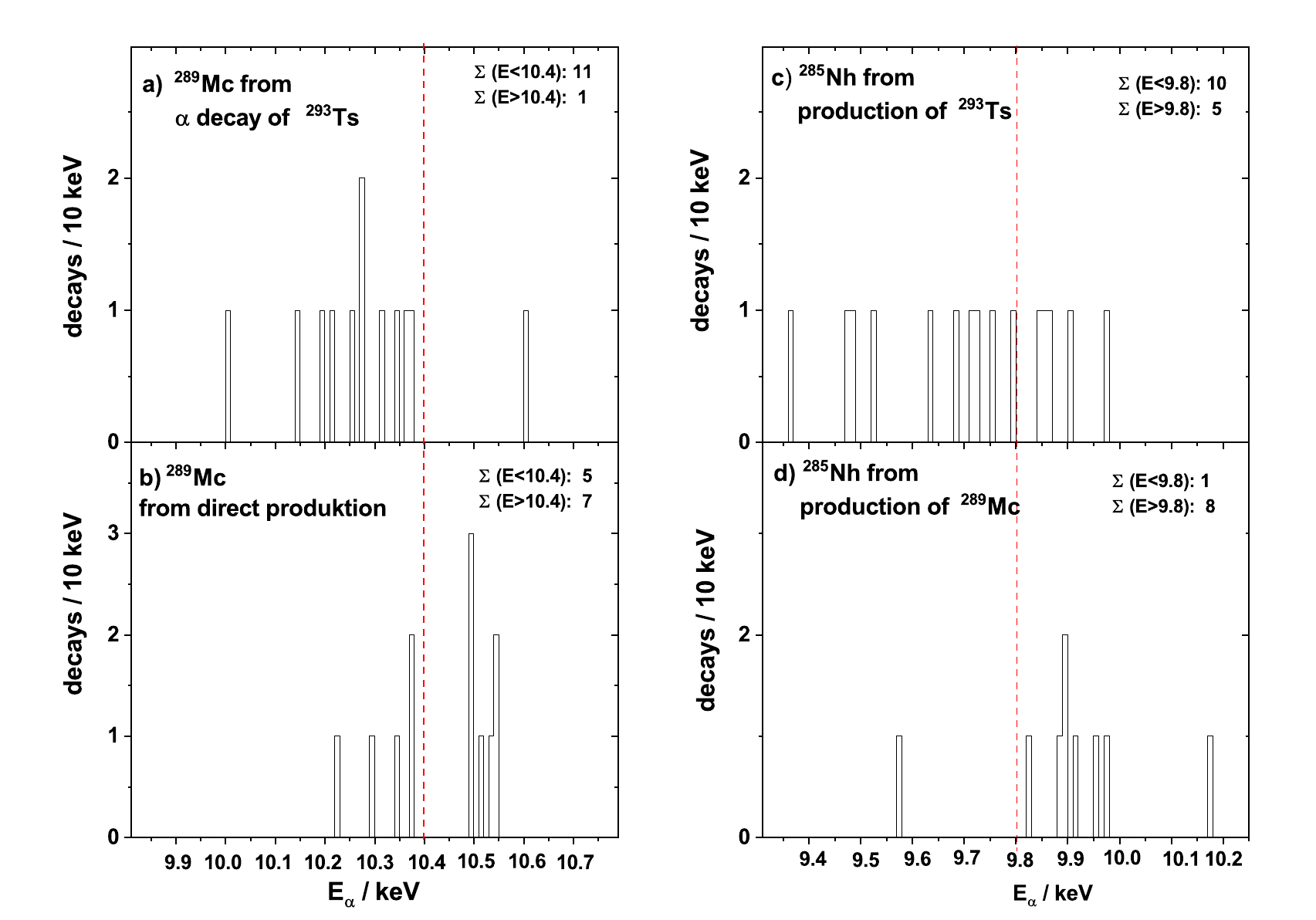}
	}
	% If not, use
	%\vspace{5cm}       % Give the correct figure height in cm
	\caption{Comparison of $\alpha$ decay spectra of $^{289}$Mc and $^{285}$Nh for different ways of production;
		a)$^{289}$Mc from decay of $^{293}$Ts; b) $^{289}$Mc from 'direct' production $^{243}$Am($^{48}$Ca,2n)$^{289}$Mc;
		c)$^{285}$Nh produced in $\alpha$ decay chains starting from $^{293}$Ts;
		d)$^{285}$Nh produced in $\alpha$ decay chains starting from $^{289}$Mc.
	}
	\label{fig:13}       % Give a unique label
\end{figure}

The data are shown in table 1. Evidently the events $\alpha_{4}$ and 'SF' would represent $^{287}$Fl and $^{283}$Cn if the chain starts at $^{299}$120.
$\alpha_{4}$ is recorded as an $\alpha$ particle escaping the detector, releasing only an energy loss of $\Delta$E\,=\,0.353 MeV in it. Using the measured lifetime (20 s) 
and a hindrance factor HF\,=\,104, as derived from the full energy $\alpha$ event (10.29 MeV) attributed to $^{287}$Fl in \cite{Ogan99}, the authors calculated a full $\alpha$ decay energy 
for $\alpha_{4}$ of {\it{E}}\,=\,10.14$^{+0.09}_{-0.27}$ MeV. Using the same procedure they obtained a full $\alpha$ energy {\it{E}}\,=\,10.19$^{+0.10}_{-0.28}$ MeV for the {\it{E}}\,=\,2.31 MeV - 'escape' event in \cite{Ogan99}.
The time differences {\it{$\Delta$T($\alpha_{3}$ - $\alpha_{4}$)}} and  {\it{$\Delta$T($\alpha_{4}$ -SF)}} resulted
in lifetimes {\it{$\tau$($\alpha_{4}$)}}\,=\,20$^{+89}_{-9}$ s 
({\it{T$_{1/2}$}}\,=\,14$^{+62}_{-6}$ s) and   {\it{$\tau$(SF)}}\,=\,12$^{+56}_{-5}$ min 
({\it{T$_{1/2}$}}\,=\,500$^{+233}_{-208}$ s) \cite{HofH16} and thus were inline with the 'VASSILISSA - data' for $^{287}$Fl ({\it{E$_{\alpha}$}}\,=\,10.29$\pm$0.02 MeV, {\it{T$_{1/2}$}}\,=\,5.5$^{+9.9}_{-2.1}$s)
and $^{283}$Cn ({\it{T$_{1/2}$}}\,=\,308$^{+212}_{-89}$s).\\
This finding was seen as a 'mutual support' of the data strengthen the (tentative) assignments of the chains in \cite{Ogan99,HofH16}, although the authors in \cite{HofH16}
could not give a reasonable explanation, why these data were only seen in the VASSILISSA experiment and could not be reproduced in other laboratories. As it was shown in 
\cite{HesA17} that the decay chain in \cite{HofH16} represents just a sequence of background events, it becomes clear, that blinkered data analysis may lead to 
correlations between background events even if they are obtained in different experiments.\\

\begin{figure}
%\begin{figure*}
\resizebox{0.95\textwidth}{!}{%
  \includegraphics{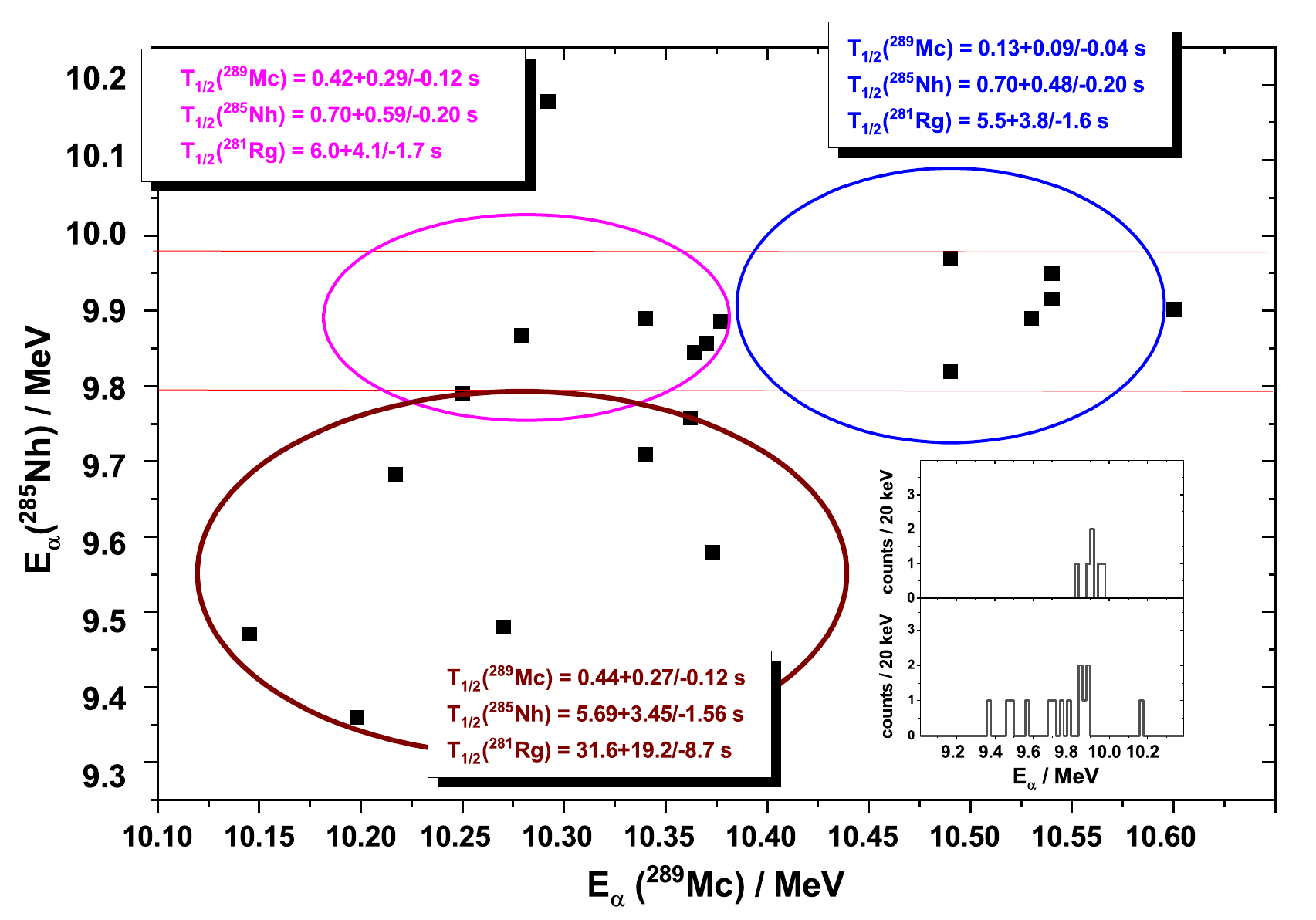}
}
% If not, use
%\vspace{5cm}       % Give the correct figure height in cm
\caption{ $\alpha$ - $\alpha$ - correlations between decays of $^{289}$Mc and $^{285}$Nh; the insert shows the 
energy distribution of $^{285}$Nh (E$_{\alpha}$\,=\,9.75-10.0 MeV), either correlated to $^{289}$Mc 
E$_{\alpha}$\,$>$\,10.4 MeV (upper figure) or  E$_{\alpha}$\,$<$\,10.4 MeV (lower figure).
}
\label{fig:14}       % Give a unique label
\end{figure}

\begin{figure}
%\begin{figure*}
\resizebox{0.95\textwidth}{!}{%
  \includegraphics{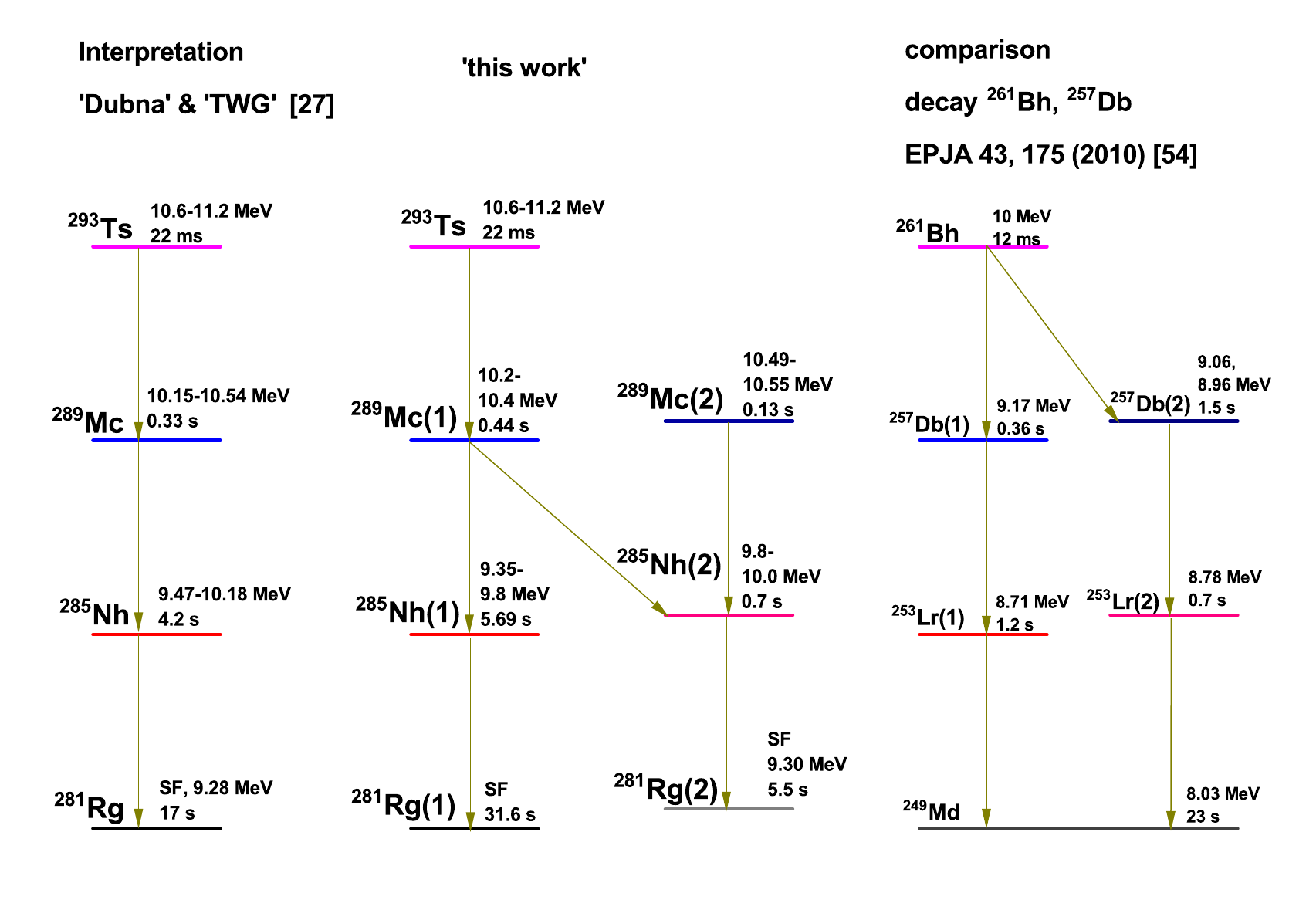}
}
% If not, use
%\vspace{5cm}       % Give the correct figure height in cm
\caption{ Suggestion for decay schemes of $^{293}$Ts, $^{289}$Mc(1) and $^{289}$Mc(2).
}
\label{fig:15}       % Give a unique label
\end{figure}

\subsection{\bf{6.5 Alpha decay chain of N-Z = 59 nuclei}}
Within the so far assigned superheavy nuclei, the decay properties of the {\it{N-Z}}\,=\,59 nuclei are  of specific importance and
interest, as the acceptance of discovery of elements {\it{Z}}\,=\,117 (tenessine) and  {\it{Z}}\,=\,115 (moscovium) is based on them.
The heaviest known nucleus of that chain, $^{293}$Ts, was produced by the reaction $^{249}$Bk($^{48}$Ca,4n)$^{293}$Ts
\cite{Ogan13,Ogan12,OgA10,OgA11,Khu14,Khu19}. A second entry point into this  chain is $^{289}$Mc produced in the reaction
$^{243}$Am($^{48}$Ca,2n)$^{289}$Mc \cite{OgA13,OgA12,FoR16}. It was stated in \cite{Ogan13} 'decay energies and halflives of 
the nuclei $^{289}$115 $\rightarrow$ $^{285}$113 $\rightarrow$ $^{281}$Rg decay chains observed in the $^{243}$Am + $^{48}$Ca reaction
agree within the statistical uncertainties  with the decay properties of the daughter nucleus of the $^{293}$117 nucleus produced in the $^{249}$Bk($^{48}$Ca, 4n)$^{293}$117 reaction (10 events) [...]. Such agreement provides indirect identification  and consistency checks via cross - bombardment production of the same nuclei in different fusion reactions of $^{243}$Am and $^{249}$Bk targets with $^{48}$Ca projectiles'.
The fourth IUPAC/IUPAP Joint Working Party (JWP) on the priority of claims to the discovery of the new elements Z\,=\,113, 115 and 117 followed that statement and regarded it as a central issue to have met the criteria for the elements  {\it{Z}}\,=\,115 and  {\it{Z}}\,=\,117 using the following phrasing \cite{KarB16}:
a) Element 115: 'JWP ASSESSMENT: The 2010 [...] jointly with the 2013 [...] collaborations of Oganessian et al. have met the Criteria for discovery of the element with the atomic number  {\it{Z}}\,=\,115 in as much as the reproducibility of the alpha chain energies and lifetimes of $^{289}$115 in a cross reaction comparison is very convincing.'\\
b) Element 117: 'JWP ASSESSMENT: A convincing case in cross reaction producing $^{289}$115 and $^{285}$113 from both $^{48}$Ca + $^{249}$Bk and $^{243}$Am is demonstrated in the top of the precious table. Thus, the 2010 [...], 2012 [...] and 2013 [...] jointly with the 2013 [...] collaborations of Oganessian et al. have met the criteria for discovery of the elements with atomic numbers  {\it{Z}}\,=\,115 and  {\it{Z}}\,=\,117.'\\
This assessment was critizised and a different interpretation of the results was suggested \cite{FoR16a}. This issue will be illuminated in the
following discussion. A final solution of the problem, however, cannot be presented on the basis of the available data.\\
The so far published decay data for the members of the  {\it{N-Z}}\,=\,59 chain  \cite{Oga16} are summarized in table 2. It should be noted, however,
that they are based solely on the results from the DGFRS - experiments. A complete list of the decay data published so far 
for the  {\it{N-Z}}\,=\,59 chain members $^{293}$Ts, $^{289}$Mc, $^{285}$Nh and $^{281}$Rg are given in table 3. 
Altogether eighteen decay chains assigned to start from $^{293}$Ts are reported so far, sixteen from experiments at the DGFRS (Dubna) \cite{Ogan13,Ogan12,OgA11} and two from an experiment at TASCA (GSI) \cite{Khu19}. Evidently only one of the TASCA chains was complete, in the second one the first two members ($^{293}$Ts, $^{289}$Mc) were missing. Eleven chains starting 
from $^{289}$Mc were reported, four from experiments performed at the DGFRS \cite{OgA13}, seven from a TASCA experiment
\cite{FoR16}. Also included in table 3 are three events observed in an irradiation of $^{243}$Am with $^{48}$Ca at the BGS, Berkeley \cite{GaG15}, although they were not explicitly assigned to $^{289}$Mc.\\ 
At first glance three items are striking:\\
a) four of the decay chains interpreted to start from $^{289}$Mc consist of an $\alpha$\,-\,sf correlation, i.e. sf decay of $^{285}$Nh, while in none of the eighteen chains starting from $^{293}$Ts fission of $^{285}$Nh was observed;\\
b) in chain no. D4 of \cite{OgA13} $^{285}$Nh has an $\alpha$ - decay energy more 0.2 MeV higher than that of all other decays where the $\alpha$ particle was registered with full energy (23 cases), which all had values {\it{E$_{\alpha}$}}\,$<$\,10 MeV.\\
c) $\alpha$ decay of $^{281}$Rg was only observed in decay chains starting from $^{293}$Ts.\\

To obtain more detailed information on the decay properties of the isotopes assigned to the {\it{N-Z}}\,=\,59 chain
a closer inspection of the data listed in table 3 was performed, specifically the results from the 'entry' into the chain
at $^{293}$Ts (reaction $^{48}$Ca + $^{249}$Bk) and 'entry' into the chain at $^{289}$Mc (reaction $^{48}$Ca + $^{243}$Am) 
were compared.
The resulting $\alpha$ spectra are shown in fig. 13. Before starting a detailed discussion
it seems, however, necessary to stress some items that could cause confusion.
First, as discussed in sect. 4.4 (individual) $\alpha$ energies measured in the experiments performed at the
different laboratories (DGFRS Dubna, TASCA Darmstadt, BGS Berkeley) vary considerably, eventually due to the 
calibration procedures applied. Differences {\it{$\Delta$E}}\,=\,76 keV were found between the DGFRS and BGS results,
and {\it{$\Delta$E}}\,=\,41 keV between the DGFRS and TASCA results for $^{272}$Bh; as $\approx$90 $\%$ of the data in Table 3
are from DGFRS or TASCA an uncertainty of $\approx$50 keV in the absolute value may be considered. As will be shown, the 
energy differences from the different production mechanisms are munch larger and thus cannot be attributed to 
calibration discrepancies.\\
Part of the $\alpha$ energies were obtained as sum events from 'stop' and 'box' detector, thus suffering from worse 
accuracy due to worse energy resolution. A few decays (three events) from the DGFRS experiments were registered as 'box - only' events
(i.e. the $\alpha$ particle escaped the 'stop' detector with an energy loss below the registration threshold). 
That means, the measured energies are too low by a few hundred keV. Due to the low number of these events also 
this feature cannot be 
the reason for the differences in the energy distribtions.\\
In the considered reaction on direct production of $^{289}$Mc also $^{288}$Mc is produced; the assigment to $^{289}$Mc
was based on the fact, that the properties of the decay chains did not fit to the decay chain of $^{288}$Mc. 
In principle a production of $^{287}$Mc by the reaction $^{243}$Am($^{48}$Ca,4n)$^{287}$Mc can also not be ruled out.
However, the properties of the decay chains attributed to $^{289}$Mc are not in-line with the decay proprties of $^{287}$Mc
and its daughter products (see e.g. \cite{Oga16}).\\

\begin{table}
	\caption{Summary of decay properties of {\it{N-Z}}\,=\,59 nuclei; data taken from \cite{Oga16}.}
	\label{tab:2}       % Give a unique label
	% For LaTeX tables use
	\begin{tabular}{llll}
		\hline\noalign{\smallskip}
		\noalign{\smallskip}\hline\noalign{\smallskip}
		Isotope & Decay mode & E$_{\alpha}$ / MeV  & half-life \\ 
		\hline\noalign{\smallskip}     
		$^{293}$Ts  & $\alpha$ &  10.60-11.20 & 22$^{+8}_{-4}$ ms  \\ 
		$^{289}$Mc  & $\alpha$ &  10.15-10.54 & 330$^{+120}_{-80}$ ms  \\
		$^{285}$Nh  & $\alpha$ &   9.47-10.18 & 4.2$^{+1.4}_{-0.8}$ s  \\ 
		$^{281}$Rg  & $\alpha$($\approx$0.12, SF (0.88$^{+0.07}_{-0.09}$) &   9.28$\pm$0.05 & 17$^{+6}_{-3}$ s  \\ 
		$^{277}$Mt  & SF &   & 5$^{+9}_{-2}$ ms  \\ 
	\end{tabular}
	% Or use
	\vspace*{0.cm}  % with the correct table height
\end{table}  

\begin{table}
\caption{Summary observed decay chains starting from either $^{293}$Ts or $^{289}$Mc.}
\label{tab:3}       % Give a unique label
% For LaTeX tables use
\footnotesize{
\begin{tabular}{lllllllll}
\hline\noalign{\smallskip}
\noalign{\smallskip}\hline\noalign{\smallskip}
 Ref. & $^{293}$Ts &   & $^{289}$Mc &  & $^{285}$Nh &   & $^{281}$Rg &   \\ 
          & E$_{\alpha}$/MeV & $\Delta$t/ms & E$_{\alpha}$/MeV & $\Delta$t/s & E$_{\alpha}$/MeV & $\Delta$t/s & E$_{\alpha}$/MeV & $\Delta$t/s \\
\hline\noalign{\smallskip}     
\cite{OgA11}  & 10.99 &  17.01 & missing &     &  9.72  & (16.17) & SF & 40.19   \\ 
\cite{OgA11}  & 11.14 &  7.89 & missing &     &  9.52**  & (2.23) & SF & 4.25   \\ 
\cite{OgA11}  & 11.08* &  4.60 & 10.34 & 0.0175 &  9.71**  &  1.17 & SF & 12.3   \\ 
\cite{OgA11}  & 10.91 &  53.0 & 10.25 & 0.5118 &  9.79  & 0.238 & SF & 31.66   \\ 
\cite{OgA11}  & 11.00 &  20.24 & 10.27 & 0.4244 &  9.48  & 13.49 & SF & 76.56   \\ 
\cite{Ogan12}  & 10.90$\pm$0.10* & 7.525 & 10.37$\pm$0.28** & 0.2665  & 9.857$\pm$0.040 & 1.5155 & SF & 9.4192   \\ 
\cite{Ogan12}  & 11.142$\pm$0.065 & 3.305 & 10.310$\pm$0.065 & 0.1819  & missing &  & SF & (7.4538)   \\ 
\cite{Ogan12}  & 10.114$\pm$0.089* & 153.948 & missing &   & 9.631$\pm$0.067 & (19.0456) & SF & 1.4809  \\ 
\cite{Ogan12}  & 10.914$\pm$0.068 & 10.547 & 10.198$\pm$0.068 & 1.4348  & 9.36$\pm$0.3** & 1.3153 & SF & 103.406   \\ 
\cite{Ogan12}  & 10.598$\pm$0.049 & 109.878 & 10.217$\pm$0.049 & 0.1510  & 9.683$\pm$0.049 & 1.5155 & SF & 42.1349   \\ 
\cite{Ogan13}  & 10.969$\pm$0.068 & 0.043 & 10.60$\pm$0.33** & 0.0136  & 9.902$\pm$0.068 & 0.421 & SF & 15.0551   \\ 
\cite{Ogan13}  & 11.183$\pm$0.048 & 2.553 & 10.364$\pm$0.048 & 0.9572  & 9.845$\pm$0.048 & 1.3712 & SF & 4.751   \\ 
\cite{Ogan13}  & 11.203$\pm$0.070 & 8.173 & 10.279$\pm$0.070 & 0.0565  & 9.867$\pm$0.070 & 4.213 & 9.36$\pm$0.30 & 3.642   \\ 
\cite{Ogan13}  & 11.059$\pm$0.050 & 7.525 & 10.362$\pm$0.050 & 0.0161  & 9.758$\pm$0.108 & 0.2589 & 9.280$\pm$0.050 & 2.9249   \\ 
\cite{Ogan13}  & missing &  & 10.145$\pm$0.066 & (0.0852)  & 9.471$\pm$0.066 & 0.2456 & SF & 21.8372   \\ 
\cite{Ogan13}  & 10.190$\pm$0.070 & 36.424 & missing &   & missing &  & SF & (4.488)   \\ 
\cite{Khu19}  & 9.70$\pm$0.08 & 8.65 & 10.00$\pm$0.03 & 0.07698  & (2.74) & 0.97 & 9.34$\pm$0.08 & 1.44   \\ 
\cite{Khu19}  & missing &  & missing &   & 9.97$\pm$0.08 & 0.49 & 9.31$\pm$0.08 & 2.57   \\ 
\hline\noalign{\smallskip}
\cite{FoR16}  &  &  & 10.51$\pm$0.01 & 0.227  & SF & 0.378 &  &    \\ 
\cite{FoR16}  &  &  & (1.45$\pm$0.01) & 0.0645  & SF & 0.366 &  &    \\ 
\cite{FoR16}  &  &  & 10.54$\pm$0.04 & 0.261  & 9.95$\pm$0.05 & 1.15 & SF & 0.343   \\ 
\cite{FoR16}  &  &  & 10.34$\pm$0.01 & 1.46  & 9.89$\pm$0.01 & 0.0262 & SF & 0.432   \\ 
\cite{FoR16}  &  &  & 10.49$\pm$0.04 & 0.345  & 9.97$\pm$0.01 & 0.369 & SF & 13.4   \\ 
\cite{FoR16}  &  &  & 10.53$\pm$0.01 & 0.210  & 9.89$\pm$0.05 & 1.05 & SF & 8.27   \\ 
\cite{FoR16}  &  &  & (0.541$\pm$0.03) & 0.815  & (3.12$\pm$0.01) & 2.33 & SF & 2.89  \\ 
\cite{OgA13}  &  &  & 10.377$\pm$0.062 & 0.2562  & 9.886$\pm$0.062 & 1.4027 & SF & 1.977   \\ 
\cite{OgA13}  &  &  & 10.540$\pm$0.123 & 0.0661  & 9.916$\pm$0.072 & 1.55 & SF & 2.364   \\ 
\cite{OgA13}  &  &  & 10.373$\pm$0.050 & 2.3507  & 9.579$\pm$0.005 & 22.5822 & SF & 60.185   \\ 
\cite{OgA13}  &  &  & 10.292$\pm$0.170*(D4) & 0.0536  & 10.178$\pm$0.055 & 0.4671 & SF & 0.0908   \\ 
\cite{GaG15}  &  &  & 10.49$\pm$0.05 & 0.214  & 9.82$\pm$0.02 & 1.54 & (SF)*** & 7.57   \\ 
\cite{GaG15}  &  &  & 10.49$\pm$0.02 & 0.0591  & SF & 0.824 &  &    \\ 
\cite{GaG15}  &  &  & 10.22$\pm$0.02 & 0.0455  & SF & 0.142 &  &    \\ 
\end{tabular}
}
% Or use
\vspace*{0.cm}  % with the correct table height
\end{table}

\normalsize
\begin{table}
\caption{Summary of half-life measurements.}
\label{tab:4}       % Give a unique label
% For LaTeX tables use
\begin{tabular}{llll}
\hline\noalign{\smallskip}
\noalign{\smallskip}\hline\noalign{\smallskip}
Isotope & E$_{/alpha}$ / MeV & T$_{1/2}$ / s  & correlation \\ 
\hline\noalign{\smallskip}     
$^{289}$Mc  & 10.51$\pm$0.02 & 0.13$^{+0.09}_{-0.04}$ & corr. to $^{285}$Nh (9.8\,-\,10.0 MeV) \\ 
            & 10.49\,-\,10.55  &                         &                                     \\
$^{289}$Mc  & 10.20\,-\,10.40 & 0.42$^{+0.29}_{-0.12}$ & corr. to $^{285}$Nh (9.8\,-\,10.0 MeV) \\ 
$^{289}$Mc  & 10.20\,-\,10.40 & 0.44$^{+0.27}_{-0.12}$ & corr. to $^{285}$Nh (9.3\,-\,9.8 MeV) \\ 
\hline\noalign{\smallskip}
$^{285}$Nh  & 9.8\,-\,10.0 & 0.70$^{+0.48}_{-0.20}$ & corr. to $^{289}$Mc (10.49\,-\,10.55 MeV) \\
$^{285}$Nh  & 9.8\,-\,10.0 & 0.70$^{+0.59}_{-0.20}$ & corr. to $^{289}$Mc (10.2\,-\,10.4 MeV) \\
$^{285}$Nh  & 9.97, 9.87, 9.76 & 1.02$^{+0.xx}_{-0.xx}$ & corr. to $^{281}$Rg $\alpha$ decay (9.32$\pm$0.04) \\
$^{285}$Nh  & 9.3\,-\,9.8 & 5.69$^{+3.45}_{-1.56}$ & corr. to $^{289}$Mc (10.2\,-\,10.4 MeV) \\
\hline\noalign{\smallskip}
$^{281}$Rg  & SF & 5.5$^{+3.8}_{-1.6}$ & corr. to $^{285}$Nh (9.8\,-\,10.0 MeV) and/or  \\ 
            &    &                     & corr. to $^{289}$Mc (10.49\,-\,10.55 MeV)   \\ 
$^{281}$Rg  & SF & 6.0$^{+4.1}_{-1.7}$ & corr. to $^{285}$Nh (9.8\,-\,10.0 MeV) and/or  \\ 
                        &    &                     & corr. to $^{289}$Mc (10.2\,-\,10.4 MeV)   \\ 
$^{281}$Rg  & SF & 31.6$^{+19.2}_{-8.7}$ & corr. to $^{285}$Nh (9.3\,-\,9.8 MeV) and/or  \\ 
                                                &    &                     & corr. to $^{289}$Mc (10.2\,-\,10.4 MeV)   \\ 
$^{281}$Rg  & $\alpha$ decays & 1.8$^{+1.8}_{-0.6}$ &   \\  
\end{tabular}
% Or use
\vspace*{0.cm}  % with the correct table height
\end{table}  

As seen from fig. 13 the $\alpha$ energy spectra of $^{289}$Mc (figs. 13a,b) and $^{285}$Nh (fig. 13c,d) exhibit significant differences for the different production mechanisms. In the production by $\alpha$ decay of $^{293}$Ts nearly all $^{289}$Mc 
events (in 11 of 12 cases) have energies {\it{E$_{\alpha}$}}\,$<$\,10.4 MeV, while in the direct production seven of twelve events 
are concentrated in the energy range {\it{E$_{\alpha}$}}\,=\,(10.49-10.55)\,MeV. Also halflives are different; for the group at 
{\it{E$_{\alpha}$}}\,$<$\,10.4 MeV one abtains {\it{T$_{1/2}$}}\,=\,0.39$^{+0.14}_{-0.08}$ s, for the group at 
{\it{E$_{\alpha}$}}\,$>$\,10.4 MeV one abtains {\it{T$_{1/2}$}}\,=\,0.11$^{+0.06}_{-0.03}$ s. An analogue situation is found for $^{285}$Nh;
about two third (10 of 15 cases) of the $\alpha$ events from chains starting at $^{293}$Ts have energies {\it{E$_{\alpha}$}}\,$<$\,9.8 MeV, while for decays within the chains starting at $^{289}$Mc only one of nine events is located in this energy interval. \\
This behavior is also evident in the $\alpha$ - $\alpha$ correlation spectrum (fig. 14); the {\it{E$_{\alpha}$}}\,$>$\,10.4 MeV 
component of $^{289}$Mc is exclusively correlated to $^{285}$Nh events in the energy interval E$_{\alpha}$\,=\,(9.8-10.0) MeV,
while for $^{289}$Mc of {\it{E$_{\alpha}$}}\,$<$\,10.4 MeV only about half of the events are correlated
to $^{285}$Nh decays in that energy interval. But as seen in the insert of fig. 14 the energy distributions are not the same.
The $^{285}$Nh decays correlated to $^{289}$Mc at {\it{E$_{\alpha}$}}\,$>$\,10.4 MeV (upper figure) have somewhat higher energies 
of {\it{E(mean)}}\,=\,9.91$\pm$0.05 MeV than those correlated to $^{289}$Mc at {\it{E$_{\alpha}$}}\,$<$\,10.4 MeV (lower figure), 
which have {\it{E(mean)}}\,=\,9.86$\pm$0.04 MeV.\\
All these differences indicate that in the direct production a $^{289}$Mc component (it also could be a different isotope)
having an energy {\it{E$_{\alpha}$}}\,=\,10.53$\pm$0.04 MeV and a hallife of {\it{T$_{1/2}$}}\,=\,0.11$^{+0.06}_{-0.03}$ s is produced 
which is not present in the decay chain of $^{293}$Ts, i.e. eventually an isomeric state is populated by deexcitation of the compound nucleus in 'direct' production reaction, which is not populated by $\alpha$ decay of $^{293}$Ts. That would be nothing surprising,
as a couple of such examples are known in the transfermium region, e.g. $^{251}$No, $^{257}$Rf. \\
The assumption of an isomeric state in $^{289}$Mc does not vitiate the JWG assassment, but it clearly shows, how fragile 
such conclusions may be on the basis of very low statistics. \\

So we have to face the following situation considerating all decays listed in table 3:
\begin{itemize}
\item Within the production $^{289}$Mc via $\alpha$ decay of $^{293}$Ts the $\alpha$ particles of $^{289}$Mc are located in an
energy interval {\it{E$_{\alpha}$}}\,=\,(10.2\,-\,10.4) MeV; the resulting half-life is {\it{T$_{1/2}$}}\,=\,0.28$^{+0.13}_{-0.07}$ s; 
\item Within the production of $^{289}$Mc 'directly' via the reaction $^{243}$Am($^{48}$Ca,2n)$^{289}$Mc we observe two components 
in the $\alpha$ energies: one at {\it{E$_{\alpha}$}}\,=\,(10.2\,-\,10.4) MeV with a half-life of {\it{T$_{1/2}$}}\,=\,0.71$^{+0.71}_{-0.24}$ s; and
one at {\it{E$_{\alpha}$}}\,=\,(10.49\,-\,10.55) MeV with a half-life of {\it{T$_{1/2}$}}\,=\,0.12$^{+0.13}_{-0.07}$ s;
\item The $^{285}$Nh events from the production via $^{293}$Ts $\rightarrow$ $^{289}$Mc $\rightarrow$ $^{285}$Nh are spread over
an energy range {\it{E$_{\alpha}$}}\,=\,(9.35\,-\,10.0) MeV and exhibit a half-life  {\it{T$_{1/2}$($^{285}$Nh)}}\,=\,2.44$^{+0.94}_{-0.53}$ s;
\item The $^{285}$Nh events from the 'direct production' of $^{289}$Mc  via $^{289}$Mc $\rightarrow$ $^{285}$Nh are in the 
an energy range {\it{E$_{\alpha}$}}\,=\,(9.8\,-\,10.0) MeV (except the two events at 9.58 MeV and 10.18 MeV ) and exhibit a half-life  {\it{T$_{1/2}$($^{285}$Nh)}}\,=\,0.76$^{+0.38}_{-0.19}$ s;
\item Taking into account in addition the $\alpha$ - $\alpha$ - correlations $^{289}$Mc $\rightarrow$ $^{285}$Nh, we tentatively can distinguish
three groups
\begin{enumerate}
\item $^{289}$Mc ({\it{E$_{\alpha}$}}\,=\,(10.49-10.55) MeV) $\rightarrow$ $^{285}$Nh  ({\it{E$_{\alpha}$}}\,=\,(9.8-10.0) MeV),
with T$_{1/2}$($^{289}$Mc)\,=\,0.13$^{+0.09}_{-0.04}$ s and T$_{1/2}$($^{285}$Nh)\,=\,0.70$^{+0.48}_{-0.20}$ s; the fission events terminating
the decay chain have a half-life {\it{T$_{1/2}$($^{281}$Rg)}}\,=\,5.5$^{+3.8}_{-1.6}$ s;\\ 
\item $^{289}$Mc ({\it{E$_{\alpha}$}}\,=\,(10.2\,-\,10.4) MeV) $\rightarrow$ $^{285}$Nh  (E$_{\alpha}$\,=\,(9.8-10.0) MeV),
with {\it{T$_{1/2}$($^{289}$Mc)}}\,=\,0.42$^{+0.29}_{-0.12}$ s and {\it{T$_{1/2}$($^{285}$Nh)}}\,=\,0.70$^{+0.49}_{-0.20}$ s; the fission events terminating
the decay chain have a half-life {\it{T$_{1/2}$($^{281}$Rg)}}\,=\,6.0$^{+4.1}_{-1.7}$ s;\\ 
\item $^{289}$Mc ({\it{E$_{\alpha}$}}\,=\,(10.15\,-\,10.4) MeV) $\rightarrow$ $^{285}$Nh  ({\it{E$_{\alpha}$}}\,=\,(9.35\,-\,9.8)MeV),
with {\it{T$_{1/2}$($^{289}$Mc)}}\,=\,0.44$^{+0.27}_{-0.12}$ s and {\it{T$_{1/2}$($^{285}$Nh)}} = 5.69$^{+3.45}_{-1.56}$ s; the fission events terminating
the decay chain have a half-life {\it{T$_{1/2}$($^{281}$Rg)}}\,=\,31.6$^{+19.2}_{-8.7}$ s.\\ 
\end{enumerate}
\end{itemize}

Under these circumstances we tentatively can distinguish the following decay chains.
\begin{itemize}
\item $^{289}$Mc ({\it{E$_{\alpha}$}}\,=\,(10.4\,-\,10.6) MeV, {\it{T$_{1/2}$}}\,=\,0.13$^{+0.09}_{-0.04}$ s) $\rightarrow$
$^{285}$Nh  ({\it{E$_{\alpha}$}}\,=\,(9.8\,-\,10.0 )MeV, {\it{T$_{1/2}$}}\,=\,0.70$^{+0.48}_{-0.20}$ s)  $\rightarrow$
$^{281}$Rg  (SF, {\it{T$_{1/2}$}}\,=\,5.5$^{+3.8}_{-1.6}$ s);
\item $^{289}$Mc ({\it{E$_{\alpha}$}}\,=\,(10.2\,-\,10.4) MeV, {\it{T$_{1/2}$}}\,=\,0.42$^{+0.29}_{-0.12}$ s) $\rightarrow$
$^{285}$Nh  ({\it{E$_{\alpha}$}}\,=\,(9.8\,-\,10.0 )MeV, T$_{1/2}$ = 0.70$^{+0.49}_{-0.20}$ s)  $\rightarrow$
$^{281}$Rg  (SF, {\it{T$_{1/2}$}}\,=\,6.0$^{+4.1}_{-1.7}$ s);\\
\item $^{289}$Mc ({\it{E$_{\alpha}$}}\,=\,(10.2\,-\,10.4) MeV, {\it{T$_{1/2}$}}\,=\,0.44$^{+0.27}_{-0.12}$ s) $\rightarrow$
$^{285}$Nh  ({\it{E$_{\alpha}$}}\,=\,(9.35-9.8) MeV, {\it{T$_{1/2}$}}\,=\,5.69$^{+3.45}_{-1.56}$ s)  $\rightarrow$
$^{281}$Rg  (SF, {\it{T$_{1/2}$}}\,=\,31.6$^{+19.2}_{-8.7}$ s);\\
\end{itemize}

This at first glance somewhat puzzling seeming decay pattern can qualitatively explained by existence 
of low lying long lived isomeric states in $^{289}$Mc, $^{285}$Nh and $^{281}$Rg decaying by $\alpha$ emission or
spontaneous fission. The existence of such states is due to existence of Nilsson states with low and high spins placed closely
at low excitation energies; the decay by internal transitions of such states is hindered by large spin differences and thus lifetimes become
long and $\alpha$ decay can compete with internal transitions. That is a well known phenomena in the transfermium regions.
In direct production both states are usually populated; in production by $\alpha$ decay the population of the states 
depend on the decay of the mother nucleus. If there are two longlived isomeric states in the mother nucleus, also two longlived
states in the daughter nucleus may populated, see e.g. decay of $^{257}$Db $\rightarrow$ $^{253}$Lr \cite{HesH01}; if there is only one
$\alpha$ emitting state populated by the deexcitation process two cases are possible; either only one state in the daughter nucleus is populated
as e.g. in the decay $^{261}$Sg $\rightarrow$ $^{257}$Rf \cite{Streich10}, or both long-lived states in the daughter nucleus may be
populated, as known for $^{261}$Bh $\rightarrow$ $^{257}$Db \cite{HesA10a}.\\
Under this circumstances the puzzling behavior can be understood in the following way:
decay of $^{293}$Ts populates one state $^{289}$Mc(1) (10.2-10.4 MeV), while direct production populates two states
$^{289}$Mc(1) and $^{289}$Mc(2) (10.4-10.6 MeV); $^{289}$Mc(2) decays exclusively into one state $^{285}$Nh(2)(9.8-10.0 MeV), which then
decays into $^{281}$Rg(2) which undergoes fission and $\alpha$ decay. $^{289}$Mc(1) on the other side partly populates  $^{285}$Nh(2) and $^{285}$Nh(1) (9.35-9.8 MeV) which then decays into $^{281}$Rg(1) which undergoes probably nearly exclusively spontaneous fission. 
The resulting tentative decay scheme is shown in fig. 15.\\

In addition there might be other contributions, e.g.
the chain marked as D4 in Table 3, which does not fit to the other ones.\\
Also the very short chains in Table 3 consiting of $\alpha$ $\rightarrow$ SF seemingly may have a different origins
The halflives of the $\alpha$ events, T$_{1/2}$($\alpha$) = 0.069 $^{+0.069}_{-0.23}$ s, and of the fission events. 
and T$_{1/2}$(SF) = 0.3 $^{+0.30}_{-0.1}$ are lower than the values for  $^{289}$Mc(2) and $^{285}$Nh(2),
but considering the large uncertainties they are not in disagreement. So they could indicate a fission branch of 
$^{285}$Nh in the order of b$_{SF}$$\approx$0.4. We remark here also the short half-life of 
T$_{1/2}$\,=\,1.8$^{+1.8}_{-0.6}$s of
the $\alpha$ events assigned  to $^{281}$Rg which is considerably shorter than the half-life of the fission events. Despite this fact we tentatively
assign them to $^{281}$Rg(1). \\
The joint analysis of the data presented for the decay chains interpreted to start either from $^{293}$Ts or from 
$^{289}$Mc seem to shed some light into the 'puzzeling' decay data reported so far and suggests a solution. It should be noted,
however, the conclusions drawn here must be confirmed by more sensitive measurements before they can be finally 
accepted.

 \begin{figure}
	%\begin{figure*}
	\resizebox{0.95\textwidth}{!}{%
		\includegraphics{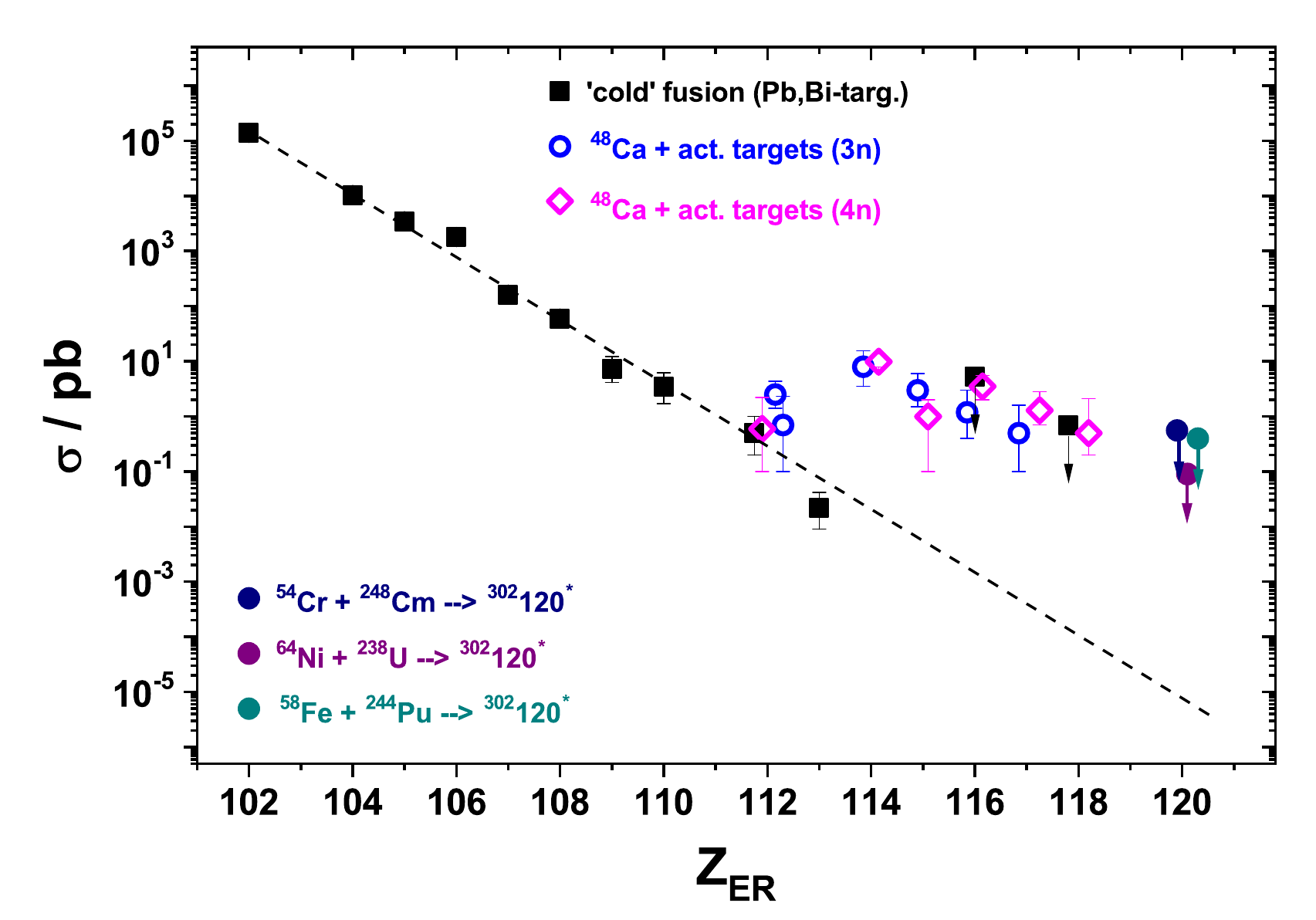}
	}
	% If not, use
	\vspace{0cm}       % Give the correct figure height in cm
	\caption{Systematics of maximum production cross-sections in cold fusion reactions and reactions using actinide targets 
	}
	\label{fig:16}       % Give a unique label
\end{figure}

 \begin{figure}
	%\begin{figure*}
	\resizebox{0.95\textwidth}{!}{%
		\includegraphics{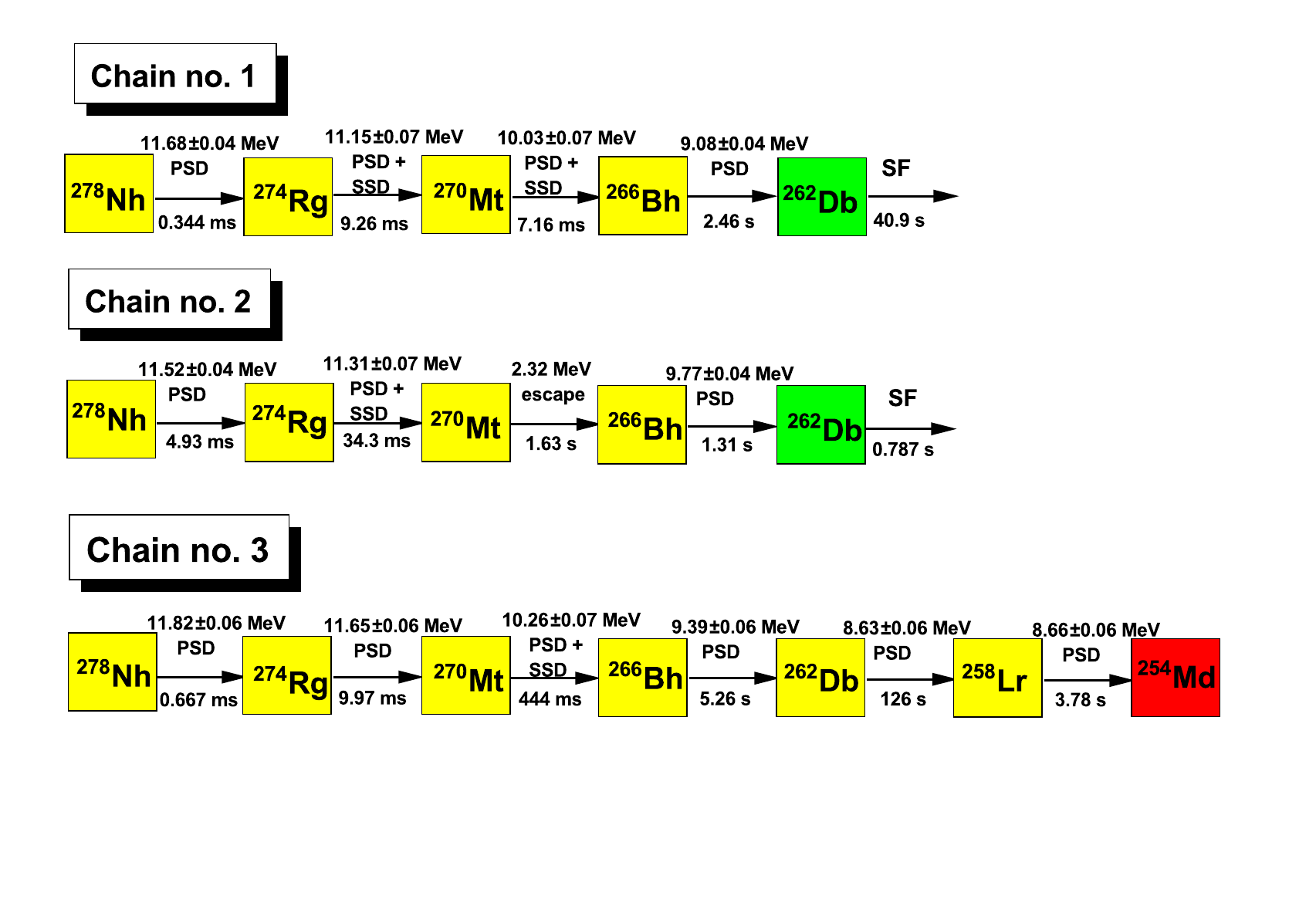}
	}
	% If not, use
	\vspace{-1cm}       % Give the correct figure height in cm
	\caption{Decay chains attributed to start from $^{278}$Nh \cite{Morita15}
	}
	\label{fig:17}       % Give a unique label
\end{figure}
 
  \begin{figure}
   %\begin{figure*}
   \resizebox{0.90\textwidth}{!}{%
     \includegraphics{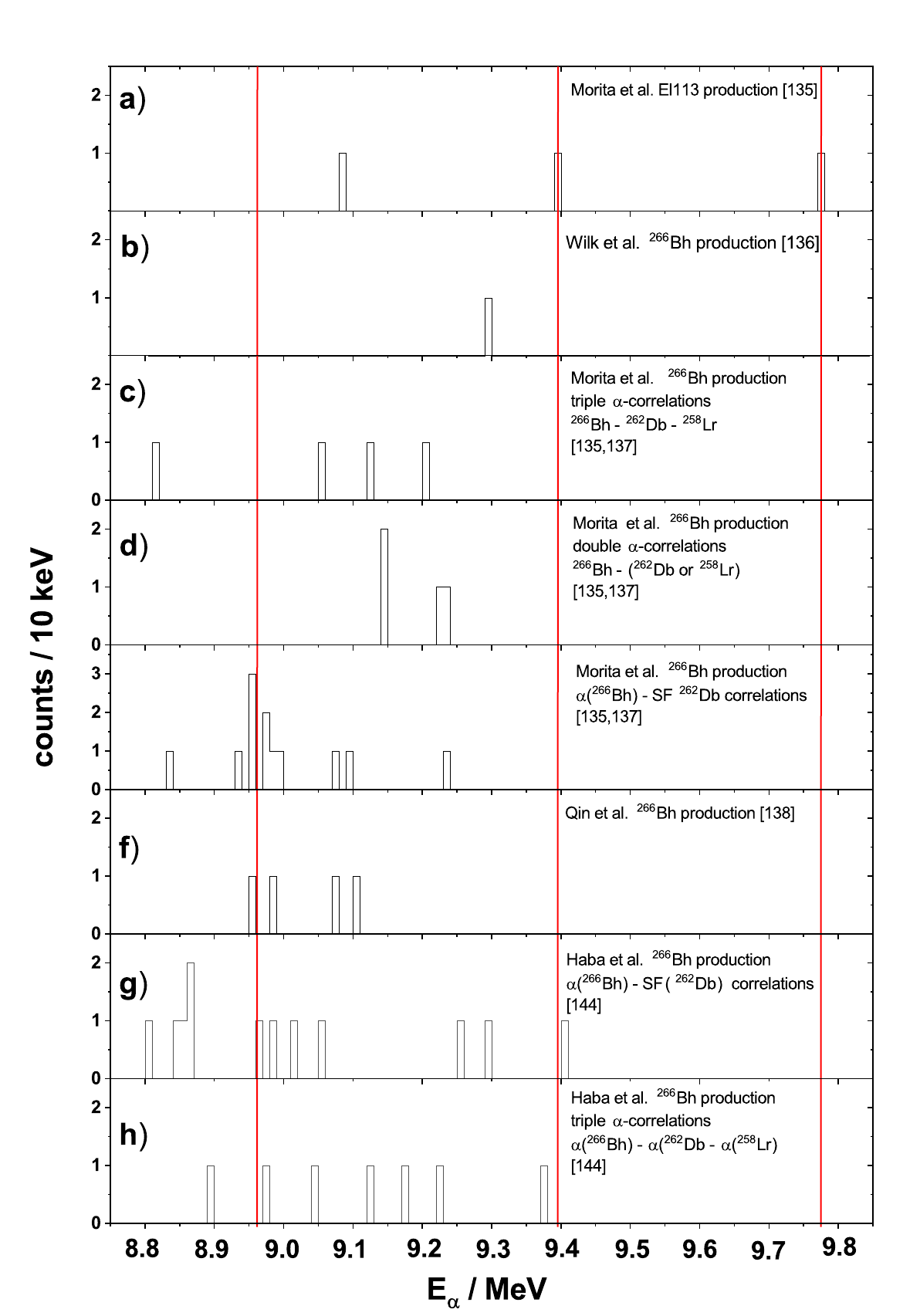}
   }
   % If not, use
   %\vspace{5cm}       % Give the correct figure height in cm
   \caption{$\alpha$ decay energies of $^{266}$Bh as reported by different authors;
   a) from decay of $^{278}$Nh \cite{Morita15},
   b) production via  $^{249}$Bk($^{22}$Ne,5n)$^{266}$Bh \cite{Wilk00},
   c) production via $^{248}$Cm($^{23}$Na,5n)$^{266}$Bh \cite{Morita09,Morita15}, $\alpha$ energies from triple $\alpha$
     correlations $\alpha_{1}$($^{266}$Bh) - $\alpha_{2}$($^{262}$Db) - $\alpha_{3}$($^{258}$Lr),
   d) production via $^{248}$Cm($^{23}$Na,5n)$^{266}$Bh \cite{Morita09,Morita15}, $\alpha$ energies from double $\alpha$
     correlations $\alpha_{1}$($^{266}$Bh) - $\alpha_{2}$($^{262}$Db or $^{258}$Lr),
   e) production via $^{248}$Cm($^{23}$Na,5n)$^{266}$Bh \cite{Morita09,Morita15}, $\alpha$ energies from  $\alpha$  - SF
     correlations,
   f) production via $^{243}$Am($^{26}$Mg,3n)$^{266}Bh$ \cite{Qin06},
   g) production via $^{248}$Cm($^{23}$F,5n)$^{266}$Bh, $\alpha$($^{266}$Bh) - SF($^{262}$Db) correlations \cite{Haba20}, 
   h) production via $^{248}$Cm($^{23}$F,5n)$^{266}$Bh, triple correlations $\alpha_{1}$($^{266}$Bh) - $\alpha_{2}$($^{262}$Db) - $\alpha_{3}$($^{258}$Db) correlations \cite{Haba20}. 
   }
   \label{fig:18}       % Give a unique label
   \end{figure}
   
    \begin{figure}
         %\begin{figure*}
         \resizebox{0.95\textwidth}{!}{%
           \includegraphics{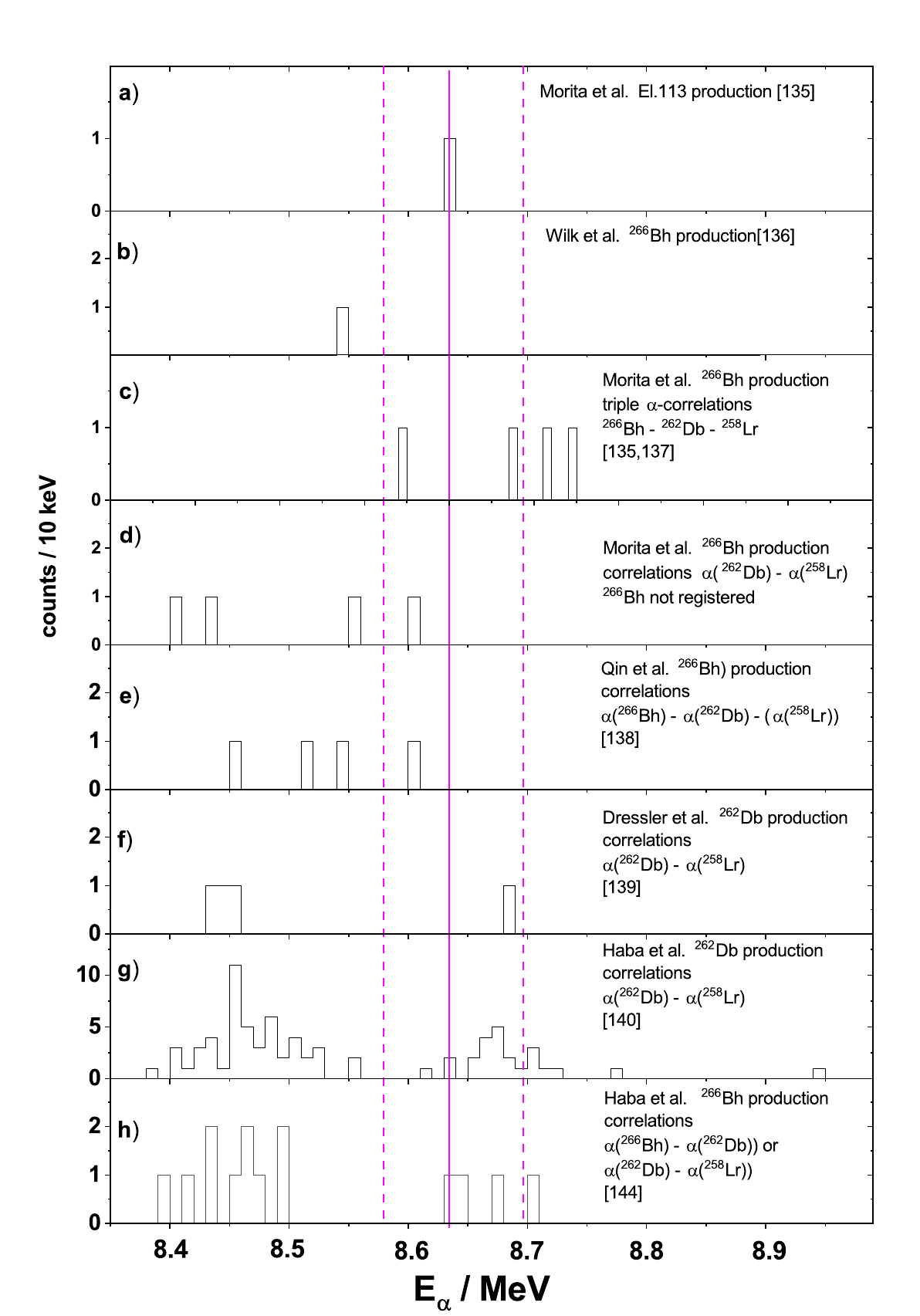}
         }
         % If not, use
         %\vspace{5cm}       % Give the correct figure height in cm
         \caption{$\alpha$ decay energies of $^{262}$Db as reported by different authors;
          a) from decay of $^{278}$Nh \cite{Morita15},
                                 b) production via  $^{249}$Bk($^{22}$Ne,5n)$^{266}$Bh \cite{Wilk00},
                                 c) production via $^{248}$Cm($^{23}$Na,5n)$^{266}$Bh \cite{Morita09,Morita15}, $\alpha$ energies from triple $\alpha$
                                   correlations $\alpha_{1}$($^{266}$Bh) - $\alpha_{2}$($^{262}$Db) - $\alpha_{3}$($^{258}$Lr),
                                 d) production via $^{248}$Cm($^{23}$Na,5n)$^{266}$Bh \cite{Morita09,Morita15}, $\alpha$ energies from double $\alpha$ correlations $\alpha_{1}$($^{262}$Db) - $\alpha_{2}$($^{258}$Lr), with $\alpha$ decay of $^{266}$Bh
                                 not recorded;
                                 e) production via $^{243}$Am($^{26}$Mg,3n)$^{266}$Bh \cite{Qin06},
                                 f) production via $^{248}$Cm($^{19}$F,5n)$^{262}$Db \cite{Dress99},
                                 g) production via $^{248}$Cm($^{19}$F,5n)$^{262}$Db \cite{Haba14},   
                                 h) production via $^{248}$Cm($^{23}$Na,5n)$^{266}$Bh \cite{Haba20}.                      
                                             }
         \label{fig:19}       % Give a unique label
     \end{figure}
         
    \begin{figure}
               %\begin{figure*}
               \resizebox{0.95\textwidth}{!}{%
                 \includegraphics{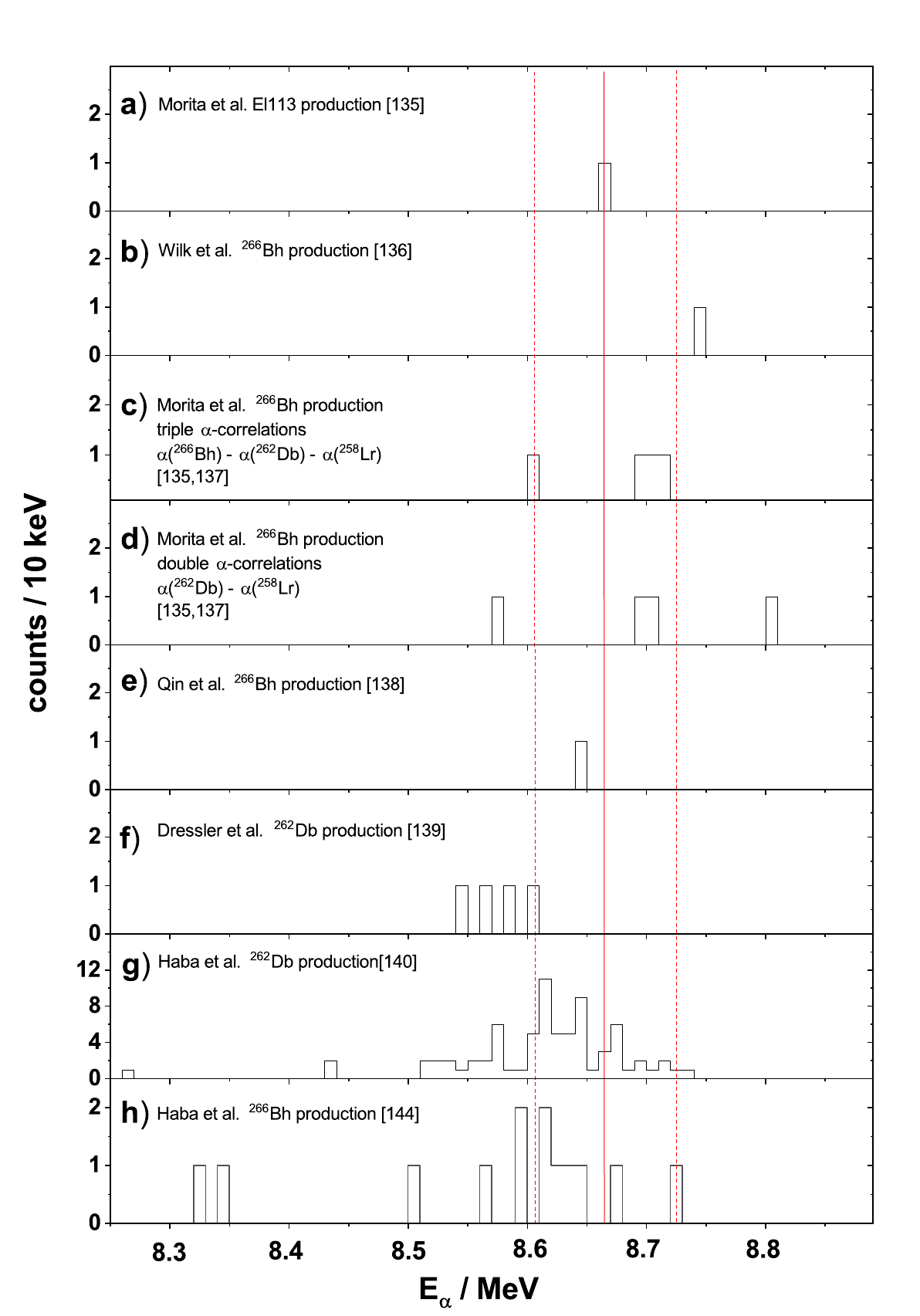}
               }
               % If not, use
               %\vspace{5cm}       % Give the correct figure height in cm
               \caption{$\alpha$ decay energies of $^{258}$Lr as reported by different authors;
                        a) from decay of $^{278}$Nh \cite{Morita15},
                                               b) production via  $^{249}$Bk($^{22}$Ne,5n)$^{266}$Bh \cite{Wilk00},
                                               c) production via $^{248}$Cm($^{23}$Na,5n)$^{266}$Bh \cite{Morita09,Morita15}, $\alpha$ energies from triple $\alpha$
                                                 correlatiions $\alpha_{1}$($^{266}$Bh) - $\alpha_{2}$($^{262}$Db) - $\alpha_{3}$($^{258}$Lr),
                                               d) production via $^{248}$Cm($^{23}$Na,5n)$^{266}$Bh \cite{Morita09,Morita15}, $\alpha$ energies from double $\alpha$ correlations $\alpha_{2}$($^{262}$Db) - $\alpha_{2}$($^{258}$Lr),
                                               with $\alpha$ decay of $^{266}$Bh not recorded,
                                               e)production via $^{243}$Am($^{26}$Mg,3n)$^{266}$Bh \cite{Qin06},
                                               f)production via $^{248}$Cm($^{19}$F,5n)$^{262}$Db \cite{Dress99},
                                               g)production via $^{248}$Cm($^{19}$F,5n)$^{262}$Db \cite{Haba14},
                                               h)production via $^{248}$Cm($^{23}$Na,5n)$^{266}$Bh \cite{Haba20}.   
               }
               \label{fig:20}       % Give a unique label
               \end{figure}
  
     \begin{figure}
                %\begin{figure*}
                \resizebox{0.95\textwidth}{!}{%
                  \includegraphics{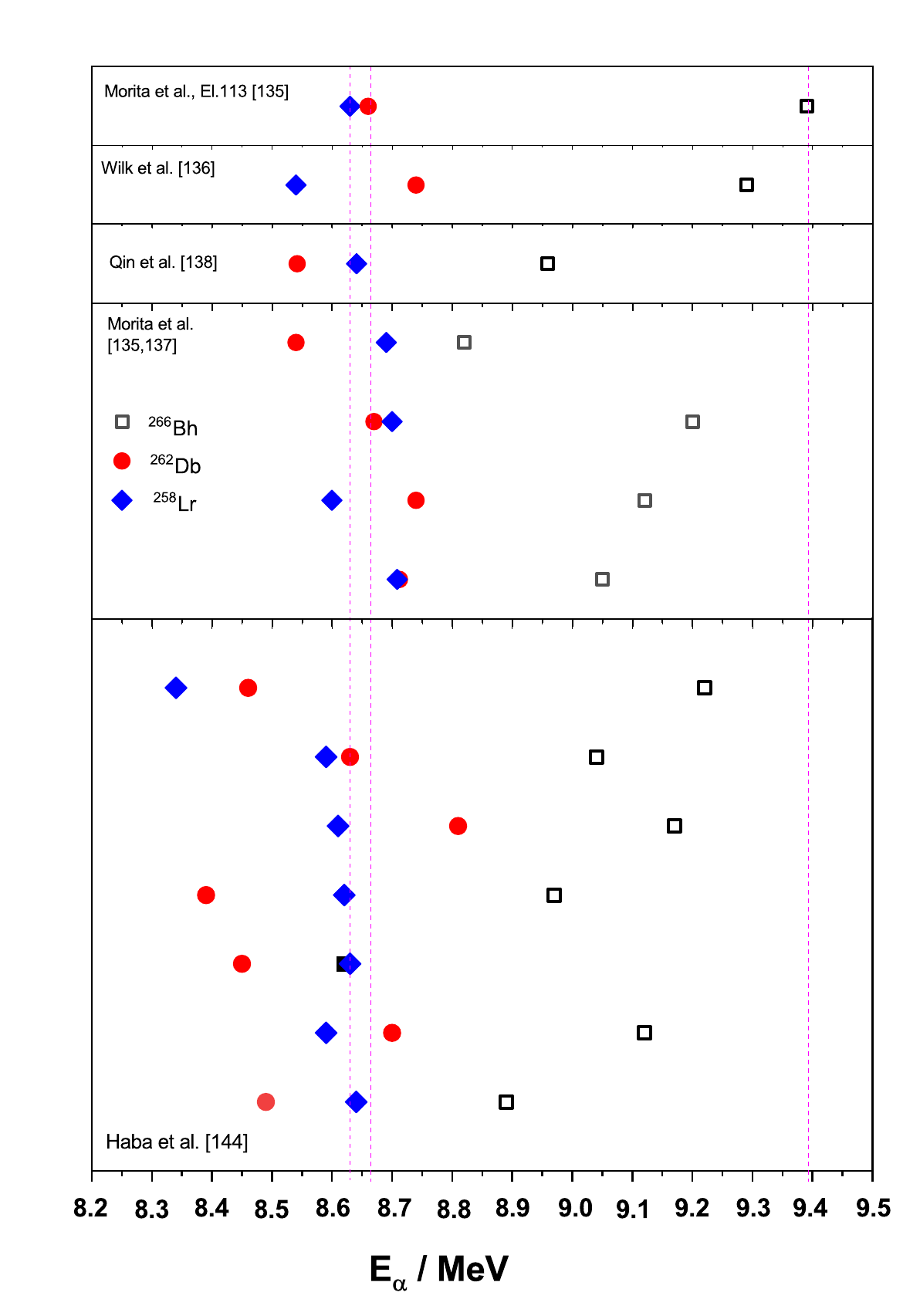}
                }
                % If not, use
                %\vspace{5cm}       % Give the correct figure height in cm
                \caption{Comparison of all published triple correlations $^{266}$Bh $^{\alpha}_{\rightarrow}$  $^{262}$Db $^{\alpha}_{\rightarrow}$
                 $^{258}$Lr $^{\alpha}_{\rightarrow}$  
                }
                \label{fig:21}       % Give a unique label
                \end{figure}

 \newpage
 \subsection{\bf{6.6 Discovery of element 113 -  Alpha decay chain of $^{278}$Nh}}

 \begin{table}
 \begin{center}
 \begin{tabular}{| l l l l l l l l l l |}
 \hline
   &  \vline & chain 1 &  & \vline & chain2 &  & \vline & chain 3 &  \\
   Isotope & \vline & E$_{\alpha}$/MeV  & T$_{1/2}$ & \vline  & E$_{\alpha}$/MeV  & T$_{1/2}$ & \vline & E$_{\alpha}$/MeV  & T$_{1/2}$  \\
 \hline
  $^{278}$Nh & \vline  & 11.68$\pm$0.04  & 0.344 ms & \vline & 11.52$\pm$0.04 &  4.93 ms & \vline & 11.82$\pm$0.06 &  0.667 ms  \\
  $^{274}$Rg & \vline   & 11.15$\pm$0.07  & 9.26 ms  & \vline & 11.31$\pm$0.07 &  34.3 ms & \vline & 10.65$\pm$0.06 &  9.97 ms  \\
  $^{270}$Mt & \vline   & 10.03$\pm$0.07  & 7.16 ms  & \vline & 2.32 (esc)     &  1.63 s  & \vline & 10.26$\pm$0.07 &  444 ms  \\
  $^{266}$Bh & \vline   &  9.08$\pm$0.04  & 2.47 s   & \vline & 9.77$\pm$0.04  &  1.31 s  & \vline & 9.39$\pm$0.06  &  5.26 s \\
  $^{262}$Db & \vline   &       sf        & 40.9 s   & \vline & sf             &  0.787 s & \vline & 8.63$\pm$0.06  &  126 s  \\
  $^{258}$Lr & \vline   &                 &          & \vline &                &          & \vline & 8.66$\pm$0.06 &   3.78 s  \\
 
 \hline
 \end{tabular}
 \end{center}
 \caption{Decay chains observed at GARIS, RIKEN in the reaction $^{70}$Zn + $^{209}$Bi and interpreted to start from $^{278}$Nh \cite{Morita15}.
 'esc' denotes that the $\alpha$ particle escaped the 'stop' detector and only an energy loss signal was recorded.} \label{tab5}
 \end{table}

The first report on discovery of element 113 was published by Oganessian et al. \cite{OganU04} in 2004. In an irradiation of $^{243}$Am with $^{48}$Ca performed at the DGFRs three decay chains were observed which were interpreted to start from $^{288}$115; the isotope
$^{284}$113 of the new element 113 was thus produced as the $\alpha$ decay descendant of $^{288}$115. The data published in 2004 
were confirmed at DGFRS \cite{OgA13}, at TASCA \cite{RuF13} and also later - after giving credit to the discovery of element 113 - at the BGS \cite{GaG15}. Nevertheless the 'fourth IUPAC/IUPAP joint working Party (JWP) did not accept these results
as the discovery of elemenent 113, as they concluded that the discovery profiles were not fulfilled: 'The 2013 Oganessian
collaboration \cite{OgA13} and the 2013 Rudolph collaboration \cite{RuF13} provide redundancy to the three $^{288}$113 chains
observed in 2004 with the $\alpha$ energies being in excellent agreement among most of the events. [...] However, the criteria (q.v. \cite{Wapstra91}) have not been met as there is no mandatory identification of the chain atomic numbers neither through a known descendant nor by cross reaction. Chemical determination as detailed in the subsequent profile of Z\,=\,115 where they are documented, serving the important role of assigning atomic number  are insufficiently selective although certainly otherwise informative' \cite{KarB16}.\\
Instead credit for discovery of element 113 was given to Morita et al. on the basis of three decay chains observed in the 'cold' fusion reaction $^{70}$Zn + $^{209}$Bi \cite{KarB16}.\\
Although cold fusion had been the successful method to synthesize elements {\it{Z}}\,=\,107 to {\it{Z}}\,=\,112, due to the steep decrease of the
cross-sections by a factor of three to four per element, it did not seem to be straightforward to assume it would be the silver bullet to the
SHE (see fig. 16). Nevertheless after the successful synthesis of element 112 in bombardments of $^{208}$Pb with $^{70}$Zn \cite{Hofm96}
it seemed straightforward to attempt to synthesize element 113 in the reaction $^{209}$Bi($^{70}$Zn,n)$^{278}$113. Being optimistic and
assuming a drop in the cross section not larger than a factor of five, as observed for the step from element 110 to element 111 
(see fig. 16), a cross section of some hundred femtobarn could be expected.\\
First attempts were undertaken at SHIP, GSI, Darmstadt, Germany in 1998 \cite{HofH99} and in 2003 \cite{HofA04}. No $\alpha$ decay chains
that could be attributed to start from an element 113 isotope were observed. Merging the projectile doses collected in both experiments
an upper production cross section limit $\sigma$\,$\le$\,160 fb was obtained \cite{Hofm11}.\\
More intensively this reaction was studied at the GARIS separator, Riken, Wako-shi, Japan. Over a period of nine years (from 2003 to 2012)
with a complete irradiation time of 575 days altogether three dcay chains interpreted to start from $^{278}$113 were observed
\cite{Morita04,Morita07a,Morita13,Morita15}.
The collected beam dose was 1.35$\times$10$^{20}$ $^{70}$Zn - ions, the formation cross-section was $\sigma$\,=\,22$^{+20}_{-13}$ fb 
\cite{Morita13}.\\
The chains are shown in fig. 17 and the data are presented in table 5. Chain 1 and chain 2 consist of four $\alpha$ particles and are
terminated by a fission event, while chain 3 consists of six $\alpha$ decays. Already at first glance the large differences in the 
$\alpha$ energies of members assigned to the same isotope is striking, especially for the events $\alpha_{2}$($^{274}$Rg) in chains 2 and 3
with {\it{$\Delta$E}}\,=\, 0.68 MeV and $\alpha_{4}$($^{266}$Bh) in chains 1 and 2 with {\it{$\Delta$E}}\,=\, 0.69 MeV. Although it is known 
that $\alpha$ - decay energies can vary in a wide range for odd - odd nuclei in the region of heaviest nuclei, as was shown, e.g.,
for $^{266}$Mt, where $\alpha$ energies were found to vary in the range {\it{E$_{\alpha}$}}\,=\,(10.456\,-\,11.739) MeV \cite{HofH97}, 
the assignment of such different energies to the decay of the same isotope or the same nuclear level can be debated ({\it{see e.g.}} \cite{Hess13}), as specifcally 
concerning the latter case it is known, that in odd - odd (and also in odd - mass) nuclei often low lying isomeric states exist, which
decay by $\alpha$ emission with energies and halflives similar to those of the ground state (see e.g. \cite{Vost15} 
for the cases of $^{258}$Db, $^{254}$Lr, and $^{250}$Md).
In the present case large $\alpha$ energy differences of {\it{$\Delta$E}}\,$>$\,0.1 MeV are evident for the corresponding members in 
all chains as shown in table 6.

 \begin{table}
 \begin{center}
 \begin{tabular}{l l l l l l l}
 \hline
 isotope & E$_{\alpha}$/MeV & E$_{\alpha}$/MeV & E$_{\alpha}$/MeV & $\mid$$\Delta$E$_{\alpha}$$\mid$/MeV & $\mid$$\Delta$ E$_{\alpha}$$\mid$/MeV & 
 $\mid$$\Delta$E$_{\alpha}$$\mid$/MeV   \\
         & Chain 1 & Chain 2 & Chain 3 & Ch.1 - Ch.2 & Ch.1 - Ch.3 & Ch.2 - Ch. 3 \\
 \hline
  $^{278}$Nh & 11.68$\pm$0.04 & 11.52$\pm$0.04 &  11.82$\pm$0.06 & 0.16 & 0.14 & 0.30  \\
  $^{274}$Rg & 11.15$\pm$0.07  & 11.31$\pm$0.07 &  10.65$\pm$0.06 & 0.16 & 0.50 & 0.68 \\
  $^{270}$Mt & 10.03$\pm$0.07  & 2.32 (esc)     &  10.26$\pm$0.07 & - & 0,23 & -  \\
  $^{266}$Bh & 9.08$\pm$0.04  & 9.77$\pm$0.04   &  9.39$\pm$0.06  & 0.69 & 0.31 & 0.48  \\
 
 \hline
 \end{tabular}
 \end{center}
 \caption{$\alpha$ energy differences of the individual chain members of the three decay chains interpreted to start from
 $^{278}$Nh. } \label{tab6}
 \end{table}

These differences are of specific importance for $^{266}$Bh which acts as an anchor point for identification of
the chains. The observation of $\alpha$ decay of this isotope has been reported by several authors who produced it in diffeent reactions:\\
a) Wilk et al. \cite{Wilk00} used the reaction $^{249}$Bk($^{22}$Ne,5n)$^{266}$Bh. They observed one event with an $\alpha$ energy of 
{\it{E$_{\alpha}$}}\,=\,9.29 MeV.\\
b) Morita et al. \cite{Morita15,Morita09} used the reaction $^{248}$Cm($^{23}$Na,5n)$^{266}$Bh; they observed in total 32 decay chains; 20 of 
them
were attributed (partly tentative) to the decay of $^{266}$Bh; four decay chains consisted of three $\alpha$ particles, assigned as decays
$\alpha_{1}$($^{266}$Bh) - $\alpha_{2}$($^{262}$Db) - $\alpha_{3}$($^{258}$Lr); four decay chains consisted of two $\alpha$ particles, 
interpreted as decays $\alpha_{1}$($^{266}$Bh) - $\alpha_{2}$($^{262}$Db or $^{258}$Lr); twelve decay chains consisted of an $\alpha$ particle 
followed by a fission event, interpreted as $\alpha$($^{266}$Bh) - SF($^{262}$Db)(possibly SF from $^{262}$Rf,
produced by EC decay of $^{262}$Db); in the case of four $\alpha$ energies of E$_{\alpha}$\,$<$\,9 MeV, the 
assignment was marked as 'tentative'. \\
c) Qin et al. \cite{Qin06} used the reaction $^{243}$Am($^{26}$Mg,3n)$^{266}$Bh. They observed four decay chains which they assigned 
start from $^{266}$Bh.\\
Evidently there is no real agreement for the $\alpha$ energies of $^{266}$Bh; two of the three energies of $^{266}$Bh from the 
$^{278}$113 decay chains (fig. 18a) are outside the range of energies observed in direct production, which is specifically critical 
for chain 3, as it is not terminated by fission, but $\alpha$ decay is followed by two more $\alpha$ events attributed to 
$^{262}$Db and $^{258}$Lr and thus is the anchor point for identification of the chain. Some agreement is obtained for the events
from the direct production followed by fission \cite{Morita09,Morita15} (fig. 18e), the $\alpha$ energy in chain 1 (fig. 18a)
(also follwed by fission) and the results from Qin et al. \cite{Qin06} (fig. 18f), where two groups at (9.05\,-\,9.1) MeV and (8.9\,-\,9.0) MeV
are visible. Note, that in \cite{Morita09,Morita15} the events at {\it{E$_{\alpha}$}}\,$<$\,9.0 MeV followed by fission are only assigned 
tentatively to $^{266}$Bh, while in \cite{Qin06} all $^{266}$Bh $\alpha$ decays are followed by $\alpha$ decays.\\
Unclear is the situation of the events followed by $\alpha$ decays. As seen in figs. 18c, 18d, and 18f, there are already in
the results from \cite{Morita09,Morita15} discrepancies in the $^{266}$Bh energies from triple correlations (fig. 18c) and double correlations
(fig. 18d). In the triple correlations there is one event at {\it{E}}\,=\,8.82 MeV, three more are in the interval  {\it{E}}\,=\,(9.08\,-\,9.2) MeV, while for the 
double correlations all four events are in the range  {\it{E}}\,=\,(9.14\,-\,9.23) MeV; tentatively merging the $^{266}$Bh $\alpha$ energies from events 
followed by $\alpha$ decay we find six of eight events (75 per cent) in the range  {\it{E}}\,=\,(9.14\,-\,9.23) MeV while only one of twelve events 
followed by fission is observed in that region. In this enery range none of the events observed by Qin et al. \cite{Qin06} is found, which are all below 9.14 MeV, also none of the events from the decay of $^{278}$113, and also not the event reported by Wilk et al. \cite{Wilk00} (fig. 18b) is found.\\
To conclude, the $\alpha$ decay energies of $^{266}$Bh reported from the different production reactions as well as from the different decay modes of the daughter products ($\alpha$ decay or (SF/EC) vary considerably, so there is no real experimental
basis to use $^{266}$Bh as an anchor point for identification of the chain assumed to start at $^{278}$113.\\
Discrepancies are also found for the halflives. From the $^{278}$113 decay chains a half-life of {\it{T$_{1/2}$}}\,=\,2.2$^{+2.9}_{-0.8}$ s is obtained for
$^{266}$Bh \cite{Morita15}, while Qin et al. \cite{Qin06} give a value T$_{1/2}$\,=\,0.66$^{+0.59}_{-0.26}$ s.
The discrepancy is already slightly outside the 1$\sigma$ confidential interval. No half-life value is given from the direct production
of Morita et al. \cite{Morita15}. \\
The disagreement in the decay properties of $^{266}$Bh reported by different authors renders the interpretation of the $\alpha$ decay chain 
(chain 3) quite difficult. It is therefore of importance to check the following $\alpha$ decays assigned to $^{262}$Db and $^{258}$Lr,
respectively, as they may help to clarify the situation. In order to do so, it is required to review the reported decay properties of these
isotopes and to compare the results with the data in chain 3.\\
It should also be remarked here that the differences of the $\alpha$ energies attributed to $^{266}$Bh followed by $\alpha$ decays or by SF 
in \cite{Morita15} indicates that the assignement of these events to the same isotope is not straightforward, at least not the
assignment to the decay of the same nuclear level.\\
In a previous data compilation \cite{Fire96} three $\alpha$ lines of {\it{E$_{\alpha1}$}}\,=\,8.45$\pm$0.02 MeV (i\,=\,0.75),
{\it{E$_{\alpha2}$}}\,=\,8.53$\pm$0.02 MeV (i\,=\,0.16), {\it{E$_{\alpha3}$}}\,=\,8.67$\pm$0.02 MeV (i\,=\,0.09) and a half-life of 
{\it{T$_{1/2}$}}\,=\,34$\pm$4 s are reported for $^{262}$Db. More recent data were obtained from decay studies of $^{266}$Bh \cite{Morita15,Wilk00,Morita09,Qin06}
or from direct production via the reaction $^{248}$Cm($^{19}$F,5n)$^{262}$Db \cite{Dress99,Haba14}. The results of the different studies are
compared in fig. 19.\\
The energy of the one event from the $^{278}$113 decay chain 3 is shown in fig. 19a.
The most extensive recent data for $^{262}$Db were collected by Haba et al. \cite{Haba14}. They observed two groups of $\alpha$-decay
energies, one at E$_{\alpha}$\,=\,(8.40-8.55) MeV (in the following also denoted as 'low energy component') and another one at E$_{\alpha}$\,=\,(8.60-8.80) MeV (in the following also denoted as 'high energy component') (fig. 19g). 
Mean $\alpha$ energy values and intensities  are 
{\it{E$_{\alpha}$}}\,=\,8.46$\pm$0.04 MeV ({\it{i$_{rel}$}}\,=\,0.70$\pm$0.05) and {\it{E$_{\alpha}$}}\,=\,8.68$\pm$0.03 MeV
(i$_{rel}$\,=\,0.30$\pm$0.05).
In \cite{Haba14} only 
one common half-life of {\it{T$_{1/2}$}}\,=\,33.8$^{+4.4}_{-3.5}$ s is given for both groups. A re-analysis of the data, however, indicates different
halflives: {\it{T$_{1/2}$}}\,=\,39$^{+6}_{-5}$ s for {\it{E$_{\alpha}$}}\,=\,(8.40\,-\,8.55) MeV and 
{\it{T$_{1/2}$}}\,=\,24$^{+6}_{-4}$ s for {\it{E$_{\alpha}$}}\,=\,(8.60\,-\,8.80) MeV. A similar behavior is reported by Dressler et al. \cite{Dress99}
2 events at {\it{E$_{\alpha}$}}\,=\,(8.40\,-\,8.55) MeV and one event at {\it{E$_{\alpha}$}}\,=\,(8.60\,-\,8.80) MeV (see fig. 19f).
Qin et al. \cite{Qin06} oberserved three events at E$_{\alpha}$\,=\,(8.40\,-\,8.55) MeV and one event at E$_{\alpha}$\,=\,8.604 MeV, outside
the bulk of the high energy group reported in \cite{Haba14}(see fig. 19e). A similar behavior is seen for the double correlations
($^{262}$Db\,-\,$^{258}$Lr), with missing $^{266}$Bh from the reaction $^{23}$Na + $^{248}$Cm measured by Morita et al. \cite{Morita15}
(see fig. 19d). Three of four events are located in the range 
of the low energy component, while for the triple correlations all four events are in the high energy group (see fig. 19c). 
This behavior seems somewhat strange as there is no physical reason why the
$\alpha$ decay energies of $^{262}$Db should be different for the cases where the preceding $^{266}$Bh $\alpha$ decay is recorded
or not recorded. It rather could mean that the triple ($^{266}$Bh $\rightarrow$ $^{262}$Db $\rightarrow$ $^{258}$Lr) and the 
double correlations ($^{262}$Db $\rightarrow$ $^{258}$Lr) of \cite{Morita15} do no respresent the same activities.
The $\alpha$ decay energy of the one event observed by Wilk et al. \cite{Wilk00} belongs to the low energy group (see fig. 19b). The one
event from the decay chain attributed to start from $^{278}$113 does not really fit to one of the groups. The energy is definitely 
lower than the mean value of the high energy group, an agreement with that group can only be postulated considering the
large uncertainty ($\pm$60 keV) of its energy value (see fig. 19a).\\
Halflives are {\it{T$_{1/2}$}}\,=\,44$^{+60}_{-16}$ s for the {\it{E$_{\alpha}$}}\,=\,(8.40\,-\,8.55) MeV component in \cite{Morita15}, 
T$_{1/2}$\,=\,16$^{+7}_{-4}$ s for the {\it{E$_{\alpha}$}}\,=\,(8.55\,-\,8.80) MeV in agreement with the values of Haba et al. \cite{Haba14},
and  {\it{T$_{1/2}$}}\,=\,52$^{+21}_{-12}$ s for the SF activity, which is rather in agreement with that of the low energy component. \\ 

To summarize: The assignment of the event $\alpha_{5}$ in chain 3 in \cite{Morita15} to $^{262}$Db is not unambiguos on the basis of its
energy, in addition also its 'lifetime' {\it{$\tau$}}\,=\,t$_{\alpha5}$-t$_{\alpha4}$ = 126 s is about five times of the half-life
of the high energy component of $^{262}$Db observed in \cite{Haba14}. One should keep in mind, that the probability to observe a decay
at times longer than five halflives is {\it{p}}\,$<$0.03.\\

\begin{table}
	\begin{center}
		\begin{tabular}{l l l l }
			\hline
			ref. \cite{Eskola73,Akovali01}  &  & ref. \cite{Bemis76,Akovali01} & \\
			\hline
			E$_{\alpha}$/MeV & i$_{\alpha}$ & E$_{\alpha}$/MeV &  i$_{\alpha}$ \\
			8.590$\pm$0.02 & 0.30$\pm$0.04 &  8.540$\pm$0.02 & 0.10$\pm$0.05  \\
			8.620$\pm$0.02 & 0.47$\pm$0.03 &  8.589$\pm$0.01 & 0.45$\pm$0.07  \\
			8.650$\pm$0.02 & 0.16$\pm$0.03 &  8.614$\pm$0.01 & 0.35$\pm$0.05  \\
			8.680$\pm$0.02 & 0.07$\pm$0.04 &  8.648$\pm$0.01 & 0.10$\pm$0.02  \\
			
			\hline
		\end{tabular}
	\end{center}
	\caption{ Alpha decay energies reported for $^{258}$Lr by Eskola et al. \cite{Eskola73} and by
		Bemis et al. \cite{Bemis76}.} \label{tab7}
\end{table}

 \begin{table}
 \begin{center}
 \begin{tabular}{l l l l }
 \hline
Reference & Isoptope & analysis mode &  T$_{1/2}$ / s  \\
\hline
  \cite{Morita15} & $^{266}$Bh  & decay chain $^{278}$Nh &  2.2$^{+2.9}_{-0.8}$ \\
  \cite{Qin06} & $^{266}$Bh &         & 0.66$^{+0.59}_{-0.26}$ \\
  \cite{Haba20} & $^{266}$Bh & correlated to SF & 12.8$^{+5.2}_{-2.9}$ \\
  \cite{Haba20} & $^{266}$Bh & correlated to $\alpha$ decay & 7.0$^{+3.0}_{-1.6}$ \\
  \cite{Haba20} & $^{266}$Bh & all events & 10.0$^{+2.6}_{-1.7}$ \\
\hline
  \cite{Fire96} & $^{262}$Db &      & 34$\pm$4   \\
  \cite{Morita15} & $^{262}$Db & E\,=\,(8.38-8.52) MeV & 44$^{+60}_{-16}$ \\
  \cite{Morita15} & $^{262}$Db & E\,=\,(8.55-8.80) MeV & 16$^{+7}_{-4}$ \\
  \cite{Morita15} & $^{262}$Db & corr. $^{266}$Bh - SF & 52$^{+21}_{-12}$ \\
  \cite{Haba14}  &  $^{262}$Db & E\,=\,(8.38-8.52) MeV & 39$^{+6}_{-5}$ \\
  \cite{Haba14}  &  $^{262}$Db & E\,=\,(8.55-8.80) MeV & 24$^{+6}_{-4}$ \\
  \cite{Qin06}  &  $^{262}$Db &  & 26$^{+26}_{-9}$ \\
  \cite{Dress99}  &  $^{262}$Db &  & 26$^{+26}_{-9}$ \\
  \cite{Haba20} & $^{262}$Bh & corre. $^{266}$Bh - SF & 39$^{+15}_{-8}$\\
  \cite{Haba20}  &  $^{262}$Db & E\,=\,(8.38-8.52) MeV & 32$^{+22}_{-9}$ \\
  \cite{Haba20}  &  $^{262}$Db & E\,=\,(8.55-8.80) MeV & 6.7$^{+16.2}_{-2.6}$ \\ 
\hline
   \cite{Fire96} & $^{258}$Lr &      & 3.9$^{+0.4}_{-0.3}$   \\
   \cite{Morita15} & $^{258}$Lr & triple corr. $^{266}$Bh - $^{262}$Db - $^{258}$Lr  & 4.7$^{+4.7}_{-1.6}$ \\
   \cite{Morita15} & $^{258}$Lr & double corr. $^{262}$Db - $^{258}$Lr  & 3.3$^{+3.3}_{-1.1}$ \\
   \cite{Morita15} & $^{258}$Lr & all events  & 4.0$^{+2.2}_{-1.1}$ \\
   \cite{Haba14} & $^{258}$Lr & corr. $^{262}$Db, E\,=\,(8.38-8.52) MeV  & 3.5$^{+0.6}_{-0.4}$ \\
   \cite{Haba14} & $^{258}$Lr & corr. $^{262}$Db, E\,=\,(8.55-8.80) MeV  & 4.1$^{+1.1}_{-0.7}$ \\
   \cite{Haba14} & $^{258}$Lr & all events  & 3.5$^{+0.5}_{-0.4}$ \\
   \cite{Dress99} & $^{258}$Lr &   & 3.1$^{+3.1}_{-1.0}$ \\
   \cite{Haba20} & $^{258}$Lr & all events  & 3.6$^{+1.4}_{-0.8}$ \\
 \hline
 \end{tabular}
 \end{center}
 \caption{Comparison of halflives of $^{266}$Bh, $^{262}$Db, and $^{258}$Lr published or analysed from
 published data by the author in this work.} \label{tab8}
 \end{table}

Observation of $^{258}$Lr was first reported by Eskola et al. \cite{Eskola73} and later by Bemis et al. \cite{Bemis76}. The reported
$\alpha$ energies and intensities slightly disagree \cite{Akovali01}. The data are given in table 6. The energies given in \cite{Bemis76} are 30-50 keV lower than those reported in \cite{Eskola73}.
More recent data were obtained from decay studies of $^{266}$Bh, $^{262}$Db \cite{Morita15,Wilk00,Morita09,Qin06,Dress99,Haba14}.
The results are compared in fig. 20.\\
The quality of the data is lower than that of $^{262}$Db, but not less confusing. Haba et al. got a broad energy distribution 
in the range {\it{E$_{\alpha}$}}\,=\,(8.50\,-\,8.75) MeV with the bulk at {\it{E$_{\alpha}$}}\,=\,(8.60\,-\,8.65) MeV having a mean value 
{\it{E$_{\alpha}$}}\,=\,8.62$\pm$0.02 MeV (see fig. 20g), for which a half-life of 
{\it{T$_{1/2}$}}\,=\,3.54$^{+0.46}_{-0.36}$ s is given. Dressler et al. \cite{Dress99} observed all the events at {\it{E$_{\alpha}$}}\,$\le$\,8.60 MeV
(see fig. 20f), but obtained a similar halflife of {\it{T$_{1/2}$}}\,=\,3.10$^{+3.1}_{-1.0}$ s. Each one event within the energy range of the
Haba - data was observed by Qin et al. \cite{Qin06} (fig. 20e) and Wilk et al. \cite{Wilk00} (fig. 20b).
Contrary to the energies of $^{266}$Bh and $^{262}$Db the $\alpha$ energies for $^{258}$Lr from the $^{266}$Bh decay study of 
Morita et al. \cite{Morita15} are quite in agreement for the triple (fig. 20c) and double (fig. 20d) correlations, a bulk of five
events at a mean energy of E$_{\alpha}$\,=\,8.70$\pm$0.01 MeV, further two events at a mean energy  E$_{\alpha}$\,=\,8.59$\pm$0.02 MeV,
and single one at {\it{E$_{\alpha}$}}\,=\,8.80 MeV; halflives are  {\it{T$_{1/2}$}}\,=\,4.7$^{+4.7}_{-1.6}$ s for the events from the triple correlations,
 {\it{T$_{1/2}$}}\,=\,3.3$^{+3.3}_{-1.1}$ s for the events from the double correlations, being in agreement within the error bars.
They may be merged to a single half-life of  {\it{T$_{1/2}$}}\,=\,4.0$^{+2.2}_{-1.0}$ s. 
The one decay event from chain 3 attributed to start from $^{278}$113 \cite{Morita15} of {\it{E$_{\alpha}$}}\,=\,8.66$\pm$0.06 MeV and $\tau$\,=\,t$_{\alpha6}$-t$_{\alpha5}$ = 3.78 s fairly fits into the decay properties reported for $^{258}$Lr.\\

The dilemma is evident as seen from fig. 21 where all triple correlations $^{266}$Bh $^{\alpha}_{\rightarrow}$  $^{262}$Db $^{\alpha}_{\rightarrow}$
$^{258}$Lr $^{\alpha}_{\rightarrow}$ reported so far \cite{Morita15,Wilk00,Qin06,Haba20} are shown. None of the fourteen chains agrees with any other
one. This feature may indicate the complicate $\alpha$ decay pattern of these isotopes, but it makes the assignment to the same isotope
speculative. In other words: the 'subchain' $^{266}$Bh $^{\alpha}_{\rightarrow}$  $^{262}$Db $^{\alpha}_{\rightarrow}$
$^{258}$Lr $^{\alpha}_{\rightarrow}$ of the decay chain interpreted to start from $^{278}$113 does not agree with any other so far
observed $\alpha$ decay chain
interpreted to start from $^{266}$Bh. The essential item, however, is that this triple correlation was regared as the key point
for first identification of element 113, to approve the discovery of this element and give credit to the discoverers. 
But this decision is based rather on weak probability considerations than on firm experimental facts. 
The only solid pillar is the agreement with the decay properties reported for $^{258}$Lr, which might be regarded
as rather weak. In other words, the assignment of the three decay chains to the decay of $^{278}$Nh is probable, but not firm.\\
It should be reminded that in case of element 111 in the JWP report from 2001 it was stated: 'The results of this study are definitely of high quality but there is insufficient internal redundancy to warrant certitude at this stage. Confirmation by further results is needed to assign priority of discovery to this collaboration' \cite{Karol01}.
So it seems strange that in a similar situation as evidently here, such concerns were not expressed.\\

A new decay study of $^{266}$Bh was reported recently by Haba et al. \cite{Haba20} using the same production reaction as in \cite{Morita15}. 
Alpha decays were observed correlated to fission events, assigned to the decay of $^{262}$Db and $\alpha$ decay chains $^{266}$Bh\,$^{\alpha}_{\rightarrow}$\,
$^{262}$Db\,$^{\alpha}_{\rightarrow}$ or $^{266}$Bh\,$^{\alpha}_{\rightarrow}$\,$^{262}$Db\,$^{\alpha}_{\rightarrow}$\,$^{258}$Lr\,$^{\alpha}_{\rightarrow}$. 
The $\alpha$ spectra of decays followed by fission is shown in fig. 18g, that of events followed by $\alpha$ decays of $^{262}$Db in fig. 18h. Evidently 
in correlation to fission events a concentration of events ('peak') is observed at E$_{\alpha}$\,=\,8.85 MeV, not observed in \cite{Morita15}, while in the range 
E$_{\alpha}$\,=\,8.9-9.0 MeV, where in \cite{Morita09} a peak - like structure was oberserved Haba et al. registered only a broad distribution. Also only
a broad distribution without indication of a peak - like concentration in the range E$_{\alpha}$\,=\,8.8-9.4 MeV is observed in correlation to $\alpha$ decays.
However, two $\alpha$ decays at E\,$\approx$\,9.4 MeV were now reported, close to the $\alpha$ energy of $^{266}$Bh in the $^{278}$Nh decay chain 3 \cite{Morita15}.
A remarkable result of Haba et al. \cite{Haba20} is the half-life of $^{266}$Bh; values of T$_{1/2}$\,=\,12.8$^{+5.2}_{-2.9}$ s are obtained for events 
correlated to SF, and 
T$_{1/2}$\,=\,7.0$^{+3.0}_{-1.6}$ s for events correlated to $\alpha$ decays. This finding suggests a common half-life of 
T$_{1/2}$\,=\,10.0$^{+2.6}_{-1.7}$ s  as given in \cite{Haba20} despite the discrepancies in the $\alpha$ energies. That value is, however, significantly larger
than the results from previous studies \cite{Morita15,Qin06}.\\
For $^{262}$Db in \cite{Haba20} a similar $\alpha$ decay energy distribution is observed as in \cite{Haba14}, as seen in figs. 19g and 19h, 
but again not in agreement with the results from \cite{Morita09} (figs. 19c and 19d) where the low energy component E\,=\,(8.38-8.52) MeV is practically missing. 
Halflives of SF events assigned to $^{262}$Db are in agreement with those of $\alpha$ events at E\,=\,(8.38-8.52) MeV (see table 8), but again for the events 
at E\,=\,(8.55-8.80) MeV a shorter half-life (T$_{1/2}$\,=\,6.7$^{+16.2}_{-2.6}$ s) is indicated. Interestingly all events at E\,=\,(8.55-8.80) MeV are correlated to 
$^{266}$Bh $\alpha$ decays E\,$>$\,9 MeV. These are the events no. 13, 28 in \cite{Haba20} and no. 1, 2, 3 in \cite{Morita09}. The two events of $^{266}$Bh 
in \cite{Haba20} have 
extremely low correlation times of (0.92 s and 0.33 s), the $^{262}$Db events have a half-life T$_{1/2}$\,=\,13.7$^{+11.1}_{-4.2}$ s.
Despite the above mentioned differences the same feature is observed in \cite{Morita15}. The data
are summarized in table 8. A halflife T$_{1/2}$\,=\,22.5$^{-4.9}_{-3.4}$ s is obtained clearly lower than that of the SF events and $\alpha$ - events
E\,=\,(8.38-8.55 MeV), corroborating the possible existence of two long-lived states in $^{262}$Db. Although data are scarce two other features are
indicated: a) all $^{266}$Bh energies are above the 'bulk' of the $\alpha$ energy distribution of $^{266}$Bh and b) one obtains a half-life 
of T$_{1/2}$\,=\,3.4$^{-4.7}_{-1.3}$ s, lower than the value of 10 s extracted from all events. This can be regarded as a hint for the existence of 
two long-lived states in $^{266}$Bh decaying by $\alpha$ emission, resulting in two essential decay branches $^{266}$Bh(1) $\rightarrow$ $^{262}$Db(1)
and $^{266}$Bh(2) $\rightarrow$ $^{262}$Db(2).\\
The $\alpha$ spectrum for $^{258}$Lr measured in \cite{Haba20} is shown in fig. 20h. Essentially it is in-line with the one obtained in \cite{Haba14}.\\
To summarize: the new decay study of $^{266}$Bh delivers results not really in agreement with those from previous studies concerning the decay 
energies and delivers a considerably longer halflife for that isotope. So the results do not remove the concerns on the decay chains interpreted to start
from $^{278}$Nh. But it delivers some interesting features: the $\alpha$ decays in the range (8.38-8.55) MeV and the SF events following $\alpha$ decay of $^{266}$Bh
seemingly are due to the decay of the same state in $^{262}$Db, the fission activity, however, may be due to $^{262}$Rf produced by EC of $^{262}$Db.
The $\alpha$ - decays of $^{262}$Db of E\,=\,(8.55-8,80) MeV eventually are from decay of a second long-lived level. There is also strong evidence that this
level is populated essentially  by $\alpha$ decay of a long-lived level in $^{266}$Bh, different to that populating the one in $^{262}$Db decaying by 
$\alpha$ particles in the range (8.38-8.55) MeV. Further studies are required to clarify this undoubtedly interesting feature.\\
Discussing the items above one has of course to emphasize the different experimental techniques used which may influence the measured energies. 
The important feature are the different detector resolutions which determine the widths of the distributions. 
So comparison of energies might be somewhat 'dangerous'. Another item is energy summing between $\alpha$ particles and CE. In the experiments of Wilk et al.
\cite{Wilk00} and Morita et al. \cite{Morita15}, the reaction products were implanted into the detector after in-flight separation. 
Qin et al. \cite{Qin06}, Dressler et al. \cite{Dress99} and Haba et al. \cite{Haba14,Haba20} collect the reaction products on the detector surface or on a thin foil
between two detectors. The letter procedure reduces the efficiency for energy summing of $\alpha$ particles and CE considerably. This could be  the reason for the 'shift' of 
the small 'bulk' of the $^{266}$Bh $\alpha$ energy distribution from $\approx$8.85 MeV \cite{Haba20} in fig. 18g to $\approx$8.95 MeV \cite{Morita15}
in fig. 18e. This interpretation might be speculative, but it clearly shows that such effects renders the consistency of data more difficult if different
experimental techniques are applied.

\begin{table}
	\begin{center}
		\begin{tabular}{l l l l l }
			\hline
			\hline
			Reference & E$_{\alpha}$($^{266}$Bh) / MeV  & $\Delta$t / s &  E$_{\alpha}$($^{262}$Db) / MeV  & $\Delta$t / s \\
			\hline
			\cite{Morita15}  & 9.05  &     &  8.71 & 54.91     \\
			\cite{Morita15}  & 9.12  &     &  8.74 & 13.76     \\
	        \cite{Morita15}  & 9.20  &     &  8.67 & 13.71     \\
	   		\hline
		    \cite{Haba20}  & 9.12  &  0.92  &  8.70 & 10.29     \\
		    \cite{Haba20}  & 9.04  &  0.33  &  8.63 & 9.07     \\
			\hline
		\end{tabular}
	\end{center}
	\caption{$\alpha$-$\alpha$ correlations $^{266}$Bh (E$>$9.0 MeV) - $^{262}$Db (E\,=(8.60-8.75 MeV)
		from triple correlations   $^{266}$Bh\,$^{\alpha}_{\rightarrow}$\,$^{262}$Db\,$^{\alpha}_{\rightarrow}$\,$^{258}$Lr\,$^{\alpha}_{\rightarrow}$.}
	    \label{tab9}
\end{table}

\section{\bf{7. (Exemplified) Cross-checks with Nuclear Structure Theory}}

In the following some discussion on selected decay and nuclear structure properties will be presented.

\subsection{\bf{ 7.1 Alpha-decay energies / Q-alpha values; even Z, odd-A, odd-odd nuclei}}

Alpha-decay energies provide some basic information about nuclear stability and properties. Discussing the properties
one strictly has to distinguish two cases, a) $\alpha$ decay of even-even nuclei, and b) $\alpha$ decay of isotopes 
with odd proton and/or odd neutron numbers.\\
In even-even nuclei $\alpha$ transitions occur with highest intensities between the {\it{I$^{\pi}$}} = 0$^{+}$ ground - states of mother and daughter isotopes. Still, in the region of strongly deformed heaviest nuclei ({\it{Z}} $\ge$ 90) notable population with 
relative intensities of (10-30 $\%$) is observed for transitions into the {\it{I$^{\pi}$}} = 2$^{+}$ level of the ground-state rotational band \cite{Fire96}, while band members of higher spins (4$^{+}$, 6$^{+}$ {\it{etc.}}) are populated only weakly with 
relative intensities of $<$1$\%$. Under these circumstances the $\alpha$ line of highest intensity represents the Q-value of the transition and is thus a measure for the mass difference of the mother and the daughter nucleus. It should be kept in mind, however,
that only in cases where the mass of the daughter nucleus is known, the Q-value can be used to calculate the mass of the mother nucleus, and only in those cases $\alpha$-decay energies can be used to 'directly' test nuclear mass predictions. Nevertheless, already the mass differences, i.e. the Q-values, can be used for qualitative assessments of those models. Particulary, as crossing of nucleon shells is accompanied by a strong local decrease of the {\it{Q$_{\alpha}$}} - values, existence, and by some extent also strength of such shells can be verified by analyzing systematics of {\it{Q$_{\alpha}$}} - values. That feature is displayed in fig. 22, where experimental {\it{Q$_{\alpha}$}} values for the known isotopes of even-Z elements {\it{Z}} $\ge$ 104 are compared with results of two (widely used) mass predictions based on the macropscopic - microscopic approach, the one reported by R. Smolanczuk and A. Sobiczewski \cite{Smolan95} (fig. 22a), and the one reported by P. M\"oller et al. \cite{Moller95} (fig. 22b). The neutron shells
at {\it{N}} = 152 and {\it{N}} = 162, indicated by the black dashed lines are experimentally and theoretically verified by the local minima in the {\it{Q$_{\alpha}$}} - values. But significant differences in the theoretical predictions are indicated, those of \cite{Smolan95} reproduce the experimental data in general quite fairly, while the agreement of those from \cite{Moller95} is
significantly worse.\\

\begin{figure}
%\begin{figure*}
\resizebox{0.9\textwidth}{!}{%
  \includegraphics{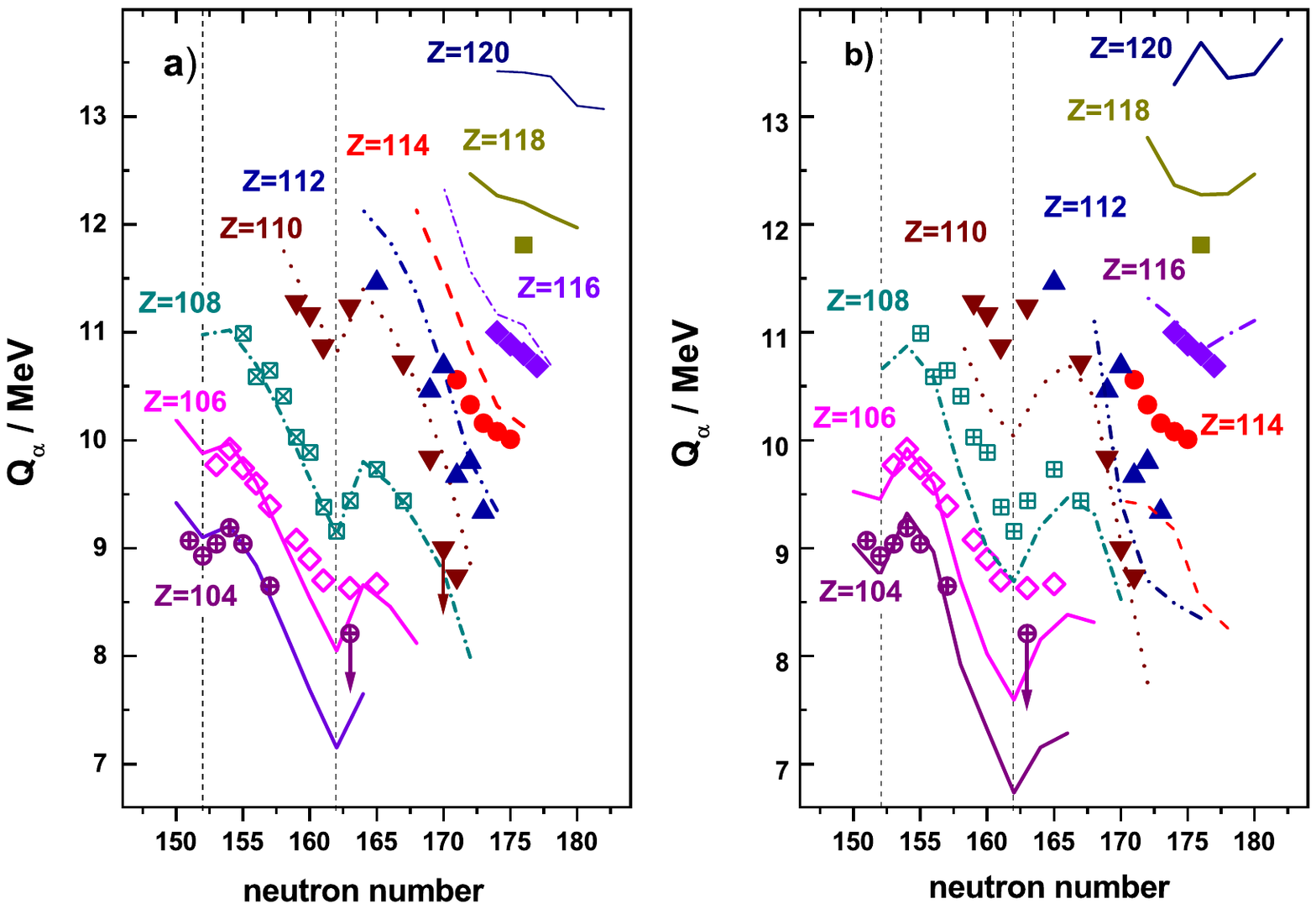}
}
% If not, use
%\vspace{5cm}       % Give the correct figure height in cm
\caption{Comparison of experimental {\it{Q$_{\alpha}$}} - values of even-Z elements {\it{Z}} $\ge$ 104 with theoretical predictions of R. Smolanczuk and A. Sobiczewski 
\cite{Smolan95} (fig. 22a) and P. M\"oller et al. \cite{Moller95} (fig. 22b). In case of isotopes with odd neutron numbers, the 
{\it{Q$_{\alpha}$}} - value was calculated from the highest reported decay energy.}
\label{fig:22}       % Give a unique label
\end{figure}
\begin{figure}
%\begin{figure*}
\resizebox{0.9\textwidth}{!}{%
  \includegraphics{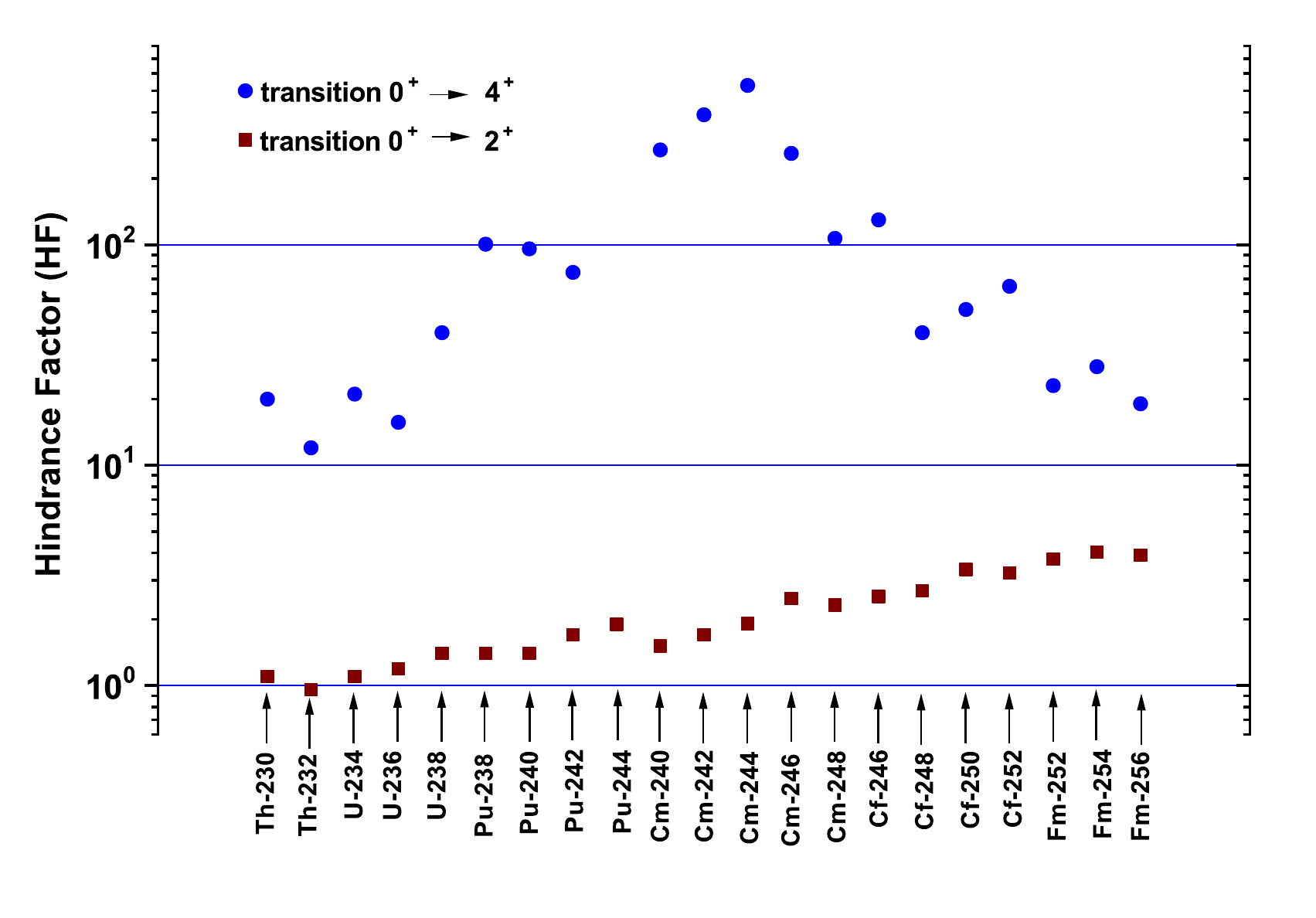}
}
% If not, use
%\vspace{5cm}       % Give the correct figure height in cm
\caption{Hindrance factors for decays into {\it{I$^{\pi}$}} = 2$^{+}$ and {\it{I$^{\pi}$}} = 4$^{+}$ daughter levels of
even-even actinide isotopes {\it{Z}} $\ge$ 90. Alpha-decay data are taken from \cite{Fire96}.}
\label{fig:23}       % Give a unique label
\end{figure}
\begin{figure}
%\begin{figure*}
\resizebox{0.9\textwidth}{!}{%
  \includegraphics{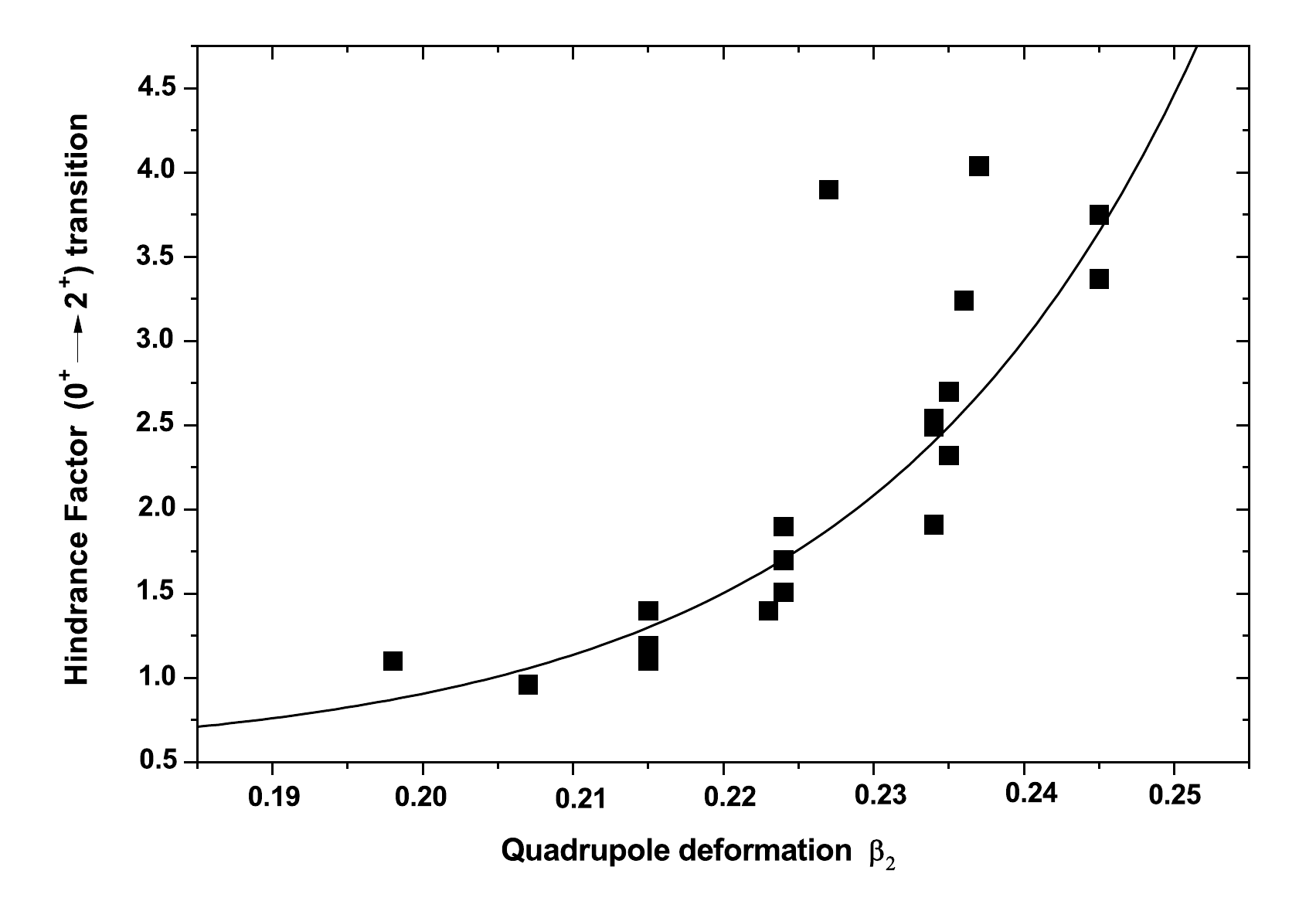}
}
% If not, use
%\vspace{5cm}       % Give the correct figure height in cm
\caption{ Hindrance factors for decays into {\it{I$^{\pi}$}} = 2$^{+}$ daughter levels of
even-even actinide isotopes {\it{Z}} $\ge$ 90 as function of the quadrupole deformation parameter $\beta_{2}$.
The line is to guide the eye. }
\label{fig:24}       % Give a unique label
\end{figure}
\begin{figure}
	%\begin{figure*}
	\vspace{-0.5cm}
	\resizebox{0.75\textwidth}{!}{%
		\includegraphics{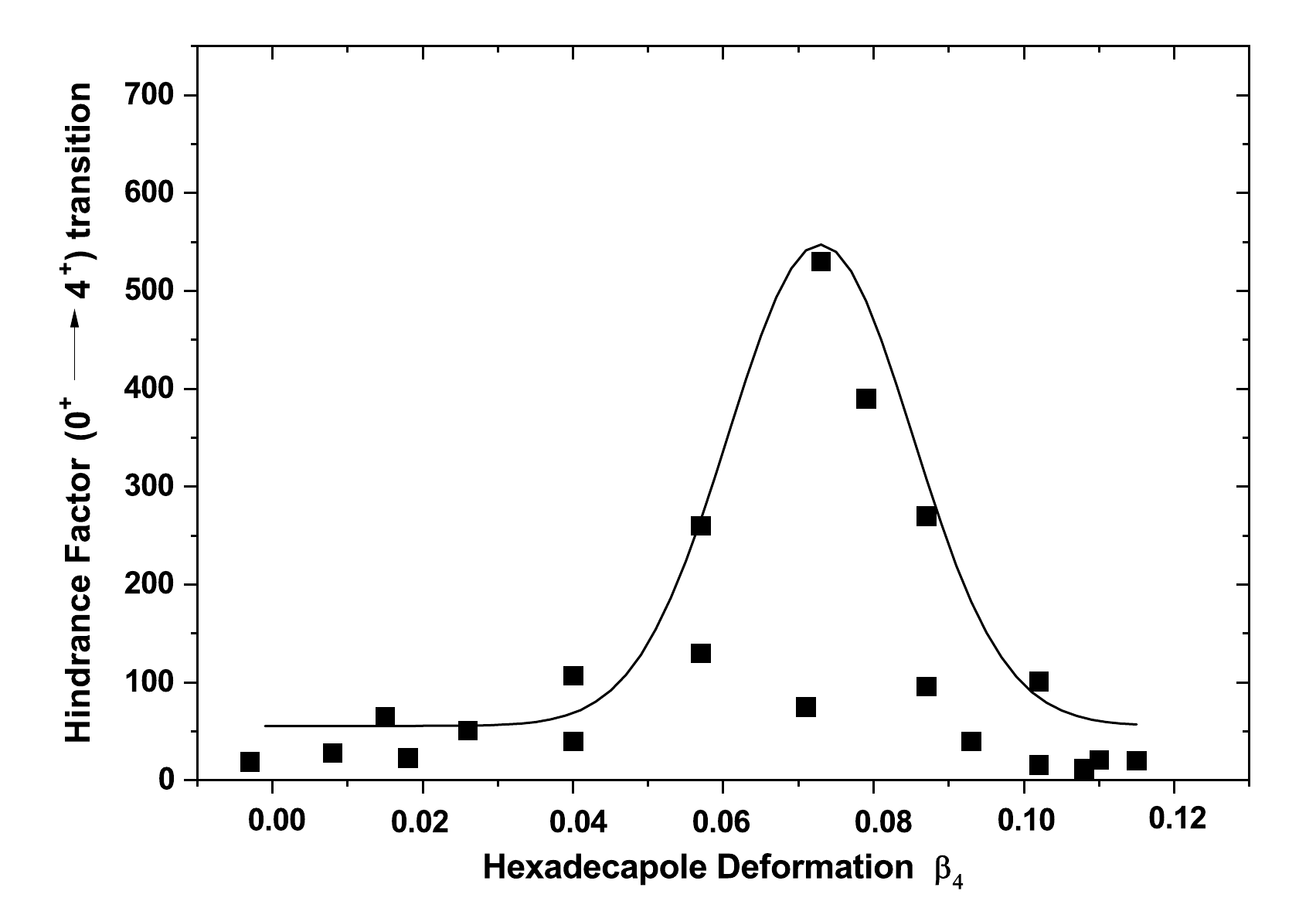}
	}
	% If not, use
	%\vspace{5cm}       % Give the correct figure height in cm
	\caption{Hindrance factors for decays into {\it{I$^{\pi}$}} = 4$^{+}$ daughter levels of
		even-even actinide isotopes {\it{Z}} $\ge$ 90 as fuction of the hexadecapole deformation parameter $\beta_{4}$.
	The line is to guide the eye. }
	\label{fig:25}       % Give a unique label
\end{figure}
\begin{figure}
	%\begin{figure*}
	\vspace{-2cm}
	\resizebox{1.0\textwidth}{!}{%
		\includegraphics{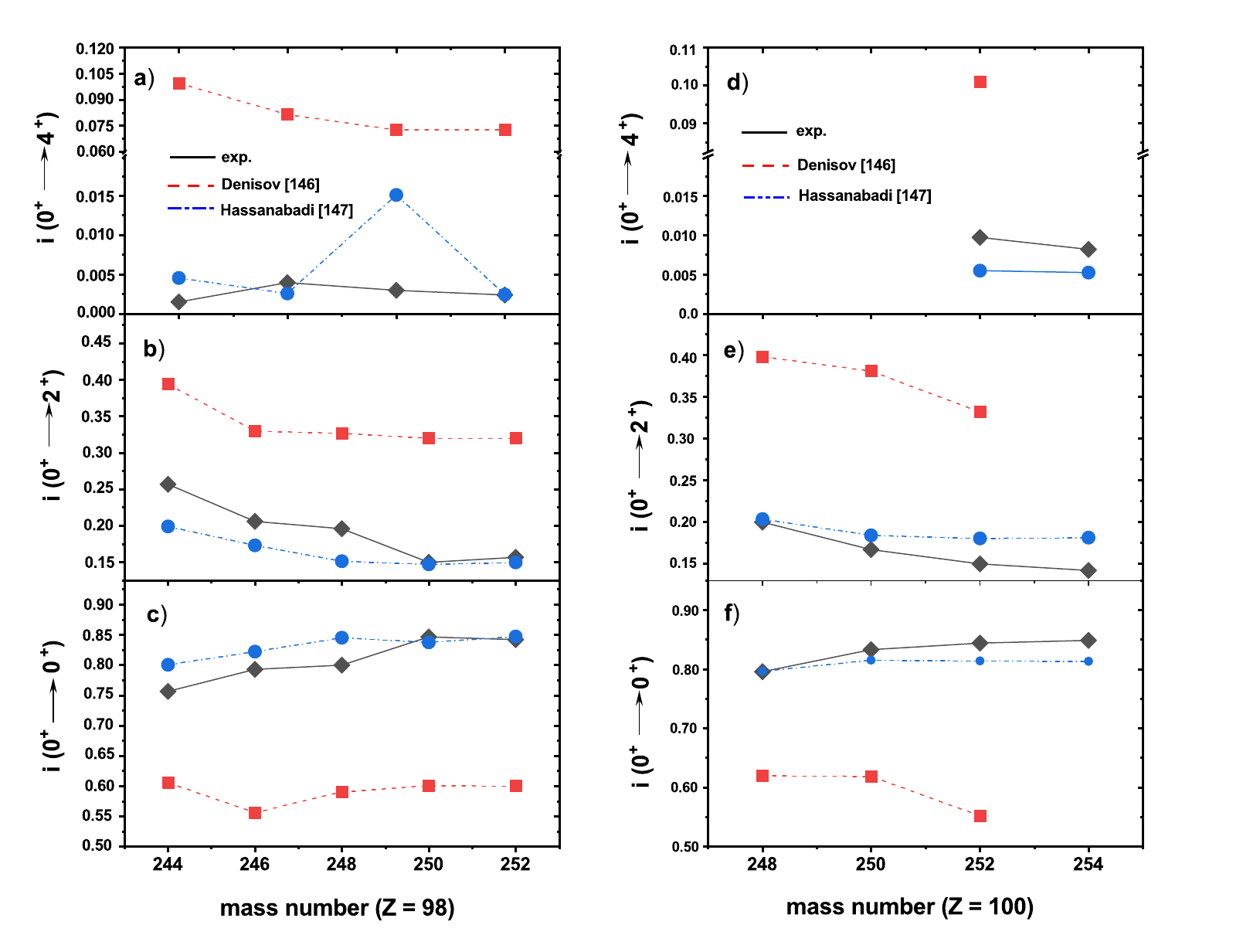}
	}
	% If not, use
	%\vspace{5cm}       % Give the correct figure height in cm
	\caption{Comparison of experimental and theoretical \cite{Denis09,Hassan18} $\alpha$ transition intensities into 
		I$^{\pi}$\,=\,0$^{+}$, 2$^{+}$, and 4$^{+}$ daughter levels for californium (a-c)and fermium isotopes (d-f).
	    Black lines and diamond represent experimental values, red dashed lines and squares represent the calculations of \cite{Denis09},
        blue dashed - dotted lines and circles represent the calculations of \cite{Hassan18}.}
	\label{fig:26}       % Give a unique label
\end{figure}
Alpha-decay between states of different spins are hindered. Quantitatively the hindrance can be expressed by a hindrance factor {\it{HF}}, defined as HF = T$_{\alpha}$(exp) / T$_{\alpha}$(theo), where {\it{T$_{\alpha}$(exp)}} denotes the experimental partial $\alpha$-decay half-life and {\it{T$_{\alpha}$(theo)}} the theoretical one. To calculate the latter a couple of (mostly empirical) relations are available. In the following will use the one proposed by D.N. Poenaru \cite{PoI80} with the parameter modification suggested by Rurarz \cite{Rur83}. This formula has been proven to reproduce experimental partial $\alpha$-decay halflives of even-even nuclei in the region of superheavy nuclei within a factor of two \cite{Hess16a}. \\
A semi-empirical relation for the hindrance due to angular momentum change was given in 1959 by J.O. Rasmussen \cite{Rasm59}.
\begin{figure}
	%\begin{figure*}
	\vspace{0cm}
	\resizebox{0.8\textwidth}{!}{%
		\includegraphics{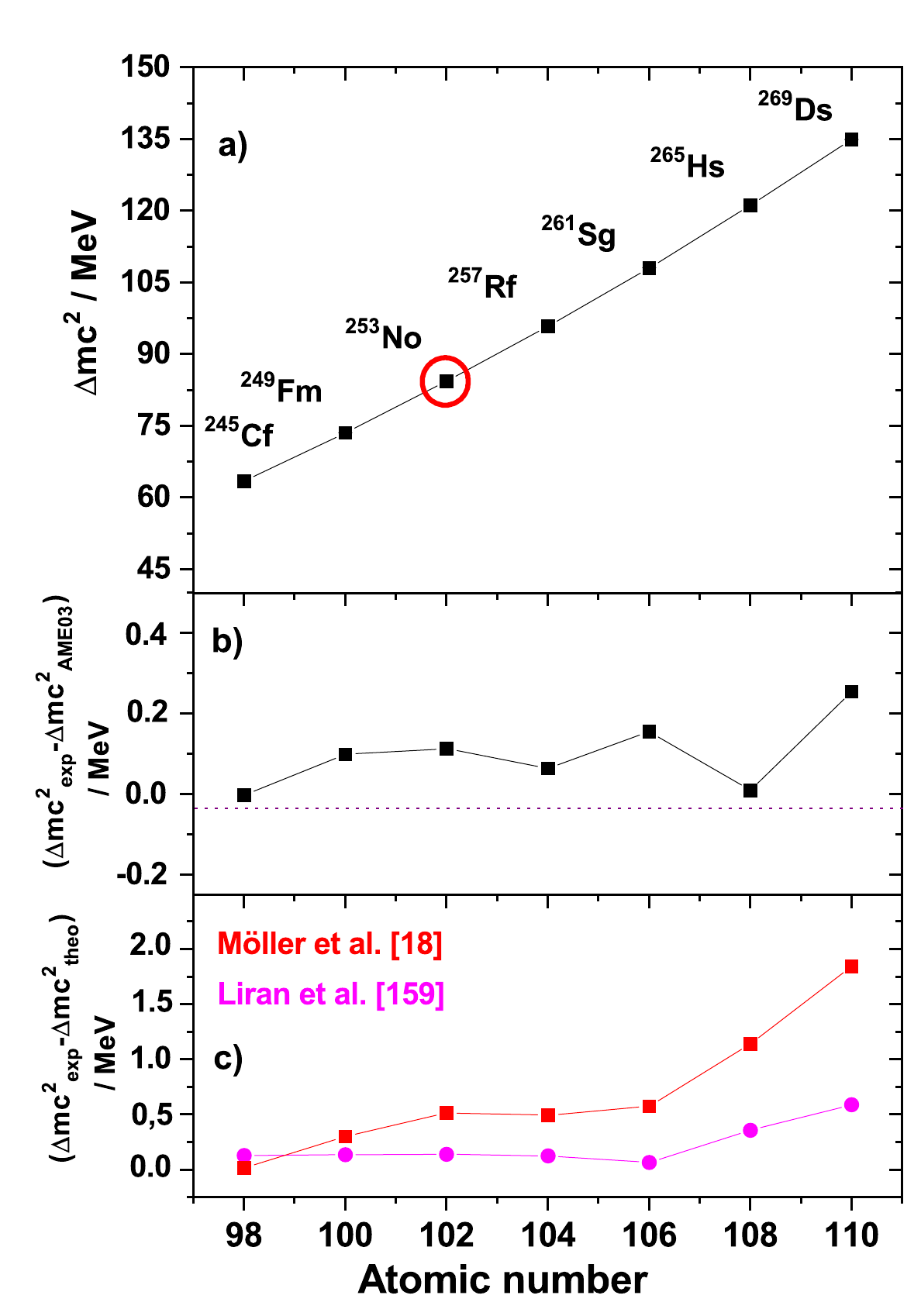}
	}
	% If not, use
	%\vspace{5cm}       % Give the correct figure height in cm
	\caption{(a) Experimental ground-state mass excesses of {\it{N-Z}} = 49 and (b) comparisons with (previously) evaluated values \cite{Audi03} 
		   and (c) theoretical predictions by P. M\"oller et al. \cite{Moller95} (squares) or Liran et al. \cite{Liran02} (circles).}
	\label{fig:27}       % Give a unique label
\end{figure}
The change of the transition probability P$_{0}$ (no angular momentum change) through the barrier, which can be equated with the inverse hindrance factor,
was given as \\

                                         P$_{L}$/P$_{0}$\,=\,exp[-0.2027L(L+1)Z$^{-1/2}$A$^{-1/6}$] \\

with {\it{L}} denoting change of the angular momentum, {\it{Z}} and {\it{A}} the atomic number of the daughter nuclei.\\
In the range of actinide nuclei where
data are available ({\it{Z}}\,$\approx$\,90\,-\,102) one expects hindrance factors {\it{HF}}\,$\approx$\,(1.6\,-\,1.7) for {\it{$\Delta$L}}\,=\,2
and {\it{HF}}\,$\approx$\,(5\,-\,6) for {\it{$\Delta$L}}\,=\,4, with a slight decrease at increasing  {\it{A}} and {\it{Z}}. 
The experimental hindrance factors for $\alpha$ decay into the {\it{I$^{\pi}$}} = 2$^{+}$ and {\it{I$^{\pi}$}} = 4$^{+}$ levels for the known cases in the actinide region {\it{Z}} $\ge$ 90 are shown in fig. 23. 
They exhibit a complete different behavior: for the {\it{$\Delta$L}}\,=\,2 transitions the experimental hindrance factor are comparable, but increase at increasing {\it{A}} and {\it{Z}}. For the {\it{$\Delta$L}}\,=\,4 transitions the hindrance factors are considerably larger and a  maximum is indicated for  curium isotopes in the mass range {\it{A}}\,=\,(240\,-\,246). 
Interestingly this  behavior can be related to the ground-state deformation as shown in fig. 24 and 25. In fig. 24 the hindrance factors for the {\it{I$^{\pi}$}} = 0$^{+}$ $\rightarrow$ {\it{I$^{\pi}$}} = 2$^{+}$ transitions are plotted as function of the quadrupole deformation parameter $\beta_{2}$ (taken from \cite{Moller95}). Evidently a strong increase of the hindrance factor at increasing quadrupole deformation is observed. In fig. 25 the hindrance factors for the {\it{I$^{\pi}$}} = 0$^{+}$ $\rightarrow$ {\it{I$^{\pi}$}} = 4$^{+}$ transitions are plotted as a function of the hexadecapole deformation parameter $\beta_{4}$ (taken from \cite{Moller95}). Here, a maximum at a deformation parameter $\beta_{4}$ $\approx$ 0.08 is indicated. 
This suggests a strong dependence of the hindrance factor on nuclear deformation and 
measuring transitions into rotational members of the ground-state band  may already deliver valuable information about the ground-state deformation of the considered nuclei.\\
Some attempts to calculate the transition probabilty into the ground - state rotational band members have been undertaken by
V. Yu. Denisov and A.A. Khudenko \cite{Denis09} as well as by H. Hassanabadi and S.S. Hosseini \cite{Hassan18}. \\
In both papers the $\alpha$ halflives were calculated in the 'standard' way as\\

        T$_{1/2}$\,=\,ln2/$\nu$P \\
        
with $\nu$ denoting the frequency of assaults on the barrier, and {\it{P}} being the penetration probability through the potential barrier using
the semiclassic WKB method.\\
In \cite{Hassan18} the $\alpha$ - nucleus potential was parameterized as a polynominal of third order for r$\le$C$_{t}$ and as sum of Coulomb {\it{V$_{C}$}},
nuclear {\it{V$_{N}$}} and centrifugal {\it{V$_{l}$}} + ($\hbar^{2}$l(l+1)/(2$\mu$r$^{2}$)) potential. For {\it{V$_{C}$}} and {\it{V$_{l}$}} 'standard expressions' were used, for {\it{V$_{n}$}} 
the 'proximity potential' of Blocki et al. \cite{Blocki77}. {\it{C$_{t}$}} is the touching configuration of daughter nucleus (d) and $\alpha$ particle ($\alpha$),
{\it{C$_{t}$}}\,=\,{\it{C$_{d}$}}\,+\,{\it{C$_{\alpha}$}}, with {\it{C$_{t}$}} denoting the Suessman central radii (see \cite{Hassan18} for details).
V.Yu. Denisov and A.A. Khudenko \cite{Denis09} use the 'unified model for $\alpha$ decay and $\alpha$ capture (UMADAC). Their potential represents the sum
of a 'standard' centrifugal potential {\it{V$_{l}$}} (see above), a Coulomb potential {\it{V$_{C}$}} including quadrupole ($\beta_{2}$) and hexadecapole ($\beta_{4}$) 
deformations, and a nuclear potential {\it{V$_{N}$}} of Woods-Saxon type (see \cite{Denis09} for further details). \\
Their results for the californium and fermium isotopes are compared with the experintal values in fig. 26.
Obviously the calculations of Denisov and Khudenko do not well reproduce the experimental data for both, 
californium and fermium isotopes; the calculated
(relative) intensities for the 0$^{+}$ $\rightarrow$ 0$^{+}$ transitions (fig. 26c) are too low  and hence too high values
for the 0$^{+}$ $\rightarrow$ 2$^{+}$ transitions (fig. 26c) and the 0$^{+}$ $\rightarrow$ 4$^{+}$ transitions (fig. 26a) are obtained.
The latter are even roughly an order of magnitude higher than the experimental data for the respective transition. 
Quite fair agreement between experimental data is evident for the calculations of Hassanabadi and Hosseini
(blue lines and symbols).\\

In odd-mass nuclei the situation is completely different as ground-state of mother and daughter nuclei usually differ in spin and
often also in parity. So ground-state to ground-state $\alpha$ decays are usually hindered. Hindrance factors significally depend on the spin difference, as well as on a possible parity change and/or a spin flip. For odd-mass nuclei an empirical classification of the hindrance factors into five groups has been established (see {\it{e.g.}} \cite{SeaL90}). Hindrance factors {\it{HF}}\,$<$\,4 characterize transitions between the same Nilsson levels in mother and daughter nuclei and are denoted as 'favoured transitions'. Hindrance factors  {\it{HF}}\,=\,(4\,-\,10) indicate a favourable overlap between the initial and final nuclear state, while values
{\it{HF}}\,=\,(10\,-\,100) point to an unfavourable overlap, but still parallel spin projections of the initial and final state. 
Factors  {\it{HF}}\,=\,(100\,-\,1000) indicate a parity change and still parallel spin projections, while  {\it{HF}}\,$>$\,1000 mean a parity change and a spin flip.\\
Thus hindrance factors already point to differences in the initial and final states, but on their own do not allow for spin and parity assignments. It is, however, known that in even-Z odd-mass nuclei nuclear structure and thus $\alpha$ decay patterns are similar along the isotone lines (see {\it{e.g.}} \cite{Asai15}), while in odd-Z odd-mass nuclei this feature is evident along the isotope lines (see {\it{e.g.}} \cite{Hess05}). So, in certain cases, based on empirical relationships tentative spin and parity
assignments can be established, as done {\it{e.g.}} in suggesting an $\alpha$ decay pattern for $^{255}$No by P. Eskola et al.
\cite{Eskola70}, which later was confirmed by $\alpha$ - $\gamma$ spectroscopy measurement \cite{Hess06,Asai11}.\\
Another feature in the case of odd-mass nuclei is the fact, that competition between structural hindrance and Q-value hindrance may lead to complex $\alpha$ - decay patterns. Nilsson levels identical to the ground-state of the mother nucleus may be excited states located at several hundred keV in the daughter nuclei, {\it{e.g.}}, 
in a recent decay study of $^{259}$Sg it was shown that the 11/2$^{-}$[725] Nilsson level assigned to the ground-state in this isotope, is located at {\it{E$^{*}$}} $\approx$ 600 keV in the daughter nucleus $^{255}$Rf \cite{AntH15}. Therefore the advantage of a low hindrance factor may be cancelled by a lower barrier transmission probability due to a significantly lower Q-value compared to the ground-state to ground-state transition. Consequently $\alpha$ transitions with moderate hindrance factors into lower lying levels may have similar or even higher intensities than the favored transition as it is the case in the above mentioned examples, $^{255}$No and $^{259}$Sg.\\
A drawback of many recent $\alpha$ decay studies of odd-mass nuclei in the transfermium region was the fact that the ground-state to ground-state transition could not be clearly identified and thus the 'total' {\it{Q$_{\alpha}$}} value could not be established. 
Another difficulty in these studies was the existence of isomeric states in several nuclei, also decaying by $\alpha$ - emission 
and having halflives similar as to the ground-state, as in the cases of $^{251}$No \cite{Hessb06} or $^{257}$Rf \cite{Hess97}, while in early studies ground-state decay and isomeric decay could not be disentangled. Enhanced experimental techniques, applying also $\alpha$ - $\gamma$ spectroscopy have overcome widely that problem in the transfermium region. An illustrative example is the 
{\it{N-Z}} = 49 - line, where based on the directly measured mass of $^{253}$No \cite{Dworschak10},
and decay data of $^{253}$No \cite{Hess12}, $^{257}$Rf, $^{261}$Sg \cite{Streich10}, $^{265}$Hs \cite{Hess09} and $^{269}$Ds \cite{Hof95} 
experimental masses could be determined up to $^{269}$Ds and could serve for a test of theoretical predictions \cite{Moller95,Liran02} and empirical evaluations \cite{Audi03}, as shown in fig. 27.
The masses predicted by M\"oller et al. \cite{Moller95} agree with the experimental value within $\approx$0.5 MeV
up to {\it{Z}}\,=\,106, while towards  {\it{Z}}\,=\,110 ($^{269}$Ds) deviations rapidly increase up to nearly 2 MeV.
A similar behavior was observed for the 
even-even nuclei of the {\it{N-Z}}\,=\,50 line \cite{Hess16a}, which was interpreted as a possible signature for a lower 
shell effect at {\it{N}}\,=\,162.
\begin{figure}
	%\begin{figure*}
	\resizebox{0.93\textwidth}{!}{%
		\includegraphics{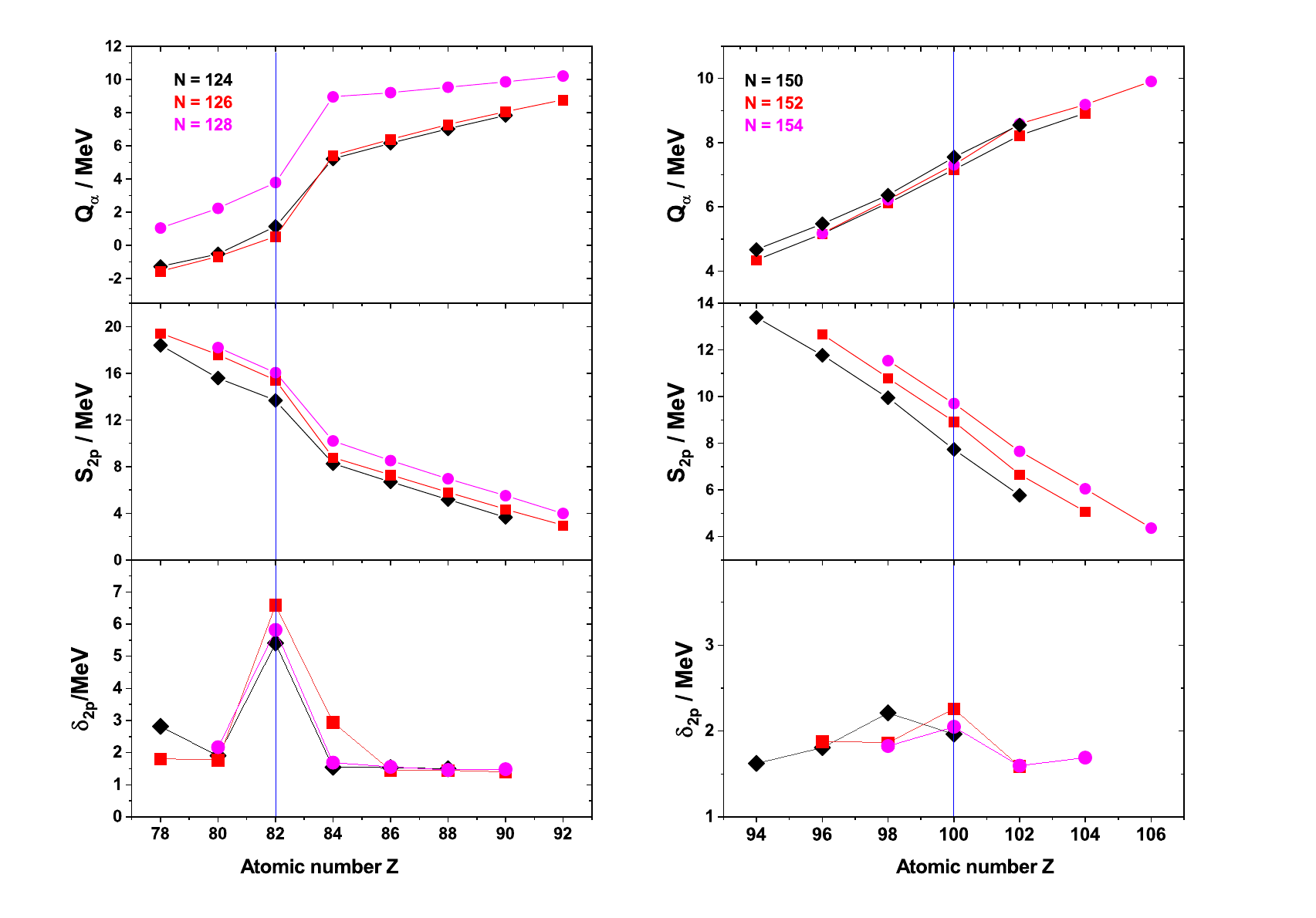}
	}
	% If not, use
	%\vspace{5cm}       % Give the correct figure height in cm
	\caption{left side: Q$_{\alpha}$ - values and 2p - binding energies (S$_{2p}$ for even - even nuclei of N\,=\,124 (diamonds), 126 (squares), 
		128 (circles) around Z\,=\,82;
		right side: Q$_{\alpha}$ - values and 2p - binding energies (S$_{2p}$ for even - even nuclei of N\,=\,150 (diamonds), 152 (squares), 
		154 (circles) around Z\,=\,100.
		Data are taken from \cite{Wang16}.}
	\label{fig:28}       % Give a unique label
\end{figure}
\subsection{\bf{7.2 Q$_{\alpha}$ values as signatures for nuclear shells}}
Historically evidence for nuclear shells was first found from the existence of specifically stable nuclei at certain 
proton and neutron numbers ({\it{Z,N}} = 2, 8, 20, 28, 50, 82 and {\it{N}} = 126) which were denoted as 'magic'. Experimental signatures
were, e.g. strong kinks in the 2p- or 2n - binding energies at the magic numbers and on the basis of enhanced nuclear decay
data also by local minima in the {\it{Q$_{\alpha}$}} values.
The existence of nuclear shells was theoretically explained by the nuclear shell model \cite{Goep48,Haxel49}, which showed
large energy gaps in the single particle levels at the 'magic' numbers, which were equated with 'shell closures'. 
This item was the basis for the prediction of 'superheavy' elements around  {\it{Z}}\,=\,114 and {\it{N}}\,=\,184  when the nuclear shell model was extended in the region
of unknown nuclei far above {\it{Z}}\,=\,82 and {\it{N}}\,=\,126 \cite{Sobi66,Meld67}.
As the shell gap is related to a higher density of single particle levels, compared to the nuclear average (expected e.g. from a Fermi gas model) 
below the Fermi level and a lower density above the Fermi level,
the large energy gaps at the magic numbers go hand in hand with large shell correction energies, leading to the
irregularities in the 2p-, 2n- separation energies and in the {\it{Q$_{\alpha}$}} values.\\
In between {\it{Z}}\,=\,82,  {\it{N}}\,=\,126 and {\it{Z}}\,=\,114 and {\it{N}}\,=\,184 a wide region of stronly deformed nuclei is existing. 
Calculations (see e.g. \cite{Chas77}) resulted in large shell gaps at {\it{N}}\,=\,152 and {\it{Z}}\,=\,100. Later theoretical studies
also showed in addition a region of large shell correction energies in between {\it{N}}\,=\,152 and {\it{N}}\,=\,184 \cite{Cwiok83,Moller86,Patyk91,Patyk91a}
the center of which is presently set at {\it{N}}\,=\,162 and {\it{Z}}\,=\,108.\\
While the (deformed) nuclear shell at N\,=\,152 is well established on the basis of the {\it{Q$_{\alpha}$}} - value as seen from fig. 22 and there is,
despite of scarce data, strong evidence for a shell at {\it{N}}\,=\,162, the quest for a shell closure at {\it{Z}}\,=\,100 is still open.
It was pointed out by Greenlees et al. \cite{Green08} that their results on nuclear structure investigation of $^{250}$Fm
are in-line with a shell gap at {\it{Z}}\,=\,100, but 2p - separation energies and {\it{Q$_{\alpha}$}} - values do not support a shell closure.
The item is shown in fig. 28. On the right hand side {\it{Q$_{\alpha}$}} values and 2p - binding energies ({\it{S$_{2p}$}}) are plotted for
three isotone chains ({\it{N}}\,=\,124, 126, 128) around {\it{Z}}\,=\,82. In all three cases a strong increase in the {\it{Q$_{\alpha}$}} values and
a strong decrease in the {\it{S$_{2p}$}} values is observed from {\it{Z}}\,=\,82 to {\it{Z}}\,=\,84. On the right hand side {\it{Q$_{\alpha}$}} values 
and {\it{S$_{2p}$}}  
values are plotted around {\it{Z}}\,=\,100 for {\it{N}}\,=\,150, 152, and 154. Here a just a straight increase for both is observed from {\it{Z}}\,=\,94 to
{\it{Z}}\,=\,106. \\
Alternatively the so-called 'shell-gap parameter'defined as the difference of the 2n ($\delta_{2n}$) - or 2p  ($\delta_{2p}$) - binding energies is used
characterize nuclear shells. In the present case we get \\
$\delta_{2p}$\,=\,S$_{2p}$(N,Z)\,-\,S$_{2p}$(N,Z+2) \\
A proton shell is indicated by a local maximum of $\delta_{2p}$.
The result for the regions around {\it{Z}}\,=\,82 and {\it{Z}}\,=\,100 are shown in the lower panels of fig. 28. We observe a clear maximum
at {\it{Z}}\,=\,82 is seen for the N\,=\,124, 126, 128 - isotones. A 'mean dífference' $\Delta$, defined as the average value for all 
$\delta$\,=\,$\mid$$\delta_{2p}$(Z=82,N)\,-\,$\delta_{2p}$(Z=80,N)$\mid$ and $\delta$\,=\,$\mid$$\delta_{2p}$(Z=82,N)\,-\,$\delta_{2p}$(Z=84,N)$\mid$
with N\,=\,124, 126, 128 of $\Delta$\,=\,3.93$\pm$0.48 MeV is obtained.\\
At {\it{Z}}\,=\,100 the situation is different. Here a maximum of $\delta_{2p}$ at {\it{Z}}\,=\,100  is observed only for the N\,=\,152, 154 isotone lines,
for N\,=\,150 the maximum is located  {\it{Z}}\,=\,98, while for the N\,=\,152, 154 we obtainin addition only a small value of $\Delta$\,=\,0.44$\pm$0.18 MeV.
So not only the location of the 'shell' is not unambiguous, but the 'strength' is much lower than at  {\it{Z}}\,=\,82. So it does not seem
to be justified to speak of  a proton shell (or a 'magic' number) at {\it{Z}}\,=\,100 as claimed recently \cite{AckT17}.

\subsection{\bf{7.3 Electron Capture decays}}
The analysis of $\alpha$ decay chains from SHN produced in reactions of $^{48}$Ca with actinide targets 
so far acted on the assumptions that the chains consisted on a sequence of $\alpha$
decays and were finally terminated by spontaneous fission \cite{Oga16}. The possibility that one of the chain members
could undergo EC - decay was not considered. Indeed, EC - decay of superheavy nuclei has been only little investigated so
far. Mainly this is due to the technical difficulties to detect EC decay at very low production rates of the isotopes.
Consequently, only very recently EC decay has been investigated successfully in the transactinide region for the cases of
$^{257}$Rf \cite{Hess16} and $^{258}$Db \cite{Hess16a}. Two ways of identifying EC - decay turned out to be successful,
a) measuring delayed coincidences between K X-rays and $\alpha$ decay or spontaneous fission of the EC - daughter, and b)
measuring delayed coincidences between implanted nuclei and conversion electrons (CE) from decay of excited states
populated by the EC or delayed coincidences between CE and decays ($\alpha$ decay or spontaneous fission) of 
the EC daughter. The latter cases, however, require population of excited level decaying by internal conversion, which is not necessarily the case.\\
Evidence for occuring EC within the decay chains of SHE gives the termination of the decay chains of odd-odd nuclei by spontaneous fission. Since spontaneous fission of odd-odd nuclei is strongly hindered, it can be assumed that it may not the odd-odd nucleus that undergoes fission, but the even-even daughter nucleus, produced by EC decay \cite{OganU15}.\\
The situation is, however, quite complicated. To illustrate we compare in fig. 29 the experimental 
(EC/$\beta^{+}$) - halflives of lawrencium and dubnium isotopes with recently calculated \cite{Karpov12} EC - halflives.
\begin{figure}
	%\begin{figure*}
	\vspace{-2.3cm}
	\resizebox{0.6\textwidth}{!}{%
		\includegraphics{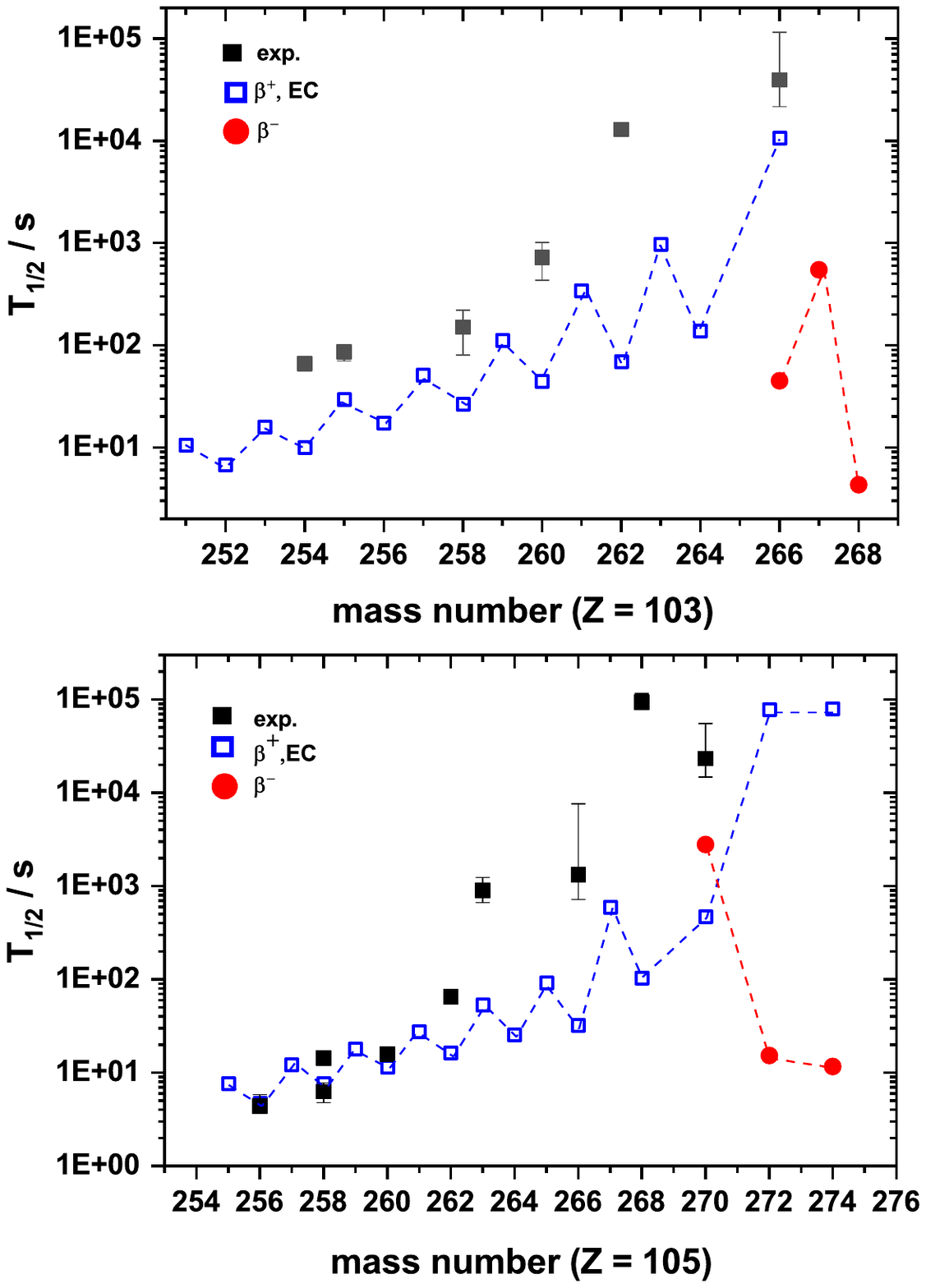}
	}
	% If not, use
	%\vspace{5cm}       % Give the correct figure height in cm
	\caption{Comparison of experimental and calculated \cite{Karpov12} EC - halflives for lawrencium (upper figure)
		and dubnium (lower figure) isotopes. For the cases $^{266,268,270}$Db it was assumed that spontaneous fission
	originates from the even-even EC - daughter. Full squares - experimental halflives, open squares - $\beta^{+}$,CE - halflives,
    circles - $\beta^{-}$ halflives.}
\label{fig:29}       % Give a unique label
\end{figure}

\begin{figure}
	%\begin{figure*}
	\vspace{-0.5cm}
	\resizebox{0.8\textwidth}{!}{%
		\includegraphics{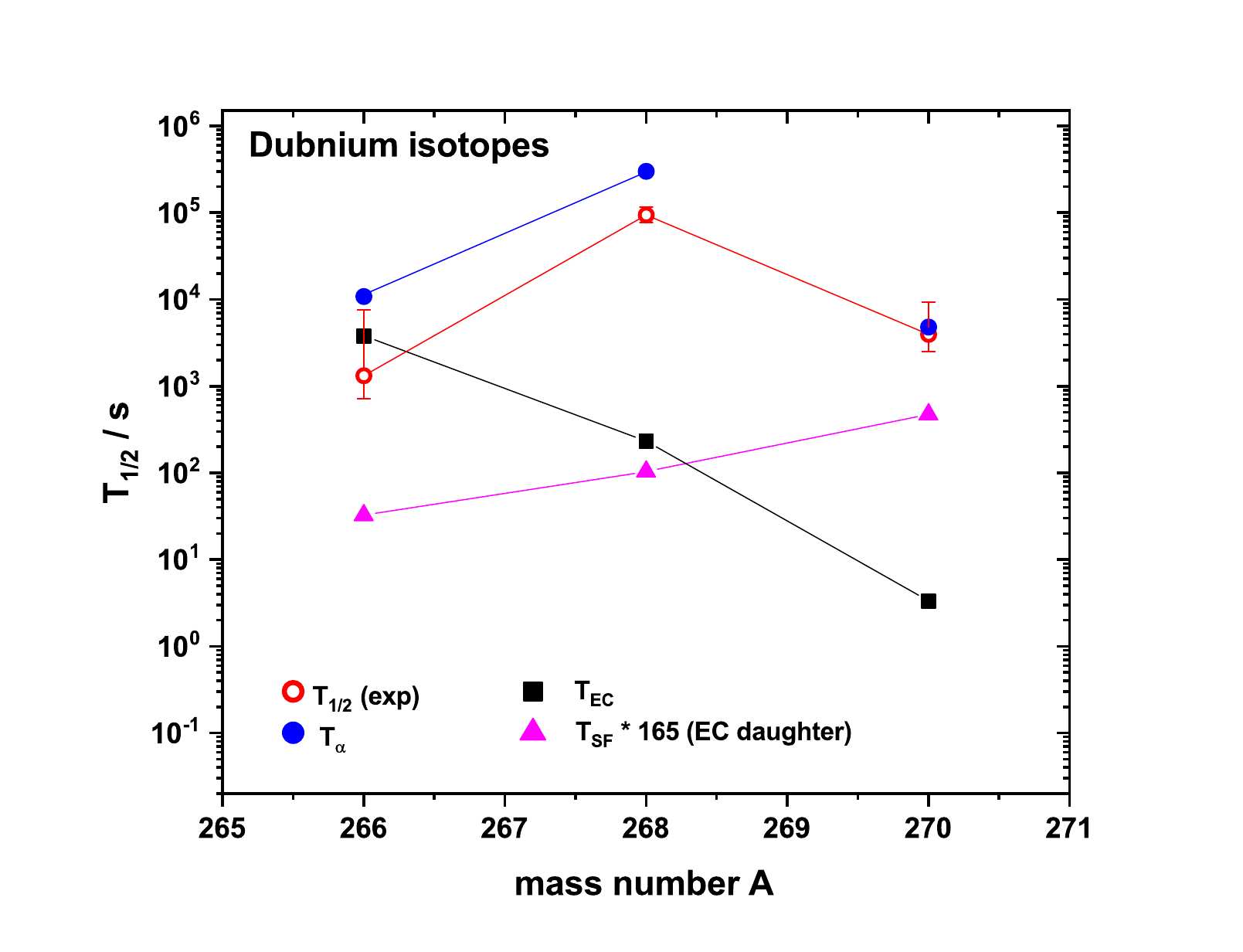}
	}
	% If not, use
	%\vspace{5cm}       % Give the correct figure height in cm
	\caption{Comparison of experimental and calculated \cite{Karpov12} EC - halflives, $\alpha$ halflives \cite{PoI80}
		of $^{266,268,270}$Db and theoretical SF halflives \cite{Smolan95a} of the EC daughters 
		$^{266,268,270}$Rf.}
	\label{fig:30}       % Give a unique label
\end{figure}
In general the agreement between experimental and calculated values is better for the dubnium, specifcally for 
{\it{A}}\,$\le$\,262, than for the lawrencium isotopes. Evidently, however, the disagreement increases at approaching
the line of beta - stability. The experimental EC halflives are up to several orders of magnitude higher than the
theoretical values, which may lead to assume that direct spontaneous of the odd-odd nuclei is observed indeed.
So there is lot of room for speculation. In this context the difficulties to make also some 'empirical'
conclusions shall be briefly discussed. So far, only in two cases, $^{260}$Md and $^{262}$Db observation 
of spontaneous fission of an odd-odd isotope is reported. $^{262}$Db seems, however, a less certain case 
(see discussion in \cite{Hess17}). In table 10 the 'fission halflives' of $^{268}$Db, $^{262}$Db and $^{260}$Md
are compared with the values obtained for their odd-mass neighbouring isotopes with {\it{A-1}} and {\it{Z-1}}, respectively.
\begin{table}
	\begin{center}
		\begin{tabular}{l l l l }
			\hline
			\hline
			Isotope & T$_{SF}$ /s  & HF \\
			\hline
			$^{268}$Db  & 93600 &      \\
			$^{267}$Db  &  4320 &  21.7 \\
			$^{267}$Rf  &  4680 &  20.0 \\
			\hline
			$^{262}$Db  &   103 &      \\
			$^{261}$Db  &    18.4 &  5.6 \\
			$^{261}$Rf  &    32.5 &  3.2 \\
			\hline
			$^{260}$Md  &   2.75x10$^{6}$ &      \\
			$^{259}$Md  &    5700 &  482 \\
			$^{261}$Rf  &    1.5 &  1.9x10$^{6}$ \\
			\hline
		\end{tabular}
	\end{center}
	\caption{ Comparison of 'fission' halflives of some selected odd-mass and odd-odd nuclei in the range
		{\it{Z}}\,=\,101\,-\,105. The 'hindrance factor' HF here means the ratio of fission halflives of the
	odd-odd nucleus and its neighbouring odd mass nuclei (see text).} \label{tab10}
\end{table}
\begin{figure}
	%\begin{figure*}
	\resizebox{0.8\textwidth}{!}{%
		\includegraphics{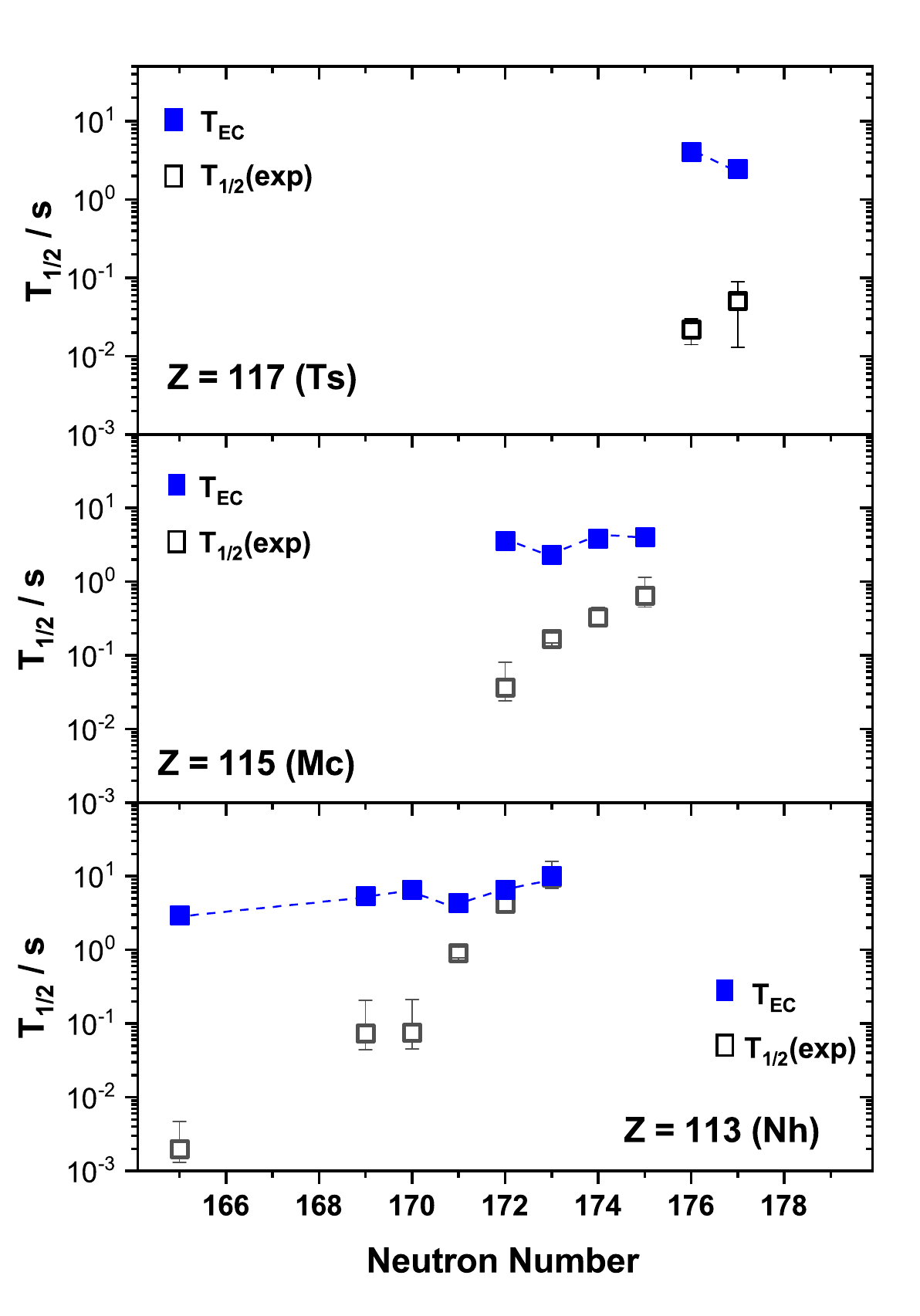}
	}
	% If not, use
	%\vspace{5cm}       % Give the correct figure height in cm
	\caption{Comparison of experimental and calculated \cite{Karpov12} EC - halflives for Z\,=\,113, 115, 117 - isotopes.
	The full squares denote the theoretical EC - halflives, the open squares the experimental halflives. The lines are to guide the eye.}
	\label{fig:31}       % Give a unique label
\end{figure}
\begin{figure}
	%\begin{figure*}
	\resizebox{0.8\textwidth}{!}{%
		\includegraphics{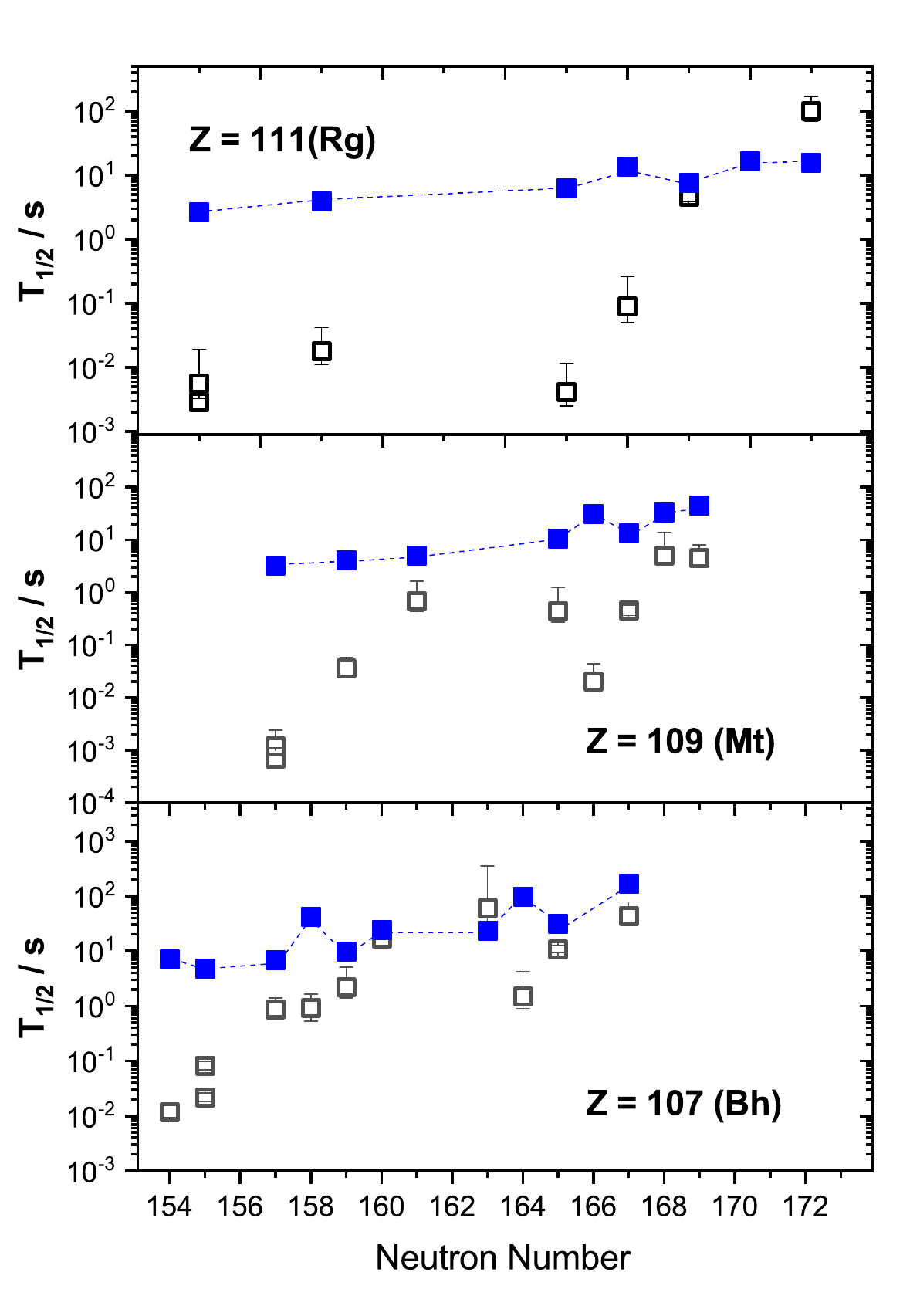}
	}
	% If not, use
	%\vspace{5cm}       % Give the correct figure height in cm
	\caption{Comparison of experimental and calculated \cite{Karpov12} EC - halflives for Z\,=\,107, 109, 111 - isotopes.
		The full squares denote the theoretical EC - halflives, the open squares the experimental halflives. The lines are to guide the eye.}
	\label{fig:32}       % Give a unique label
\end{figure}
\begin{figure}
	%\begin{figure*}
	\resizebox{0.8\textwidth}{!}{%
		\includegraphics{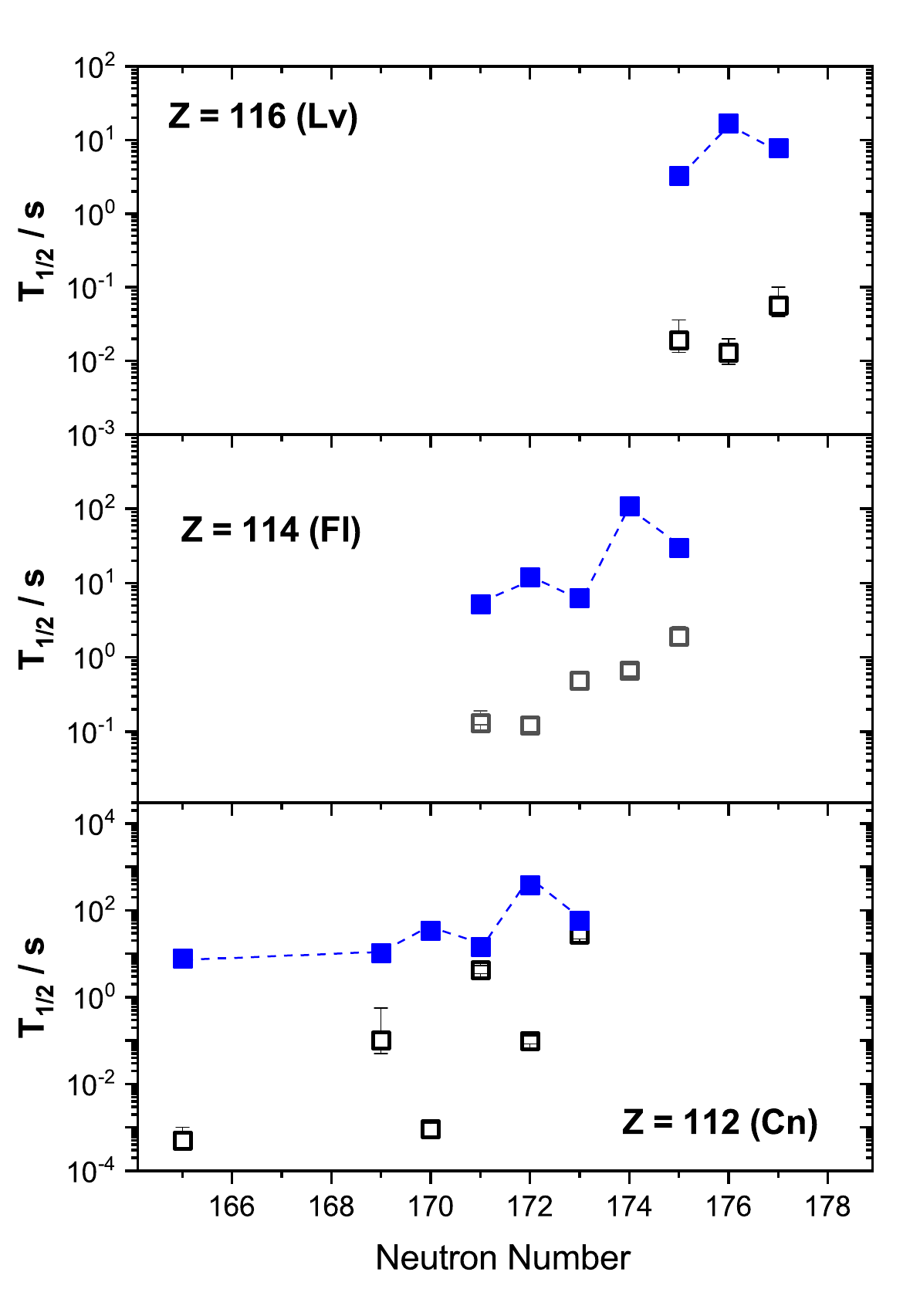}
	}
	% If not, use
	%\vspace{5cm}       % Give the correct figure height in cm
	\caption{Comparison of experimental and calculated \cite{Karpov12} EC - halflives for Z\,=\,112, 114, 116 - isotopes.
		The full squares denote the theoretical EC - halflives, the open squares the experimental halflives. The lines are to guide the eye.}
	\label{fig:33}       % Give a unique label
\end{figure}
\begin{figure}
	%\begin{figure*}
	\resizebox{0.8\textwidth}{!}{%
		\includegraphics{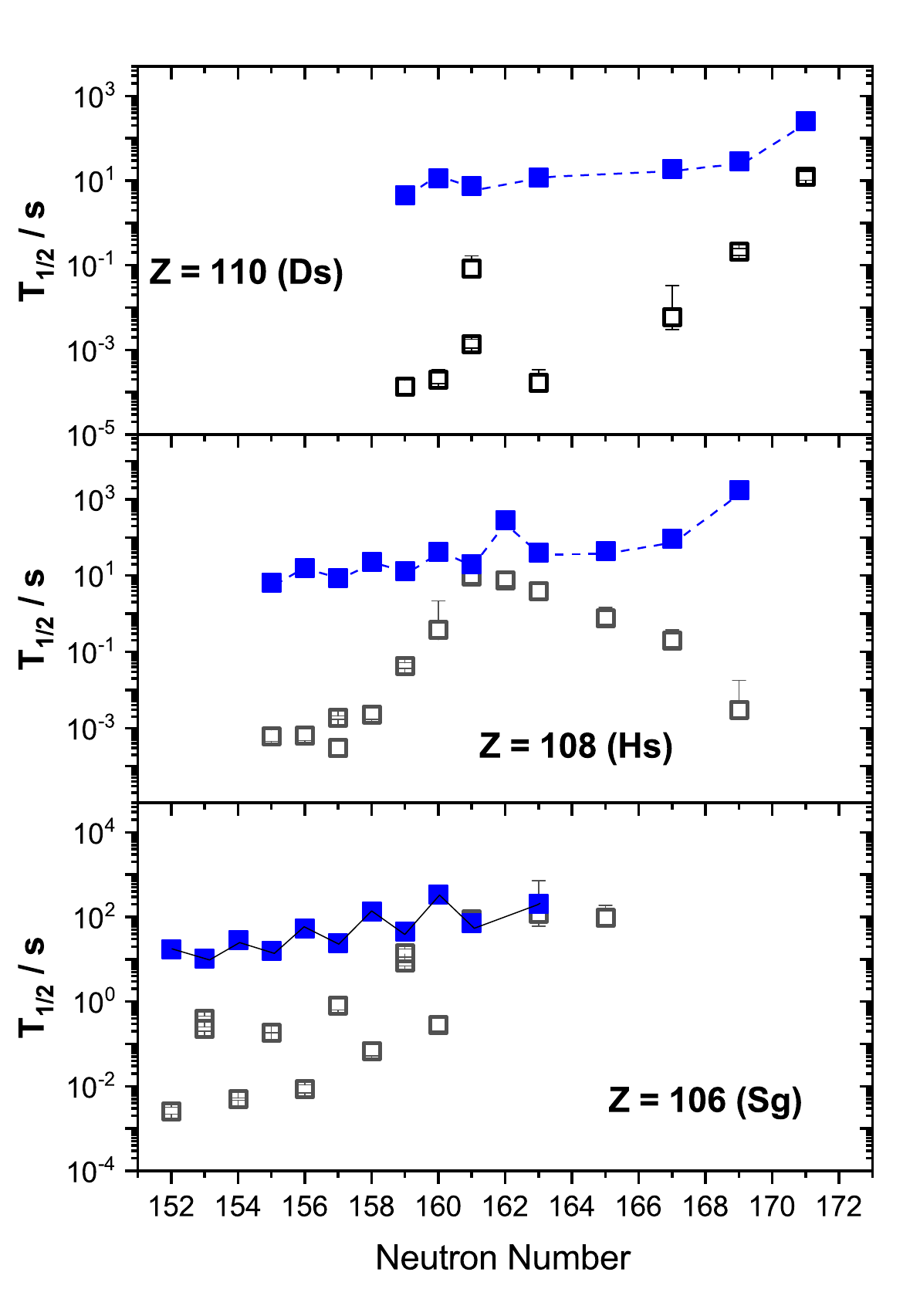}
	}
	% If not, use
	%\vspace{5cm}       % Give the correct figure height in cm
	\caption{Comparison of experimental and calculated \cite{Karpov12} EC - halflives for Z\,=\,106, 108, 110 - isotopes.
			The full squares denote the theoretical EC - halflives, the open squares the experimental halflives. The lines are to guide the eye.
		$^{271}$Sg is predicted as stable against beta - decay.}
	\label{fig:34}       % Give a unique label
\end{figure}
Evidently the resulting hindrance factors {\it{HF}}\,=\,T$_{SF}$(Z,A)/T$_{SF}$(Z,A-1) or {\it{HF}}\,=\,T$_{SF}$(Z,A)/T$_{SF}$(Z-1,A-1) are 
much lower for $^{268}$Db (21.7, 20.0) than for $^{260}$Md (482, 1.9x10$^{6}$). These low values suggest that 'fission' of $^{268}$Db originates indeed 
from the EC - daughter $^{268}$Rf. The lower hindrances factors for $^{262}$Db (see table 10), although the case is debated, puts 
some doubts in that interpretation. On the other hand it is quite common to take the ratio of the experimental 
fission half-life and an 'unhindered' fission half-life, defined as the geometric mean of the neighbouring
even - even isotopes (see \cite{Hess17} for more detailed discussion), but to estimate reliable hindrance factors the 
spontaneous fission halflives of the surrounding even-even nuclei have to be known. In the region of  $^{266,268,270}$Db
only for one even-even isotope, $^{266}$Sg the fission halflive is known, {\it{T$_{sf}$}}\,=\,58 s, while a theoretical value of
{\it{T$_{sf}$}}\,=\,0.35 s was reported \cite{Smolan95a}, which is lower by a factor of $\approx$165. Under these 
cirumstances it does not make much sense to use theoretical values to estimante hindrance factors for 
spontaneous fission.\\
Therefore final decision if the terminating odd - odd nuclei may fission directly must be left to future experiments. 
Techniques to identify EC decay have been presented at the beginnig of this section. But it has to be kept in mind that the
identification mentioned was performed for isotopes with production cross sections of some nanobarn, while in
the considered SHE region production rates are roughly three orders of magnitude lower. So technical effort to 
increase production rates and detection efficiencies are required to perform successful experiments in that direction.
From physics side such experiments may cause big problems as seen from fig. 30, where experimental halflives of 
$^{266,268,270}$Db (red circles) are compared with the calculated EC halflives from \cite{Karpov12} (black squares). 
Calculated SF halflives from \cite{Smolan95a} for the even-even EC - daughters $^{266,268,270}$Rf,  are in the 
range of $\approx$20 ms - $\approx$20 s, so the technique for identification should be applicable. 
The situation could be, however, unfavourable if there is a similar situation as in the case of $^{266}$Sg, where the
experimental SF halflife is a factor of 165 longer than the predicted one. These modified SF halflives are shown in 
fig. 30  by the margenta triangles.
For comparison in fig. 30 also the expected $\alpha$ decay halflives for $^{266,268}$Db based on the E$_{\alpha}$ values 
calculated from the mass predictions of \cite{Moller95}
({\it{E$_{\alpha}$($^{266}$Db)}}\,=\,7242 keV, {\it{E$_{\alpha}$($^{268}$Db)}}\,=\,7076 keV) are presented.
As for $^{270}$Db a value of {\it{E$_{\alpha}$}}\,=\,7721 keV is predicted in \cite{Moller95}, which is roughly 200 keV
lower than the experimental value of {\it{E$_{\alpha}$}}\,=\,7.90$\pm$0.03 MeV, conservatively 300 keV higher values were taken for 
$^{266,268}$Db. The halflives were calculated using the  formula from \cite{PoI80}. Results are shown as blue dots in fig. 30. For 
$^{270}$Db the experimental $\alpha$ - decay half-life is given. Evidently the values for $^{266,268}$Db are still about an order 
of magnitude higher, so non-observation of $\alpha$ decay of these isotopes so far is in-line with the expectations.\\
Another interesting feature is to identify candidates for EC - decay within the $\alpha$ - decay chains. With respect to the
quite uncertain predictions of EC - halflives that task is not trivial. As the experimental EC halflives are longer than the
calculated ones for the lawrencium and dubnium isotopes (see fig. 29) one tentatively may assume that conditions are 
similar for the heavier elements (it should be kept in mind that this item is not proven !). In other words, candidates for
EC decay are isotopes for which the experimental half-life is similar or even longer than the calculated \cite{Karpov12} EC - half-life.
In figs. 31-34, the experimental and calculated EC halflives are compared for the known nuclei with {\it{Z}}\,$\ge$\,106. As seen
no EC decay can be expected for isotopes of elements of {\it{Z}}\,=\,108, {\it{Z}}\,=\,110, and {\it{Z}}\,=\,114-118;
possible candidates are $^{285,286}$Nh ({\it{N}}=172,174), $^{283,285}$Cn ({\it{N}}=171,173), $^{280,281,282}$Rg ({\it{N}}=169,170,171), $^{270}$Bh (N=163), and
$^{269}$Sg. It should, however, stressed that these isotopes are just 'candidates'.\\
An example for possibly having observed EC in $\alpha$ decay chains are the 'short chains' registered in irradiations of
$^{243}$Am with $^{48}$Ca \cite{FoR16}, denoted as B1 - B3. As a possible explantion the decay sequence 
$^{288}$Mc\,$^{\alpha}_{\rightarrow}$\,$^{284}$Nh\,$^{EC}_{\rightarrow}$\,$^{284}$Cn\,$^{SF}_{\rightarrow}$ was given
in scenario 2. So far, this item is, however, is just an interesting feature, which has to be investigated thoroughly in future.\\
\\

\subsection{\bf{7.4 Spontaneous fission}}

Spontaneous fission is believed to finally terminating the charts of nuclei towards increasing proton numbers 
{\it{Z}}. The strong shell stabilization of nuclei in the vicinity of the spherical proton and neutron shells
{\it{Z}} = 114 or {\it{Z}} = 120 and {\it{N}} = 172 or {\it{N}} = 184 leads also to high fission barriers and thus
long fission halflives. Qualitatively these expections are in line with the experimental results. For all nuclei 
{\it{Z}} $>$ 114 so far only $\alpha$ decay was observed, while for {\it{Z}} = 112\,--\,114 only for five nuclei
$^{284,286}$Fl, $^{282,283,284}$Cn spontaneous fission was reported.
The spontaneous fission halflives of the even-even nuclei $^{284,286}$Fl, $^{282,284}$Cn, $^{280}$Ds agree
within two orders of magnitude, those for $^{284,286}$Fl even within one order of magnitude
with the predictions of R. Smolanczuk et al. \cite{Smolan95a}, which calculations also quite fairly reproduce the halflives 
of the even-even isotopes of rutherfordium ({\it{Z}} = 104), seaborgium ({\it{Z}} = 106), and hassium ({\it{Z}} = 108).
These results indicate that the expected high stabilization against spontaneous fission in the vicinity of the 
spherical proton and neutron shells is indeed present. For further discussion of these items we refer to the
review paper \cite{Hess17}. 

\begin{figure}
	%\begin{figure*}
	\resizebox{1.0\textwidth}{!}{%
		\includegraphics{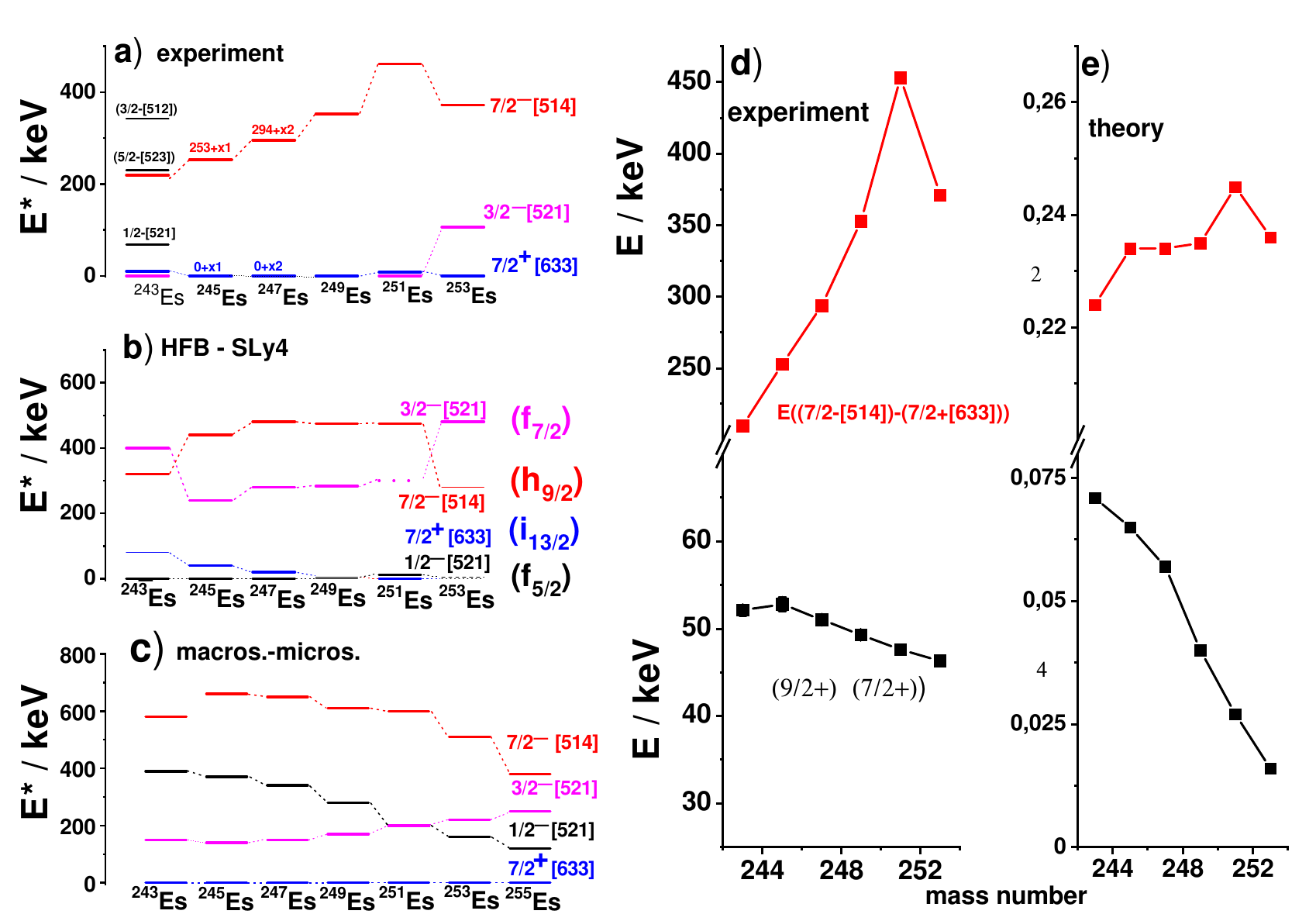}
	}
	% If not, use
	%\vspace{5cm}       % Give the correct figure height in cm
	\caption{a) experimental low lying Nilsson levels in odd-mass einsteinium iotopes (data taken from \cite{Hess05}); b) results of HFB - SLy4 calculations for odd mass einsteinium isotopes (data taken from \cite{ChaT06}); c)  results of macroscopic - microscopic calculations for odd mass einsteinium isotopes \cite{ParS04}; d) (upper panel) energy differences between the 7/2$^{-}$[514] and 7/2$^{+}$[633] Nilsson levels, (lower panel) energy difference between the 7/2$^{+}$[633] bandhead and the 9/2$^{+}$ rotational band member; e) (upper panel) quadropule deformation parameters $\beta_{2}$ for odd mass einsteinium isotopes \cite{Moller95}, (lower panel) hexadecapole deformation parameters $\beta_{4}$ for odd mass einsteinium isotopes \cite{Moller95}.  }
	\label{fig:35}       % Give a unique label
\end{figure}

\subsection{\bf{7.5 Systematics in nuclear structure -  odd-mass einsteinium isotopes}}

Detailed information on nuclear structure of heaviest nuclei provide a wide field of information for testing nuclear models with respect to their predictive power. Presently the situation, however, is not very satisfying for at least three major reasons;\\
a) 'detailed' decay studies using $\alpha$ - $\gamma$ spectroscopy are essentially only possible for nuclei with Z\,$\le$107 due to low production rates;\\
b) for many isotopes only very few Nilsson levels have been identified, while the assignment is partly only tentative;\\
c) agreement between experimental data and results from theoretical calculations is in general rather poor.\\
In \cite{Asai15} experimental data are compared with results from theoretical calculations for N\,=\,151 and N\,=\,153 isotones of even-Z elements in the range Z\,=\,94-106. Agreement in excitation energies of the Nilsson levels is often not better than a few hundred keV and also the experimentally established ordering of the levels is often not reproduced by the calculations. Thus,
for example, the existence of the low lying 5/2$^{+}$[622] - isomers in the N\,=\,151 isotones is not predicted by the calculations. These deficiencies, on the other hand, make it hard to trust in predictions of properties of heavier nuclei by these models.\\
In this study the situation is illustrated for the case of the odd-mass einsteinium isotopes (fig. 35). Experimentally only two Nilsson levels have been established in most of presented isotopes, namely 7/2$^{+}$[633] and 7/2$^{-}$[514]. In the heaviest isotopes, $^{251,253}$Es also the 3/2$^{-}$[521] was assigned. While in $^{253,249}$Es 7/2$^{+}$[633] was identified as
ground - state \cite{MooL93,AhmS70}, in case of $^{251}$Es the ground-state was assigned as 3/2$^{-}$[521] \cite{AhmC00}.
For the lighter einsteinium isotopes the situation is unclear. The Nilsson levels 7/2$^{+}$[633] and 7/2$^{-}$[514] have been established from $\alpha$ - $\gamma$ decay studies of odd-mass mendelvium isotopes \cite{Hess05,ChaT06}. However no ground-state assigment was made for $^{245,247}$Es as on the basis of the results for the heavier einsteinium isotopes as it could not be excluded, that the 
7/2$^{+}$[633] and 3/2$^{-}$[521] are close in energy and may alter as ground-state. In a more detailed decay study of
$^{247}$Md the 3/2$^{-}$[521] level was identified as the ground-state in $^{243}$Es, while the 7/2$^{+}$[633] is located at 
E$^{*}$\,=\,10 keV \cite{Hes20a}. In addition the 1/2$^{-}$[521] level and also, tentatively, the 5/2$^{-}$[523] and 3/2$^{-}$[512] Nilsson levels were
identified in that study.\\
The experimental data are compared with theoretical calculations in figs. 35b und 35c. In fig. 35b the results from a self-consistent Hartree-Fock-Bogoliubov calculation using SLy4 force (HBF - SLy4) are presented (data taken from \cite{ChaT06}), in fig. 35c the results from a macroscopic - microscopic calculation \cite{ParS04}. The HBF - SLy4 calculations only predict the ground-state of $^{253}$Es correctly, for $^{251}$Es they result in 7/2$^{+}$[633] as for $^{253}$Es. For the lighter isotopes 
the ground-state is predicted as 1/2$^{-}$[521], while the 3/2$^{-}$[521], for which strong experimental evidence exists that it is ground-state or located close to the ground-state, is located at E$^{*}$\,$\approx$400 keV, except for $^{253}$Es. The macroscopic - microscopic calculations, on the other side, predict 7/2$^{+}$[633] as a low lying level but the 3/2$^{-}$[521] one in an excitation energy range of E$^{*}$\,$\approx$\,(400-600) keV. 
As noted in fig. 35b, the 3/2$^{-}$[521], 7/2$^{-}$[514] and 7/2$^{+}$[633] Nilsson levels arise from the f$_{7/2}$, h$_{9/2}$ and i$_{13/2}$ subshells located below the shell gap at Z\,=\,114, while the 1/2$^{-}$[521] stems from the f$_{5/2}$ subshell located above it \cite{Chas77}. The  3/2$^{-}$[521], 7/2$^{-}$[514] and 1/2$^{-}$[521] decrease in energy at increasing deformation, while the 7/2$^{+}$[633] increases in energy.  At a deformation $\nu_{2}$\,$\approx$\,0.3 where a shell gap of
$\approx$1 MeV is expected at Z\,=\,100, the 3/2$^{-}$[521] and 7/2$^{+}$[633] states are located below the predicted shell gap, while the 7/2$^{-}$[514] and 1/2$^{-}$[521] are located above it. From this side one can expect that the energy difference $\Delta$E\,=\,E(7/2$^{-}$[514])\,-\,E(7/2$^{+}$[633]) gives some information about the size of the shell gap. Indeed the experimental energy difference is lower than predicted by the HFB-SLy4 calculations (typically $\approx$400 keV) and 
by the macroscopic - microscopic calcultions (typically $\approx$600 keV) as seen from figs. 35a - 35c, which hints to a lower shall gap as preticted. Indeed this could explain the non-observation of a discontinuity in the two-proton binding energies and the Q$_{\alpha}$ - values when crossing Z\,=\,100 (see sect. 7.2).\\
Two more interesting features are evident: in figs. 35d and 33e (upper panel) the energy difference $\Delta$E\,=\,E(7/2$^{-}$[514])\,-\,E(7/2$^{+}$[633] is compared with the quadrupole deformation parameter $\beta_{2}$, while in the lower panels the energy difference of the 7/2$^{-}$[514] bandhead and the 9/2- rotational level is compared with the
hexadecapole deformation parameter  $\beta_{4}$, both taken from \cite{Moller95}. Both, the experimental energy differences $\Delta$E\,=\,E(7/2$^{-}$[514])\,-\,E(7/2$^{+}$[633]) (not so evident in the calculation) and the $\beta_{2}$ values show a pronounced maximum at N\,=\,152, while as well as the energy differences E(9/2$^{-}$) - E(7/2$^{-}$ as the $\beta_{4}$ - values decrease at increasing mass number or increasing neutron number, respectively.\\

\subsection{\bf{7.6 Nuclear structure predictions for odd-odd nuclei - exemplified for $^{250}$Md and $^{254}$Lr}}

	\begin{figure}
		%\begin{figure*}
		\resizebox{0.8\textwidth}{!}{%
			\includegraphics{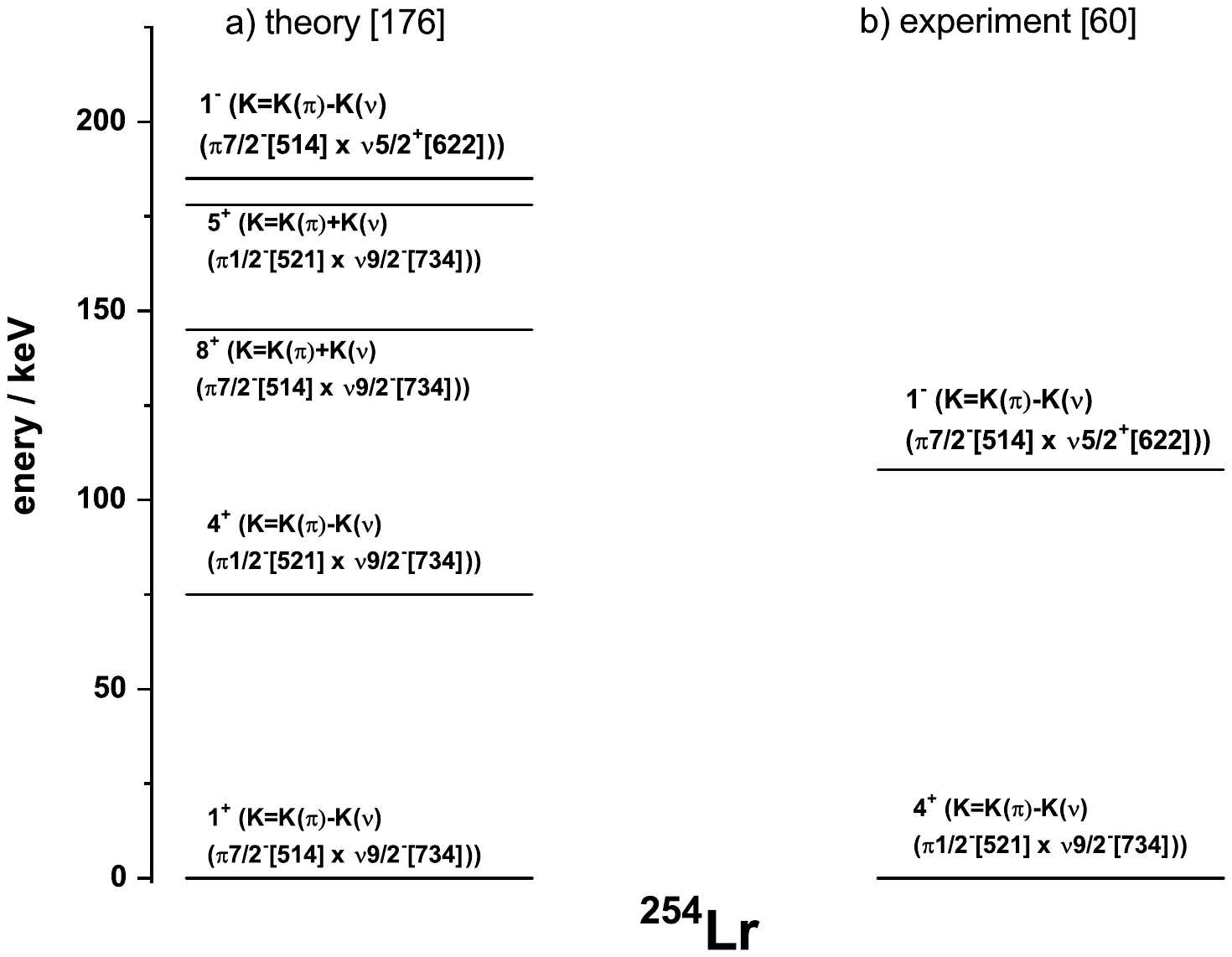}
		}
		% If not, use
		%\vspace{5cm}       % Give the correct figure height in cm
		\caption{Predicted (a) (\cite{Govri18}) and (b) experimentally (tentatively) assigned \cite{Vost15} low lying levels of $^{254}$Lr. }
		\label{fig:36}       % Give a unique label
	\end{figure}

Predictions of level schemes in heaviest odd-odd nuclei are scarce so far. Only for a couple of cases calculations have 
been performed so far. Thus we will discuss here only two cases, $^{250}$Md and $^{254}$Lr for which recently new results
have been reported \cite{Vost15}.\\
The ground-state of $^{250}$Md was predicted by Sood et al.\cite{Sood00} as K$^{\pi}$\,=\,0$^{-}$ and a long-lived isomeric state
with spin and parity K$^{\pi}$\,=\,7$^{-}$
expected to decay primarily by $\alpha$ emission or electron capture was predicted at E$^{*}$\,=\,80$\pm$30 keV.
Recently a longlived isomeric state at E$^{*}$\,=\,123 keV was identified \cite{Vost15} in quite good agreement
with the calculations. However, no spin and parity asignments have be done for the groud state and the isomeric state.\\
The other case concerns $^{254}$Lr. Levels at E$^{*}$$<$250 keV were recently calculated on the basis of a 
'Two-Quasi-Particle-Rotor-Model' \cite{Govri18}. The results 
are shown in fig. 36. The ground-state is predicted as K$^{\pi}$\,=\,1$^{+}$ and an isomeric state  K$^{\pi}$\,=\,4$^{+}$
is predicted at E$^{*}$$\approx$75 keV.	Recently an isomeric state at  E$^{*}$\,=\,108 keV was
identified in $^{254}$Lr. Tentative spin and parity assigments are, however, different. 
The ground-state was assigned as K$^{\pi}$\,=\,4$^{+}$, the isomeric state as K$^{\pi}$\,=\,1$^{-}$ (see fig. 36). This 
assignment was based on the assumed ground-state configuration K$^{\pi}$\,=\,0$^{-}$ of $^{258}$Db and the low $\alpha$-decay hindrance factor
HF$\approx$30 for the transition $^{258g}$Db $\rightarrow$ $^{254m}$Lr 
which rather favors K$^{\pi}$\,=\,1$^{-}$ than K$^{\pi}$\,=\,4$^{+}$ as the latter configuration would require  an angular momentum change
$\Delta$K\,=\,3 and a change of the parity which requires a much larger hindrance factor (see sect. 7.1). \\
Here, however, two items should be considered: \\
a) the spin-parity assigmnent of $^{258}$Db is only tentative,\\
b) the calculations are based on the energies of low lying levels in the neighboring odd mass nuclei, 
in the respecting case $^{253}$No (N = 151) and $^{253}$Lr (Z = 103). The lowest Nilsson
levels in $^{253}$No are 9/2$^{-}$[734] for the ground-state and  5/2$^{+}$[622] for a shortlived isomer at E$^{*}$\,=\,167 keV \cite{Streich10}.
In $^{253}$Lr tentative assignments of the ground-state (7/2$^{-}$[514]) and  1/2$^{-}$[521] 
for a low lying isomer are given in \cite{HesH01}. The energy of the isomer is experimentally not established, for the calculations a value 
of 30 keV  was taken \cite{Govri18}. It should be noted, however, that for the neighboring N = 152 isotope of lawrencium, $^{255}$Lr the ground-state had be determined as the Nilsson level 1/2$^{-}$[521], while 7/2$^{-}$ was attributed to a low lying isomeric
state at E$^{*}$\,=\,37 keV \cite{ChaT06}. Therefore, with respect to the uncertain starting conditions, 
the results of the calculations although not in 'perfect agreement' with the experimental results, are still
promising, and may be improved in future.\\
It should be noticed the that existence and excitation energy of the isomeric state in $^{254}$Lr has been confirmed
by direct mass measurents at SHIPTRAP \cite{Kaleja20}, and there is some confidence that spins can be determined in near future
by means of laser spectroscopy using the RADRIS technique \cite{Laati14}.

\subsection{\bf{ 7.7 Attempts to synthesize elements Z\,=\,119 and Z\,=\,120 - Quest for the spherical proton shell at Z\,=\,114 or Z\,=\,120}}

Although elements up to {\it{Z}}\,=\,118 have been syntheszied so far, the quest for the location of the spherical 'superheavy' proton and neutron shells is still open. 
Indeed synthesis of elements up  {\it{Z}}\,=\,118 in $^{48}$Ca induced reactions show a maximum in the cross sections at {\it{Z}}\,=\,114, which might be seen as an indication of a proton shell at {\it{Z}}\,=\,114 (see fig. 16). Such an interpretation, however, is not unambiguous since a complete understanding of the evaporation residue (ER) production process (capture of projectile and target nuclei, formation of the compund nucleus, deexcitation of the compound nucleus, competition between particle emission and fission)  is required to draw firm conclusions. Indeed V.I. Zagebaev and W. Greiner \cite{ZagG15} could reproduce cross-sections for elements {\it{Z}}\,=\,112 to {\it{Z}}\,=\,118 produced in $^{48}$Ca induced reaction quite fairly, but evidently a main ingredient of their calculations was quite uncertain. They approximated fission barriers as the sum of the 'shell effects' (according to \cite{Moller95}) and a 'zero-point energy' of 0.5 MeV, which resulted in quite different values than obtained from 'direct' fission barrier calculations (see e.g. \cite{Moller09,Moller15,Kowal15}). Due to these uncertainties measured cross sections are not a good argument identification for a proton shell at
{\it{Z}}\,=\,114.
Indeed on the basis of the results on decay studies of $^{286,288}$Fl and their daughter products Samark-Roth et al. \cite{Sarm21} claimed that there is not a real indication for a poton shell at {\it{Z}}\,=\,114.
Indirect evidence that the proton shell may not be located at {\it{Z}}\,=\,114 but rather at {\it{Z}}\,=\,120 comes from a recent decay study of $^{247}$Md, where
the 1/2$^{-}$[521] level was located at E$^{*}$\,=\,68$\pm$11 keV \cite{Hes20a}. As shown in fig. 35b, the Nilsson - levels 3/2$^{-}$[521], 7/2$^{+}$[633], and
7/2$^{-}$[514] Nilsson levels stem from the 2f$_{7/2}$, 1i$_{13/2}$, and 1h$_{9/2}$ sub shells, respectively, which are located below the shell gap at {\it{Z}}\,=\,114,
while the 1/2$^{-}$[521] level stems from the 2f$_{5/2}$ sub shell located above the {\it{Z}}\,=\,114 \cite{Chas77} shell gap. As seen in fig. 35 this level is predicted as 
the ground-state of $^{243}$Es by the HFB SLy4 - calculations (fig. 35b) predicting {\it{Z}}\,=\,120 as proton shell, but as an excited level at E$^{*}$$\approx$400 keV by the macroscipic-microscopic calculations (fig. 35c) predicting {\it{Z}}\,=\,114 as proton shell. The low excitation energy of the 1/2$^{-}$[521] level
might thus be a hint, that the 2f$_{5/2}$ sub-shell is lower in energy, and the enery difference E(2f$_{5/2}$) -  E(2f$_{7/2}$) which is decisive for the occurence 
of the shell gap at {\it{Z}}\,=\,114 or at {\it{Z}}\,=\,120 \cite{Ben03}, i.e. that the proton shell indeed may be located at  {\it{Z}}\,=\,120.

\begin{figure}
	%\begin{figure*}
	\resizebox{0.8\textwidth}{!}{%
		\includegraphics{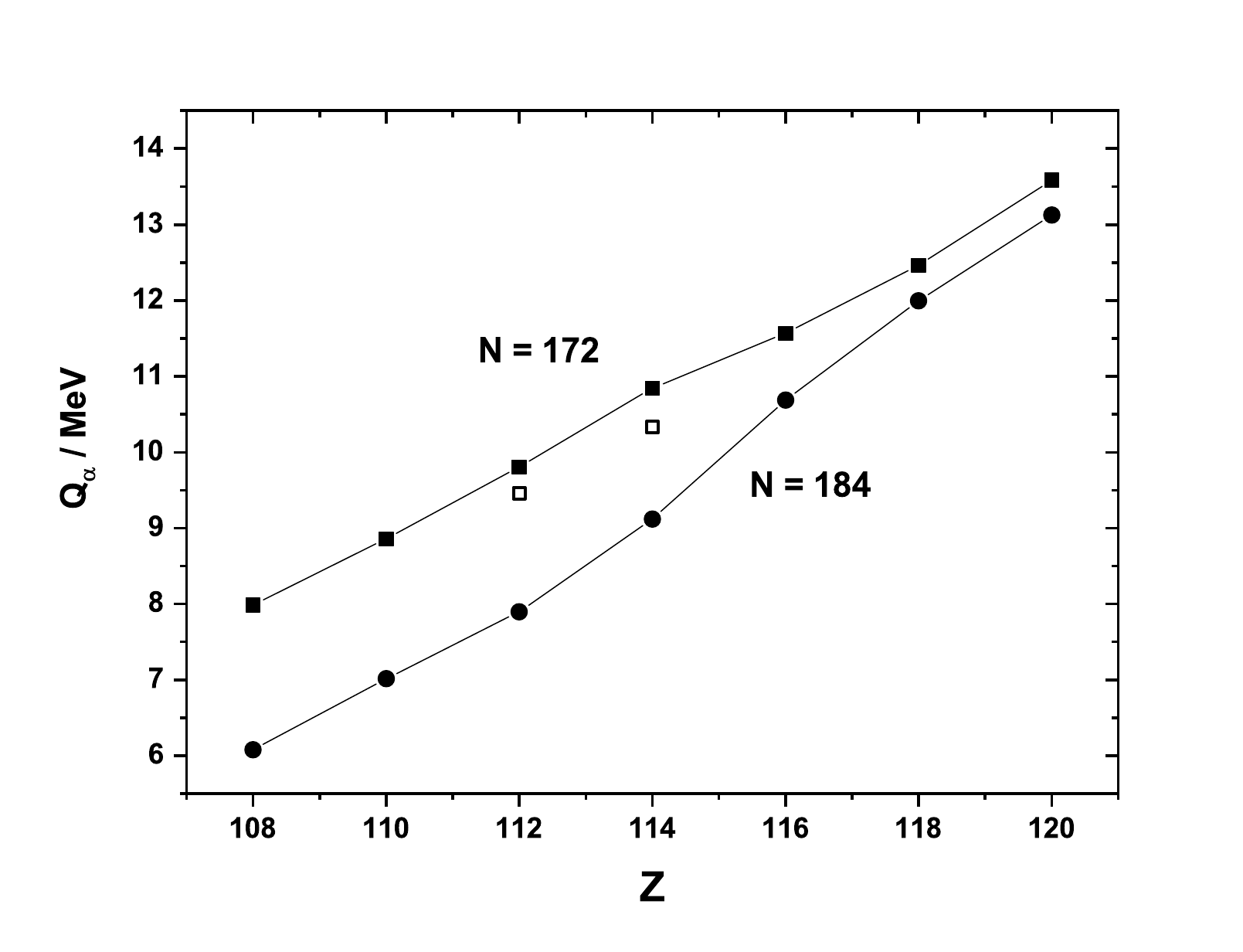}
	}
	% If not, use
	%\vspace{5cm}       % Give the correct figure height in cm
	\caption{Predicted Q$_{\alpha}$ values along the N\,=\,172 and N\,=\,184 isotones lines \cite{Smolan95}. The  
	experimental  Q$_{\alpha}$ values for $^{286}$Fl and $^{284}$Cn (data from \cite{Sarm21}) are shown by the open squares.}
	\label{fig:37}       % Give a unique label
\end{figure}

\begin{figure}
	%\begin{figure*}
	\resizebox{0.8\textwidth}{!}{%
		\includegraphics{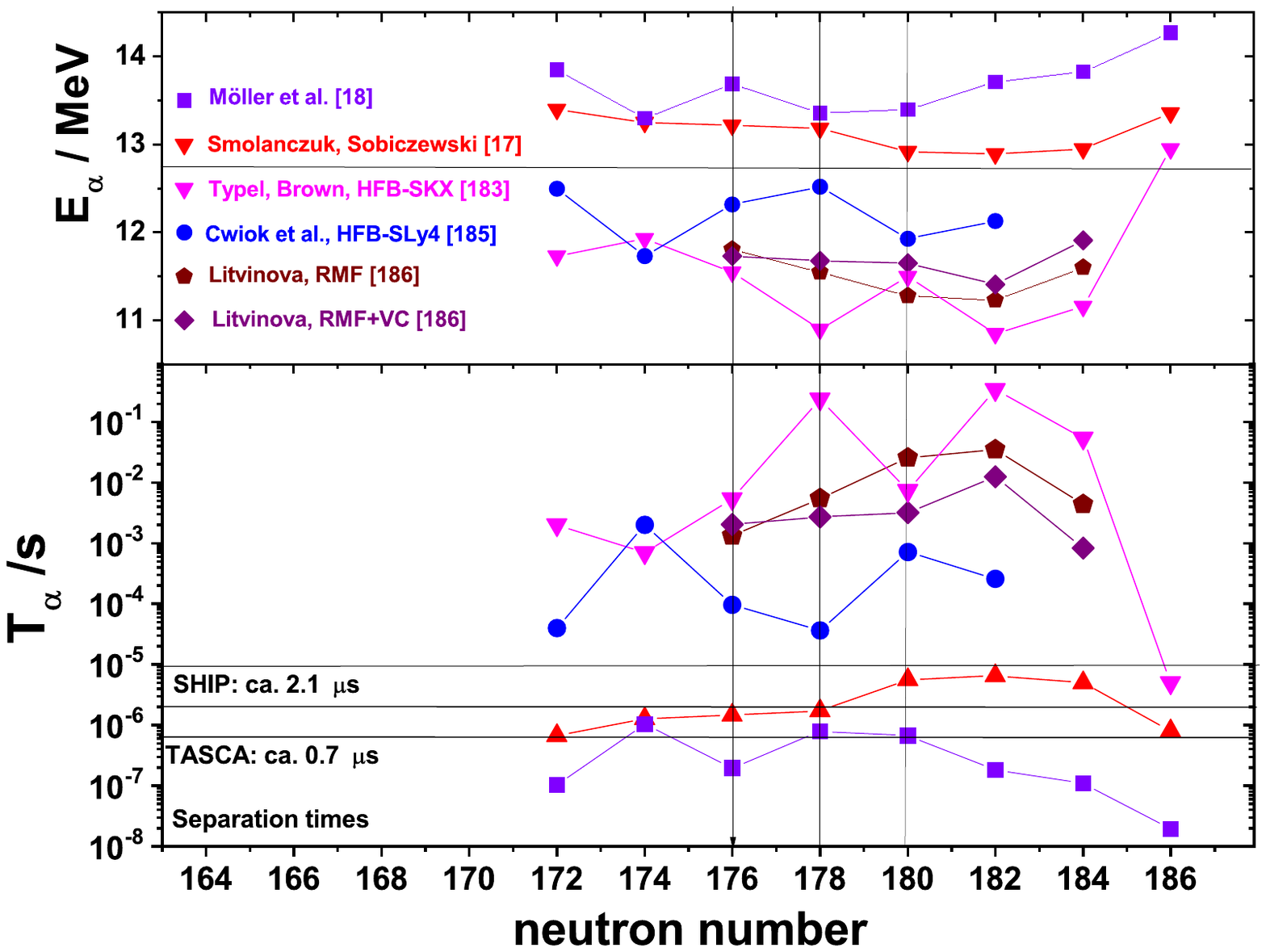}
	}
	% If not, use
	%\vspace{5cm}       % Give the correct figure height in cm
	\caption{Comparison of Q$_{\alpha}$ values and halflives for element 120 isotopes from different models. See text for details.}
	\label{fig:38}       % Give a unique label
\end{figure}

From this point of view it rather seems useful to take the $\alpha$ decay properties as a signature for a shell, as discussed in sect. 7.2. However, one has to note that strictly spoken even - even nuclei have to be considered since only for those a ground-state to ground-state transition can be assumed a priori as the strongest decay line. However one is presently not only confronted with the lack of experimental data. The situation is shown in fig. 37 where predicted Q$_{\alpha}$ values for the N\,=\,172 and N\,=\,184 isotones are presented. Different to the situation at Z\,=\,82 (see fig. 28) calculations of Smolanczuk et al. \cite{Smolan95} predicting Z\,=\,114 as proton shell
result only in a rather small change in decrease of Q$_{\alpha}$ values when crossing the shell, even at the predicted neutron shell at N\,=\,184 compared to the heavier and lighter isotones. 
At N\,=\,172 there is practically no effect any more, one gets a more or less straight decrease of the 
Q$_{\alpha}$ values. So probably Q$_{\alpha}$ values could not be
applied for identifying Z\,=\,114 as a proton shell even if more data would be available.\\
A possibility to decide whether the proton shell is located Z\,=\,114 or Z\,=\,120 results from comparison of experimental 
Q$_{\alpha}$ values and halflives with results from models predicting either Z\,=\,114 or Z\,=\,120. However, one has to consider a large straggling 
of the predicted
values, so it is required to produce and to investigate nuclei in a region, where uncertainties of the predictions are 
larger than the results from models predicting either Z\,=\,114 or Z\,=\,120 as proton shells.
An inspection of the different models shows that element 120 seems to be first one where the differences are so large that the quest of the proton shell can be 
answered with some certainty.
The situation is shown in fig. 38, where predicted Q$_{\alpha}$ values and calculated halflives are compared.
Despite the large straggling of the predicted $\alpha$  energies and halflives there is seemingly a borderline at E$_{\alpha}$\,=\,12.75 MeV evident between models precting Z\,=\,120 als proton shell \cite{Typel03,Cwiok05,Cwiok99,Litvinova12} and those predicting Z\,=\,114 as proton shell \cite{Smolan95,Moller95} while halflives $<$10$^{-5}$ s hint to Z\,=\,114, halflives $>$10$^{-5}$ s to Z\,=\,120 as proton shell. This feature makes synthesis of element 120 even more interesting than the synthesis of element 119. Suited reactions to produce an even-even isotope of element 120 seem $^{50}$Ti($^{249}$Cf,3n)$^{296}$120 (N=176), $^{54}$Cr($^{248}$Cm,2n,4n)$^{298,300}$120 (N=178,180). Expected cross-sections are, however small.  \\
V.I. Zagrebaev and W. Greiner \cite{ZagG08} predicted cross sections of $\sigma$\,$\approx$\,25 fb for $^{54}$Cr($^{248}$Cm,4n)$^{298}$120 and a slightly higher
value of $\sigma$\,$\approx$\,40 fb for $^{50}$Ti($^{249}$Cf,3n)$^{296}$120. So far only few experiments on synthesis of element 120 reaching cross sections limits
below 1 pb have been performed: $^{64}$Ni + $^{238}$U at SHIP, GSI with $\sigma$$<$0.09 pb \cite{HofA08},  $^{54}$Cr + $^{248}$Cm at SHIP, GSI with $\sigma$$<$0.58 pb \cite{HofH16}, $^{50}$Ti + $^{249}$Cf at TASCA, GSI with $\sigma$$<$0.2 pb \cite{Khuyag20}, and  $^{58}$Fe + $^{244}$Pu at DGFRS, JINR Dubna with $\sigma$$<$0.5 pb \cite{OganU09}.

\section{\bf{9. Challenges / Future}}

There are two major problems concerning the experimental techniques used in the investigation of superheavy elements.
The first is connected with the implantation of the reaction products into silicon detectors which are also used to measure
the $\alpha$-decay energy, conversion electrons and fission products. This simply means that, {\it{e.g.}} in the case of 
$\alpha$ decay not only the kinetic energy of the $\alpha$-particle is measured but also part of the recoil energy transferred by the $\alpha$ particle to the residual nucleus. Due to the high ionisation density in the stopping process of the heavy residual nucleus and partial recombination of the charge carriers, typically only about one third of the recoil energy is measured
\cite{Eyal82}. It results in energy shift of the $\alpha$-decay energy by $\approx$50 keV, which can be compensated by a proper calibration, and a deterioration of the energy resolution of the detector by typically 5\,-\,10 keV.\\
A second item is more severe. It is connected with populating excited levels in nuclei decaying promptly (with life-times of some $\mu$s or lower) by internal conversion. In these cases energy summing of $\alpha$ particles with conversion electrons (and also low energy X-rays and Auger electrons from deexcitation of the atomic shell) is observed \cite{HessH87}. The influence on the measured $\alpha$ spectra is manifold, depending also on the energy of the conversion electrons; essentially are broadening and shifting the $\alpha$ energies often washing out peak structures of $\alpha$ decay pattern. An illustrative case is $\alpha$ decay of $^{255}$No
which has been investigated using the implantation technique \cite{Hess06} and the He-jet technique with negligible probability of energy summing \cite{Asai11}. Specifically, different low lying members of the same rotational are populated, which decay by practically completely converted M1 or E2 transitions towards the band-head. These fine structures of the $\alpha$ decay spectrum cannot be resolved using the implantation technique (see also \cite{Asai15}). Although in recent years successful attempts have been untertaken to model those influences by GEANT - simulations \cite{Hess12}, direct measurements are preferred from experimental side. First steps in this direction have been recently undertaken by coupling an ion trap \cite{Rudo10} or an MRTOF - system \cite{Schury17} to a recoil separator and the BGS + FIONA system, which was used
to directly measure mass number of $^{288}$Mc \cite{Gates18}.\\ 
Also mass number measurement is an interesting feature, the ultimate goal is a save {\it{Z}} and {\it{A}} identification of a nuclide. This can be 
achieved via high precision mass measuremts, allowing for clear isobaric separation (ion traps and possibly also MRTOF - systems).
Presently limits are set by the production rate. \\
The most direct method to determine the atomic number of a nucleus is measuring characteristic X-rays in prompt or delayed coincidence with its radioactive decay.
Such measurements are, however, a gamble as they need both, highly K - converted transitions (M1, M2) with transition energies above the K - binding energy.
The latter is not a trivial problem as energies raise steadily and are in the order of 180 keV at {\it{Z}}\,=\,110.
Such measurements have been applied so far up bohrium ({\it{Z}} = 107 \cite{Hess09}). In the region of superheavy nuclei ({\it{Z}}\,$>$\,112) such attempts have been recently performed by D. Rudolph et al. \cite{RuF13} and J. Gates et al. \cite{GaG15} by investigating the $\alpha$ decay chains starting from the odd-odd nucleus $^{288}$Mc ({\it{Z}}\,=\,115), but no positive result was obtained.\\
Alternatively one can attempt to measure L - X-rays to have excess to lower energies and also to E2 transitions. Such measurement have been performed successfully 
up to {\it{Z}}\,=\,105 \cite{Bemis77}, but are more complicated due to the more complex structure of the L X-ray spectra.\\
An alternative method for X-ray identification is measuring the X-rays emitted during electron capture (EC) decay in delayed coincidence
with $\alpha$ decay or spontaneous fission of the daughter nucleus. This technique has been recently for the first time successfully applied in the transactinide region \cite{Hess16a}, by measuring K$_{\alpha}$ and 
K$_{\beta}$ X-rays from EC - decay of $^{258}$Db in delayed coincidence with spoantaneous fission and $\alpha$ decay of the daughter nucleus $^{258}$Rf.
Application in the SHN region seems possible, problems connected with that technique are discussed in sect. 7.3.

%
% BibTeX users please use
% \bibliographystyle{}
% \bibliography{}
%
% Non-BibTeX users please use

\end{document}